%% file: main_arxiv.tex
\documentclass[11pt, a4paper, logo, copyright]{googledeepmind}

\usepackage[authoryear, sort&compress, round]{natbib}
\input{math_commands.tex}

\usepackage{tabularx}
\usepackage{booktabs}
\usepackage{seqsplit}
\usepackage{times}
\usepackage{enumitem}
\usepackage{latexsym}
\usepackage[T1]{fontenc}
\usepackage[utf8]{inputenc}
\usepackage{microtype}
\usepackage{inconsolata}
\usepackage{graphicx}
\usepackage{balance}       % to better equalize the last page
\usepackage{graphics}      % for EPS, load graphicx instead 
\usepackage{hyperref}
\usepackage{color}
\usepackage{booktabs}
\usepackage{textcomp}
\usepackage{subcaption}
\usepackage{xcolor}
\usepackage{lipsum}% http://ctan.org/pkg/lipsum
\usepackage{makecell}
\usepackage{multicol}
\usepackage{multirow}
\usepackage{array}
\usepackage{verbatimbox}
\usepackage{enumitem}
\usepackage{amsmath}
\usepackage{float}
\usepackage{stfloats}
\usepackage{graphicx}
\usepackage{amsthm}
\usepackage{listings}
\usepackage{caption} 
\usepackage[export]{adjustbox}
\usepackage{xspace}
\usepackage{epsfig}
\usepackage[linesnumbered]{algorithm2e}
\usepackage{algpseudocode}
\usepackage{tabularx}
\usepackage{arydshln}
\usepackage[bottom]{footmisc}
\usepackage{tcolorbox}
\usepackage{stackengine}
\usepackage{placeins}
\usepackage{rotating}
\usepackage{booktabs}
\usepackage{multirow}
\usepackage{amsmath}
\usepackage[normalem]{ulem}
\useunder{\uline}{\ul}{}
\tcbuselibrary{breakable}

\definecolor{E+F}{RGB}{	255, 99, 71}
\definecolor{B+F}{RGB}{255, 165, 0}
\definecolor{E+I}{RGB}{	173, 216, 230}
\definecolor{B+I}{RGB}{	30, 144, 255}
\definecolor{D}{RGB}{	60, 179, 113}

\definecolor{maroon}{cmyk}{0,0.87,0.68,0.32}

\usepackage{tikz}

\definecolor{darkgreen}{rgb}{0.0, 0.5, 0.0}
\definecolor{usercolor}{RGB}{200, 230, 250} % Lighter pastel blue
\definecolor{coachcolor}{RGB}{180, 250, 180} % Lighter pastel green

\title{Cardiovascular-Kidney-Metabolic Health Age Using Wearables and Blood Biomarkers}
\reportnumber{} % Leave blank if n/a
\renewcommand{\today}{2026-02-10}

\author[1]{Zeinab Esmaeilpour}
\author[1]{A. Ali Heydari}
\author[1]{Daniel McDuff}
\author[1]{Anthony Z. Faranesh}
\author[1]{Conor Heneghan}
\author[1]{Shwetak Patel}
\author[1]{Mark Malhotra}
\author[1]{Cathy Speed}
\author[1]{Javier L. Prieto}
\author[1]{Ahmed A. Metwally}

\affil[1]{Google Research}
\correspondingauthor{zesmaeilpour@google.com, aametwally@google.com}

\input{sections/0-abstract}
\begin{document}

\maketitle
\input{sections/1-introduction}

\input{sections/2-results}

\input{sections/3-discussion}

\input{sections/4-methods}

\section{Acknowledgement}

We are deeply grateful to the Fitbit and Pixel Watch study participants who contributed their data to this research. We thank members of the Consumer Health Research Team at Google for their valuable feedback and technical support throughout this study, in particular Lindsey Sunden and Hulya Emir-Farinas. We thank the software development team who built the service that was used to recruit this large cohort and enabled remote collection of wearables and blood biomarker data; Alex Dan, Alex Badescu, Delia-Georgiana Stuparu, George-Iulian Nitroi, Silviu Grigore, Paul Navin, and Dima Trubnikov. We also thank our collaborators at Quest Diagnostics. 

\section{Data Availability}

The de-identified dataset used in this study is available to approved researchers for reproducibility purposes only. Researchers seeking dataset access must complete the Insulin Resistance Dataset Access Request Form, available here:

\url{https://docs.google.com/forms/d/e/1FAIpQLSebcfCZMQKBua7QS9MSvqr9n-fqmfJvmU4blOHXP1WZC0NFqA/viewform?usp=preview}

\section{Competing Interests}

This study was funded by Google LLC. All authors are employees of Alphabet and may own stock as part of the standard compensation package.

\bibliography{main_arxiv}
\pagebreak

\renewcommand{\thesubsection}{Supplementary Table S\arabic{subsection}}
\setcounter{subsection}{0} % Reset counter

\section*{Supplementary Figures}
% --- These two lines are the "magic" to get Figure S1, S2, etc. ---
\setcounter{figure}{0} % Resets the counter to 0
\renewcommand{\thefigure}{S\arabic{figure}} % Adds the 'S' prefix
% -------------------------------------------------------------

\begin{figure*}[ht]
    \centering
     \caption{Deviance from Normal thresholds for cardiovascular health biomarkers in the study cohort.}
    \includegraphics[width=\textwidth]{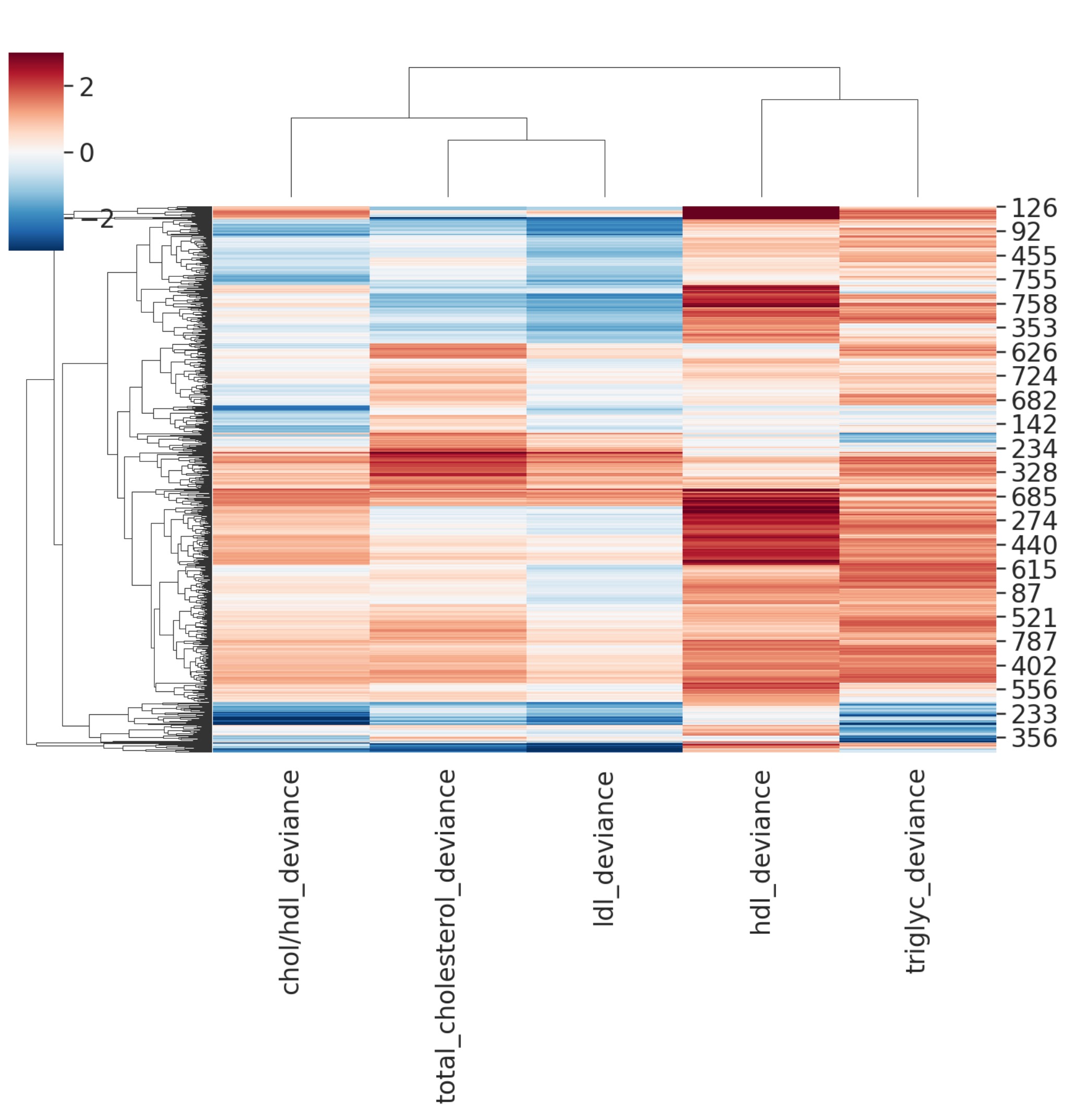}
    \label{fig:supplement_1}
\end{figure*}

\begin{figure*}[ht]
    \centering
     \caption{Distribution of each CKM biomarker.}
    \includegraphics[width=\textwidth]{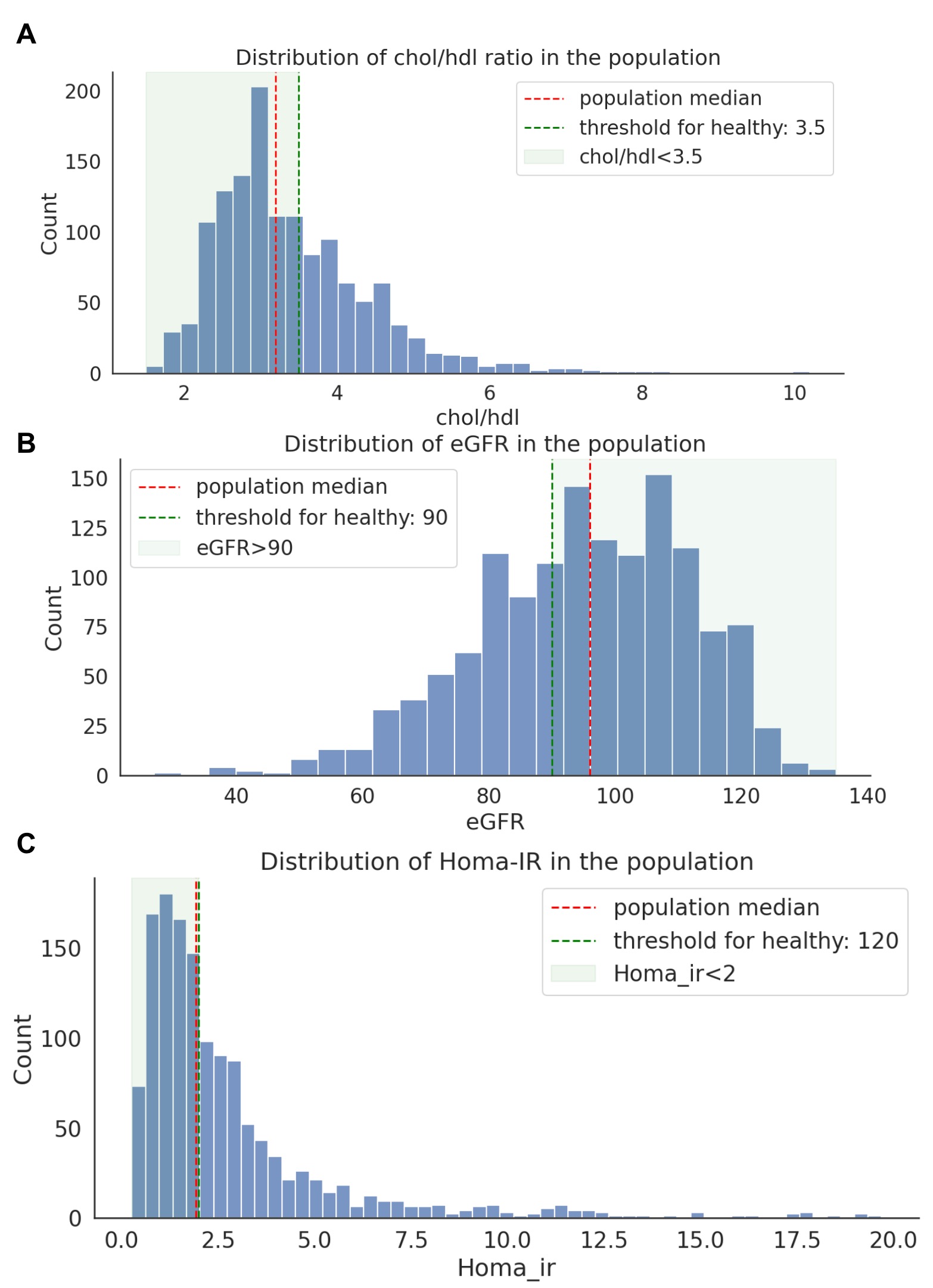}
    \label{fig:supplement_2}
\end{figure*}

\begin{figure*}[ht]
    \centering
     \caption{Most correlated features in each subsystem (Cardiovascular, Kidney and Metabolic health) adjusted by age and gender.}
    \includegraphics[width=\textwidth]{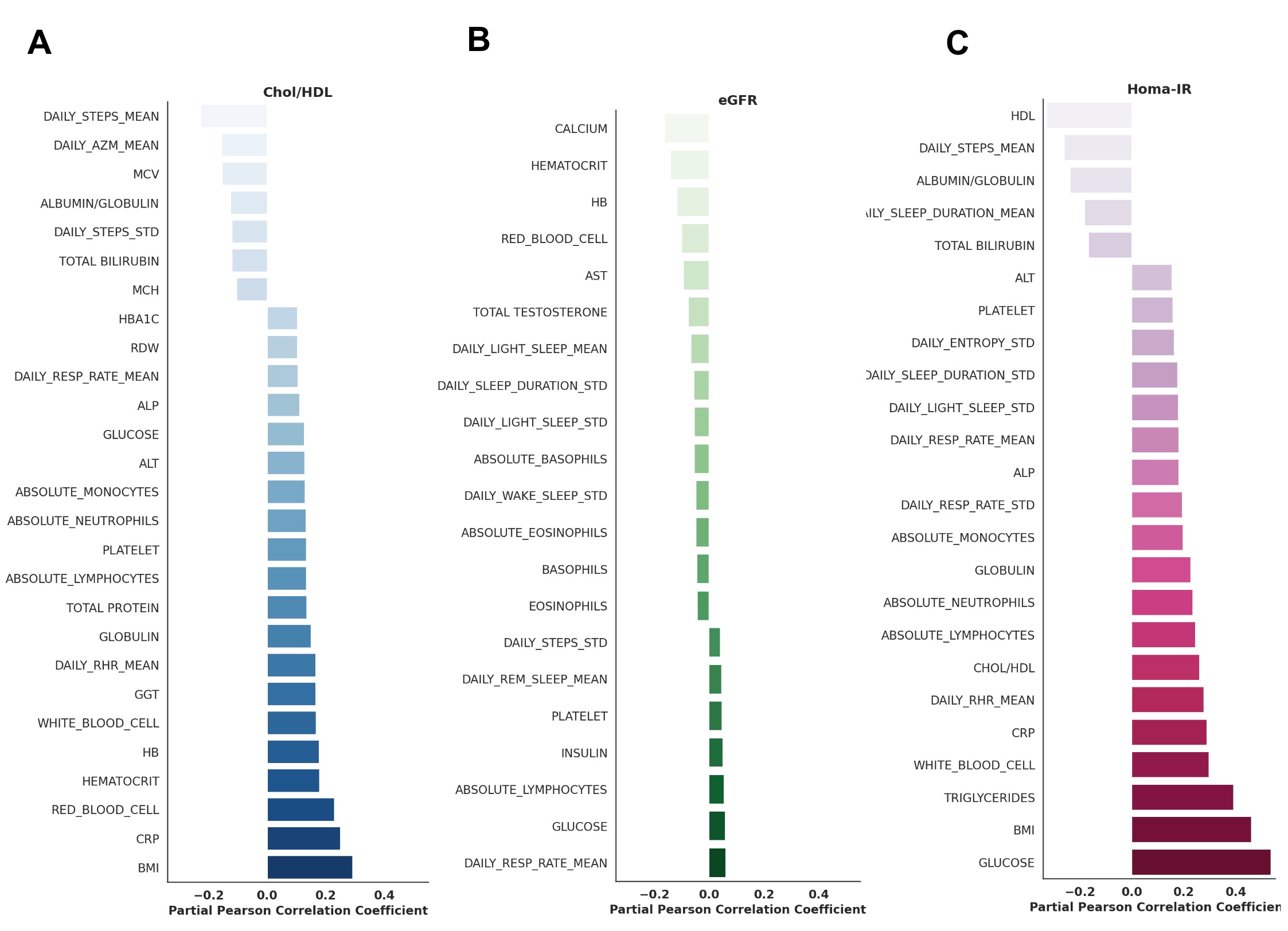}
    \label{fig:supplement_3}
\end{figure*}

\begin{figure*}[ht]
    \centering
     \caption{Relationship between CKM Metrics and Demographic Factors.}
    \includegraphics[width=\textwidth]{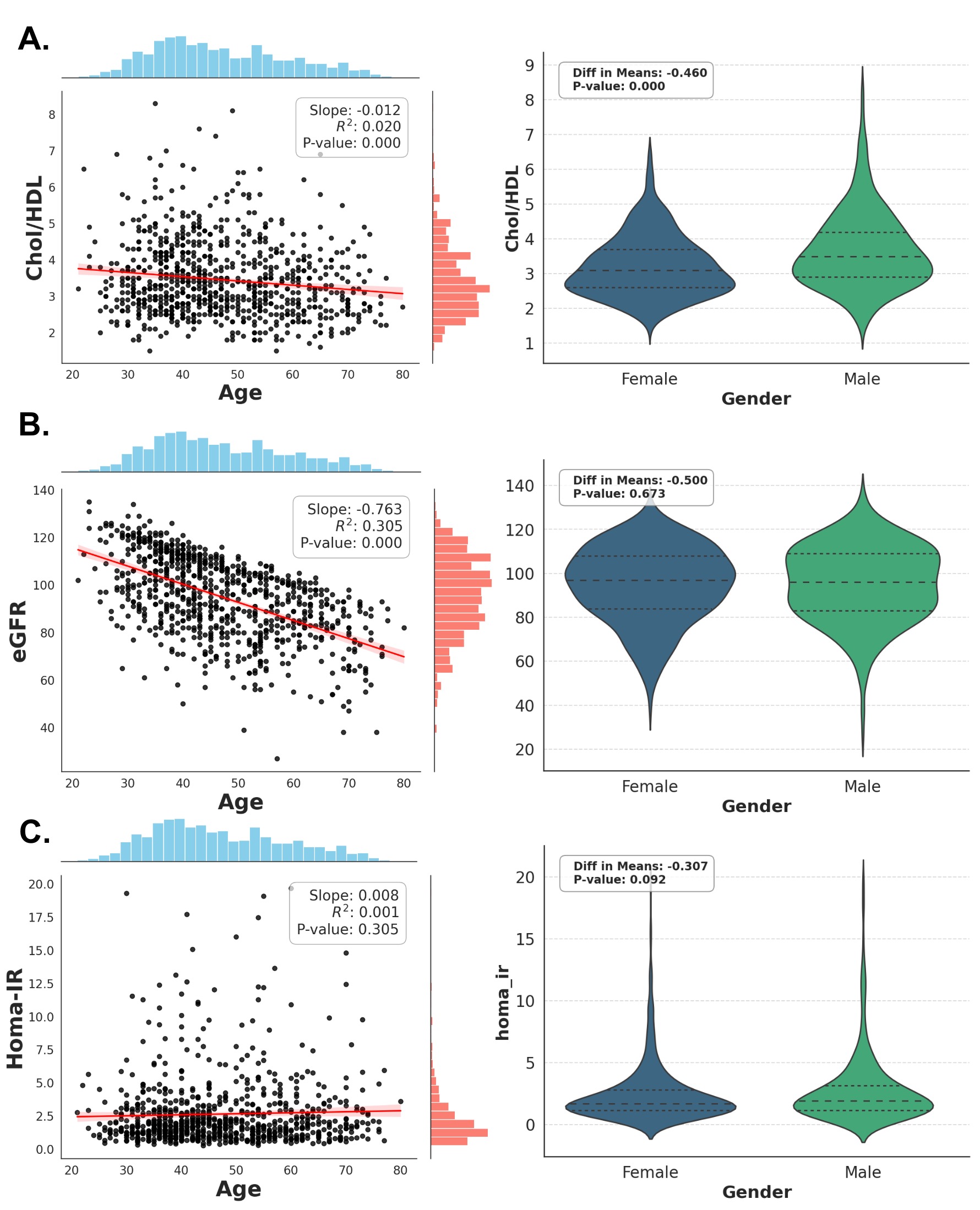}
    \label{fig:supplement_4}
\end{figure*}

\begin{figure*}[ht]
    \centering
     \caption{The Relationship Between eGFR and Deep Sleep: Age-Related Declines and Association in Adults.}
    \includegraphics[width=\textwidth]{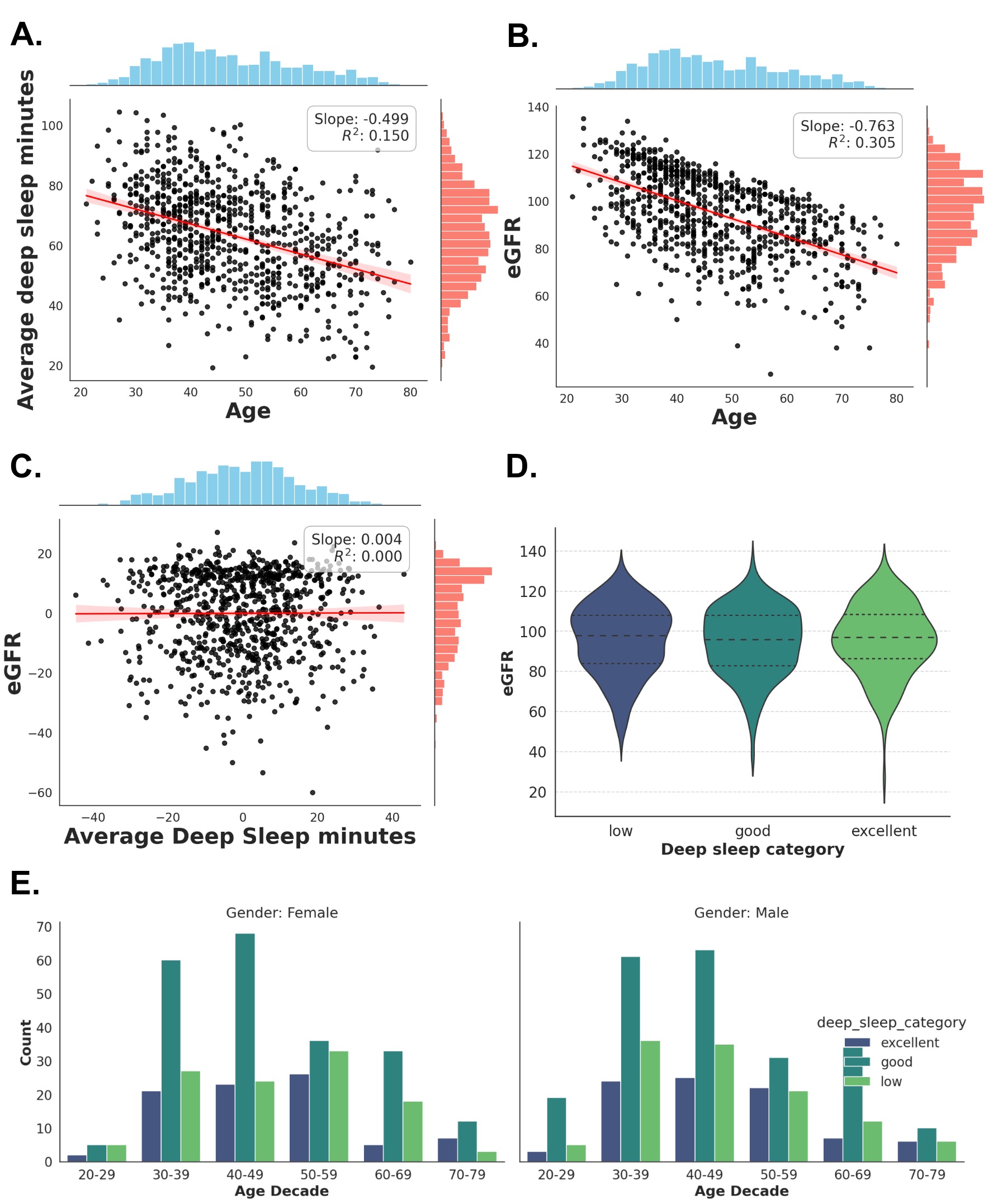}
    \label{fig:supplement_5}
\end{figure*}

\begin{figure*}[ht]
    \centering
     \caption{Population data for age and mean deep sleep.}
    \includegraphics[width=\textwidth]{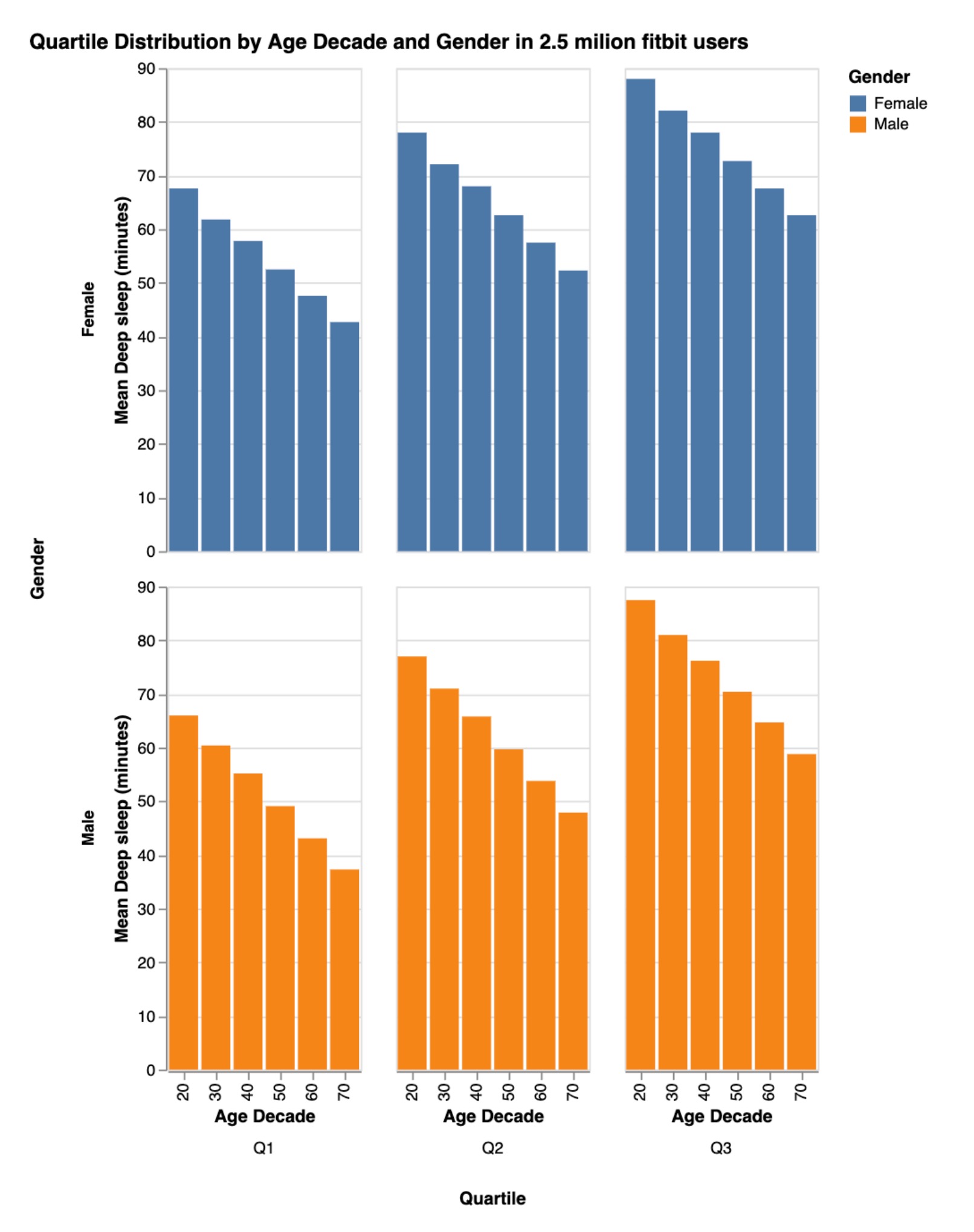}
    \label{fig:supplement_6}
\end{figure*}

\begin{figure*}[ht]
    \centering
     \caption{Population data for age and mean sleep duration.}
    \includegraphics[width=\textwidth]{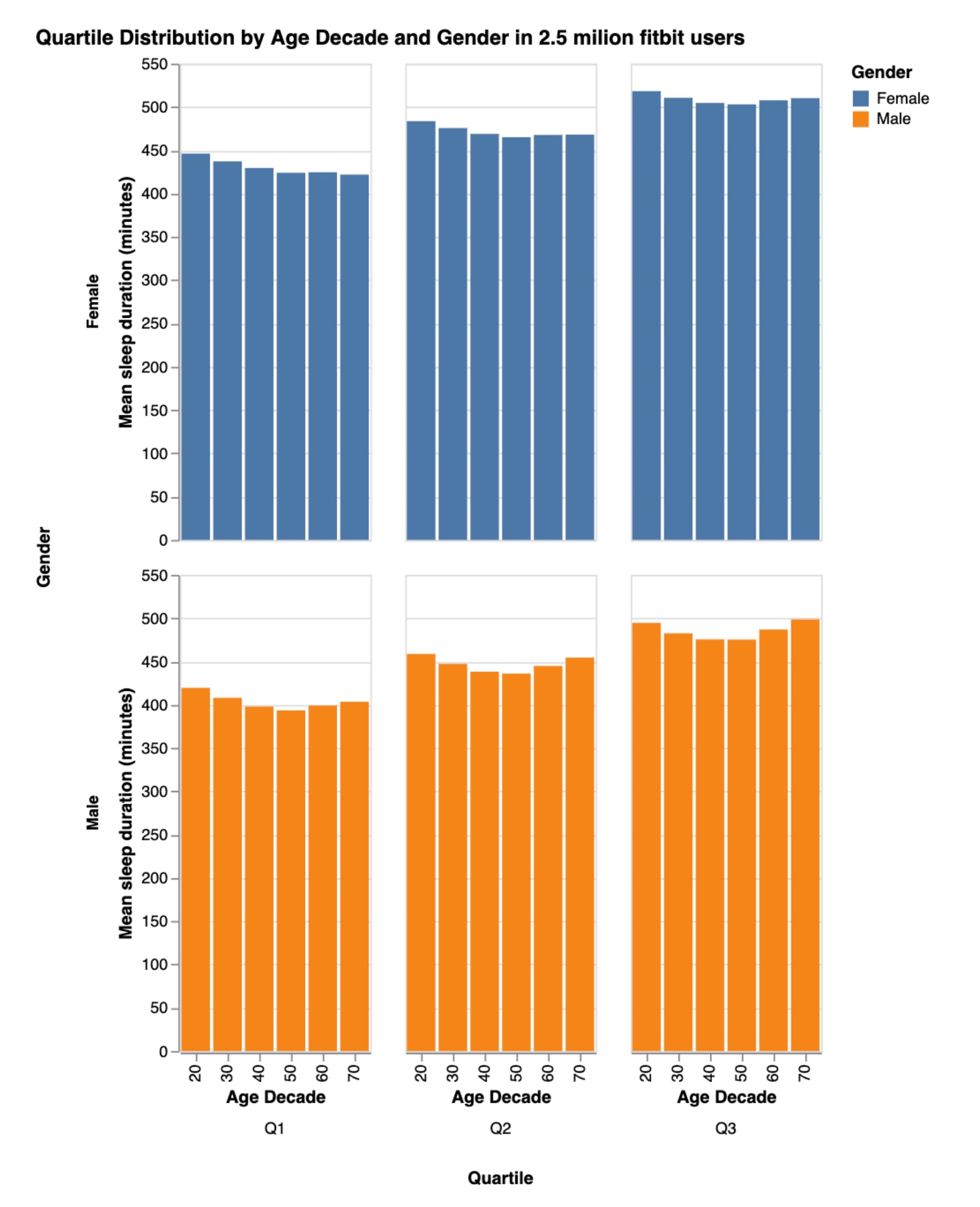}
    \label{fig:supplement_7}
\end{figure*}

\begin{figure*}[ht]
    \centering
     \caption{ Copmaring regression performance of wearable statistics versus WFM embeddings across CKM.}
    \includegraphics[width=\textwidth]{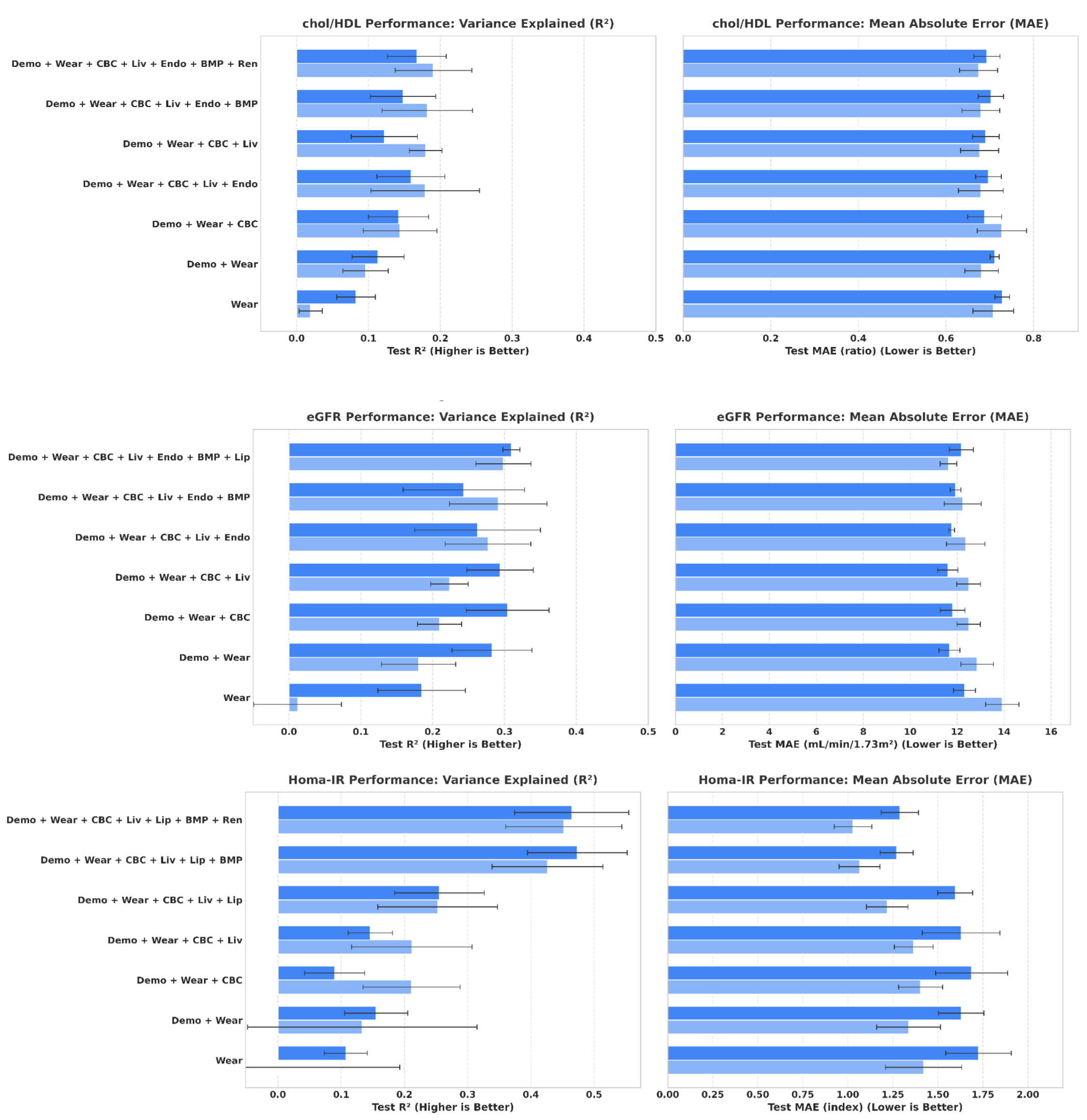}
    \label{fig:supplement_8}
\end{figure*}

\begin{figure*}[ht]
    \centering
     \caption{Classification performance using WFM embeddings and blood biomarkers.}
    \includegraphics[width=\textwidth]{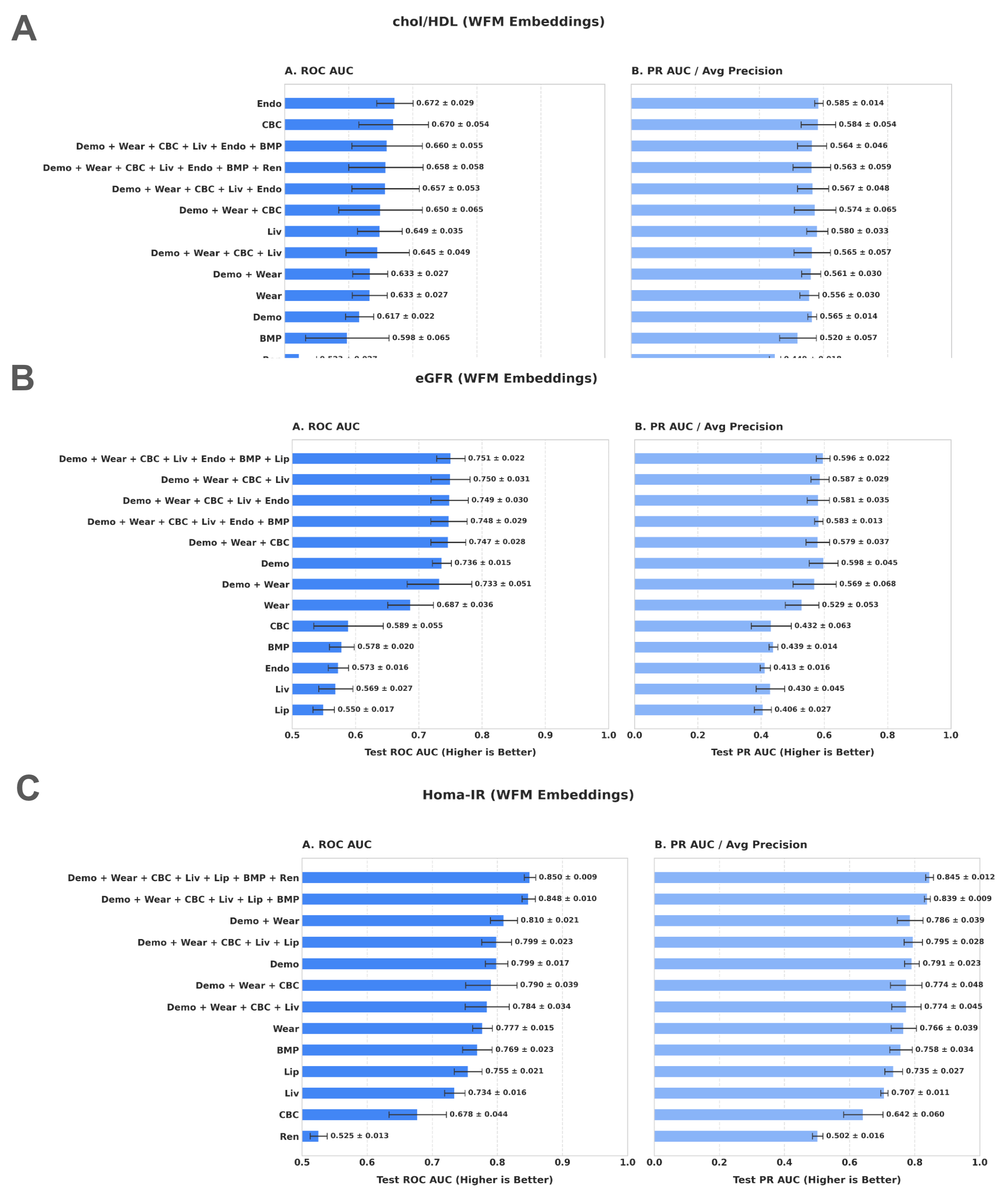}
    \label{fig:supplement_9}
\end{figure*}

\begin{figure*}[ht]
    \centering
     \caption{Classification performance using wearable statistics and blood biomarkers.}
    \includegraphics[width=\textwidth]{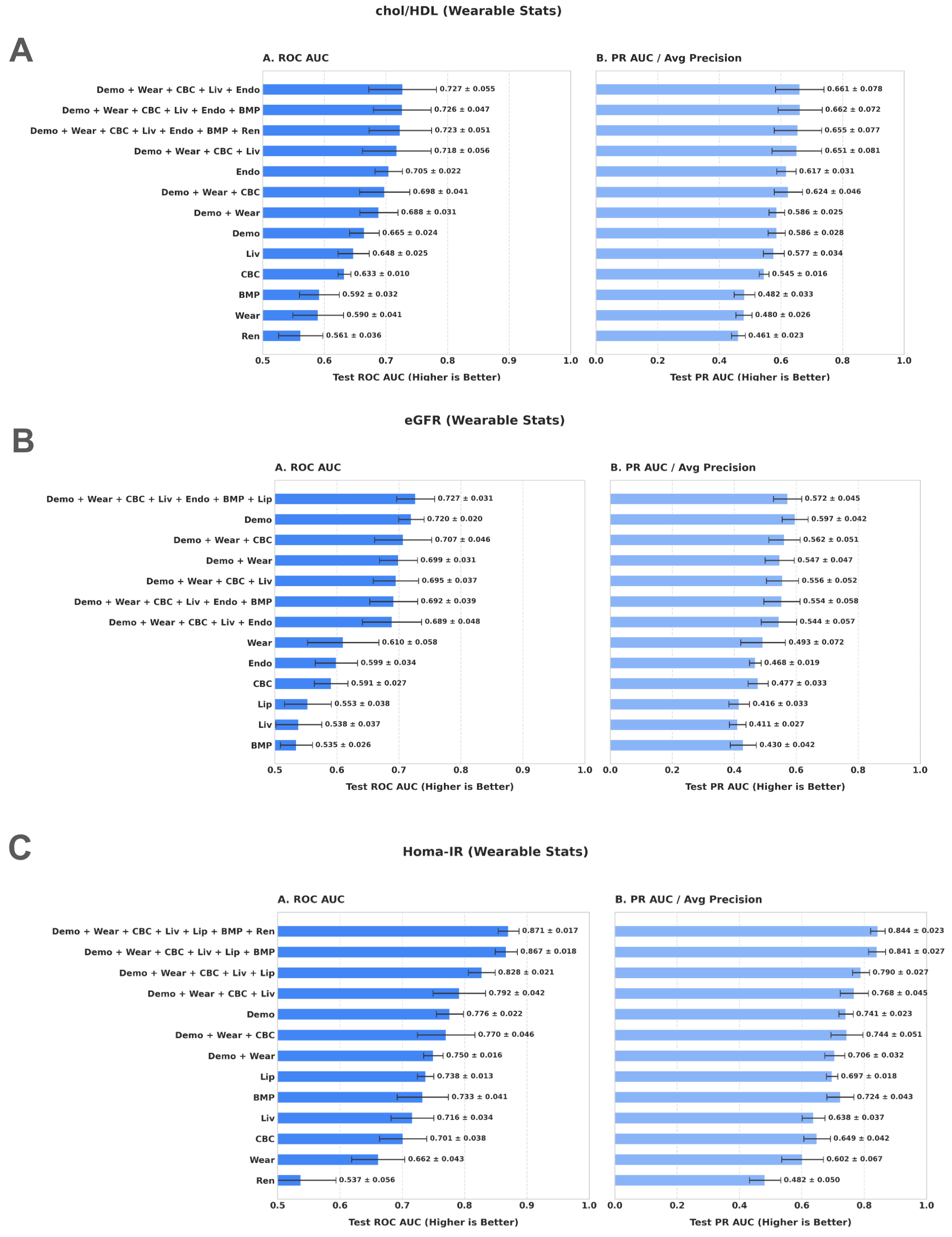}
    \label{fig:supplement_10}
\end{figure*}

\begin{figure*}[ht]
    \centering
     \caption{Comparative classification performance of wearable statistics versus WFM embeddings across CKM.}
    \includegraphics[width=\textwidth]{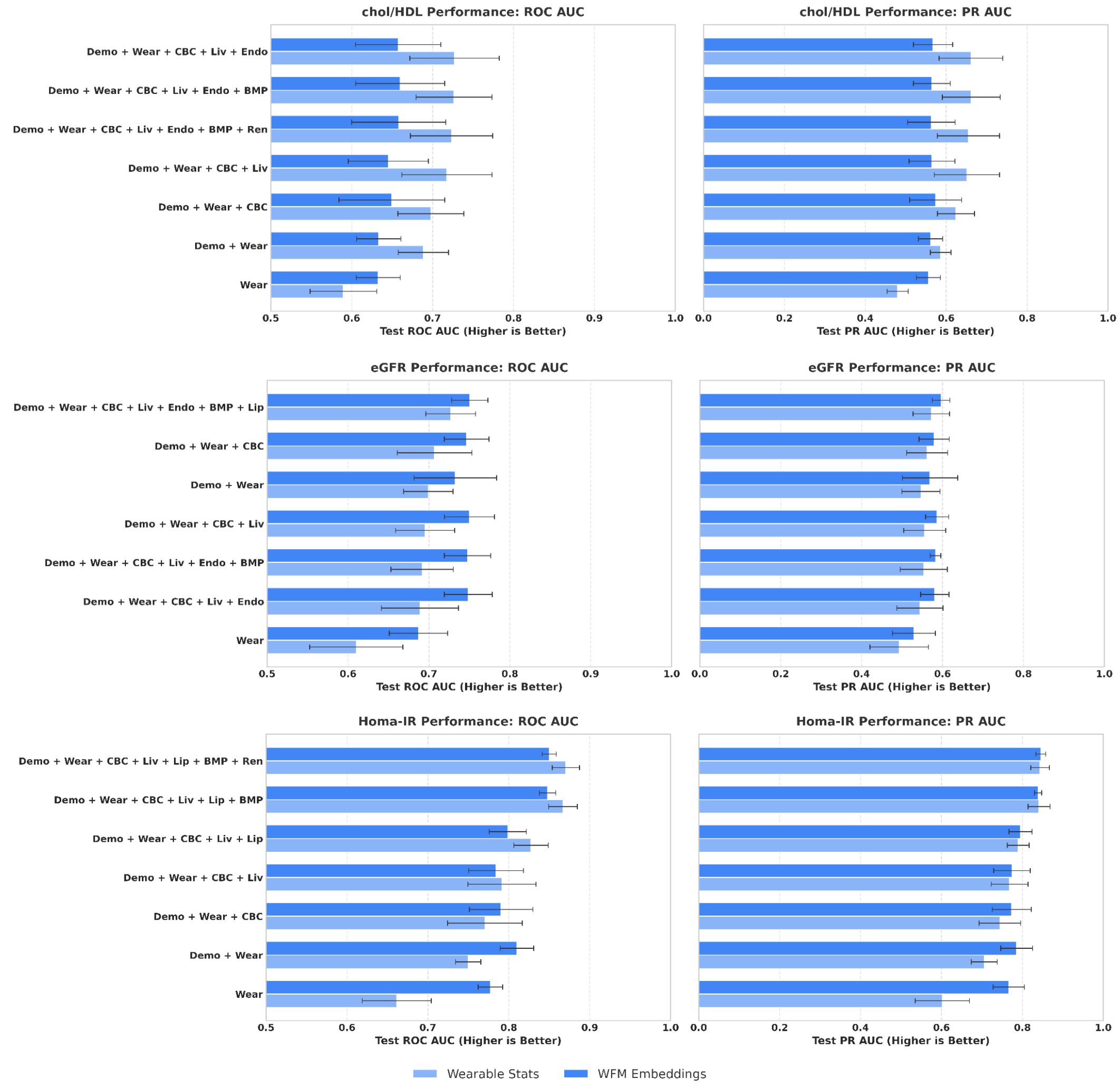}
    \label{fig:supplement_11}
\end{figure*}

\begin{figure*}[ht]
    \centering
     \caption{Impact of calculation window size on sample size, population distribution and individual stability of behavioral and physiological features}
    \includegraphics[width=\textwidth]{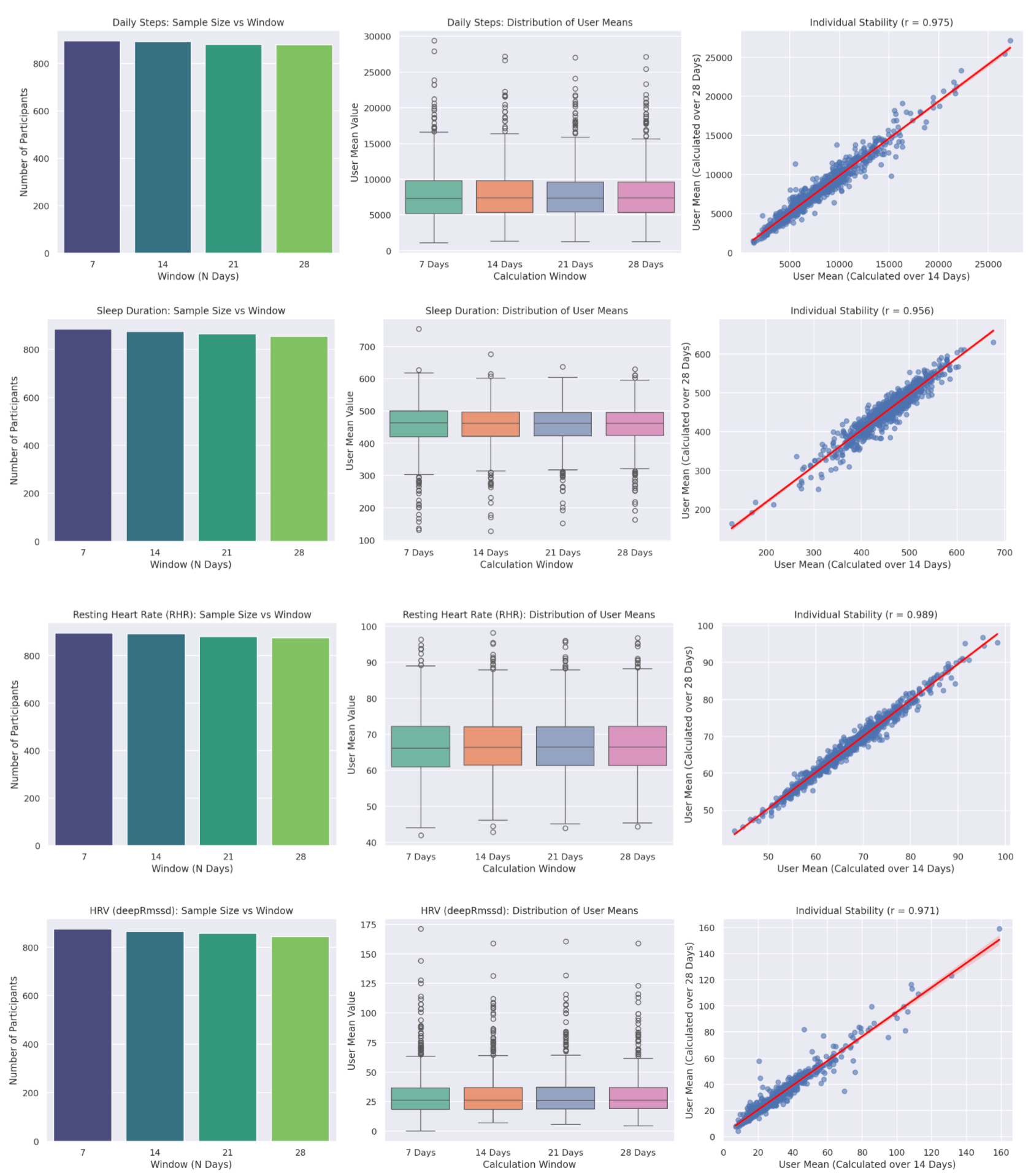}
    \label{fig:supplement_12}
\end{figure*}

\begin{figure*}[ht]
    \centering
     \caption{Cohort selection and data filtering pipeline for blood biomarkers and wearable statistical metrics dataset.}
    \includegraphics[width=\textwidth]{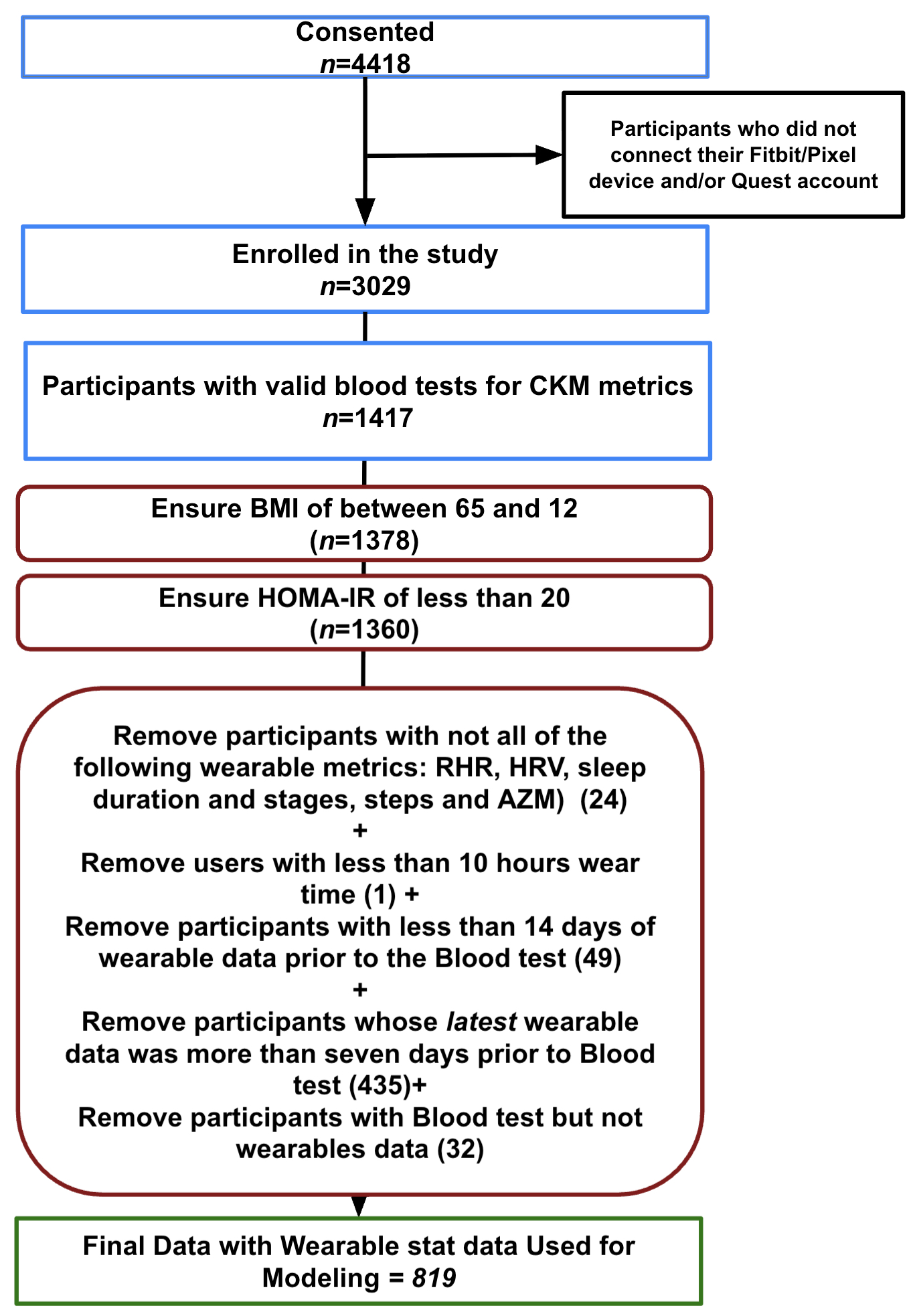}
    \label{fig:supplement_13}
\end{figure*}

\begin{figure*}[ht]
    \centering
     \caption{Cohort selection and data filtering pipeline for blood biomarkers and WFM embeddings dataset.}
    \includegraphics[width=\textwidth]{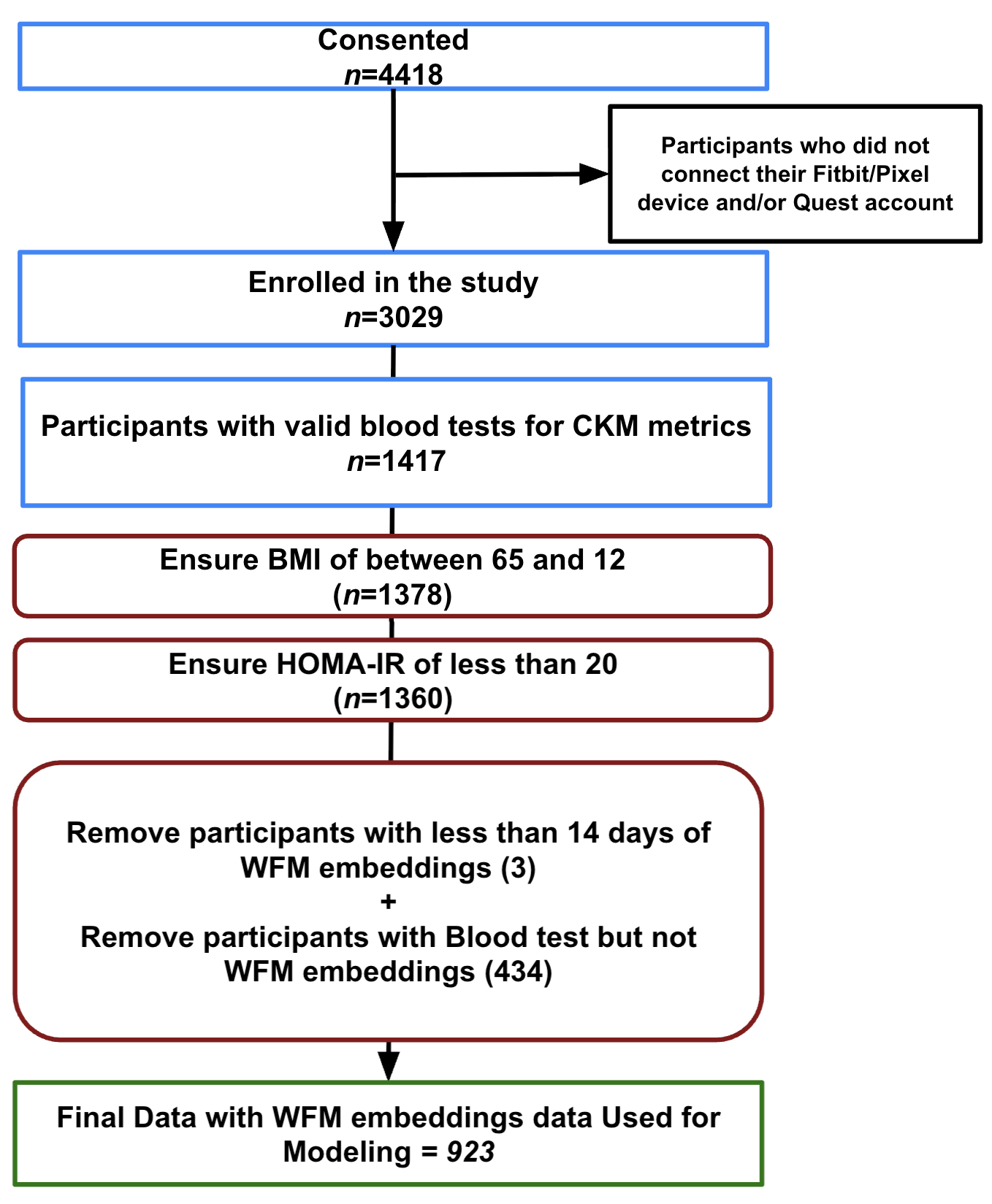}
    \label{fig:supplement_14}
\end{figure*}

\begin{figure*}[ht]
    \centering
     \caption{Comparative validation against clinical ground truth adjusted for multi-colinearity among the continuous digital health metrics in Health Age.Pearson correlation coefficients (absolute values, |r|) benchmarking baseline Chronological Age against three iterations of the delta Health Age model, Wearables-Based, CKM-Based (blood biomarkers), and a Combined (Wearables + PHR) model. Ground truths are established using the modified Life’s Essential 7 (LE7; cardiovascular baseline) and the  Metabolic Syndrome Severity Score (MSSS; metabolic baseline).}
    \includegraphics[width=\textwidth]{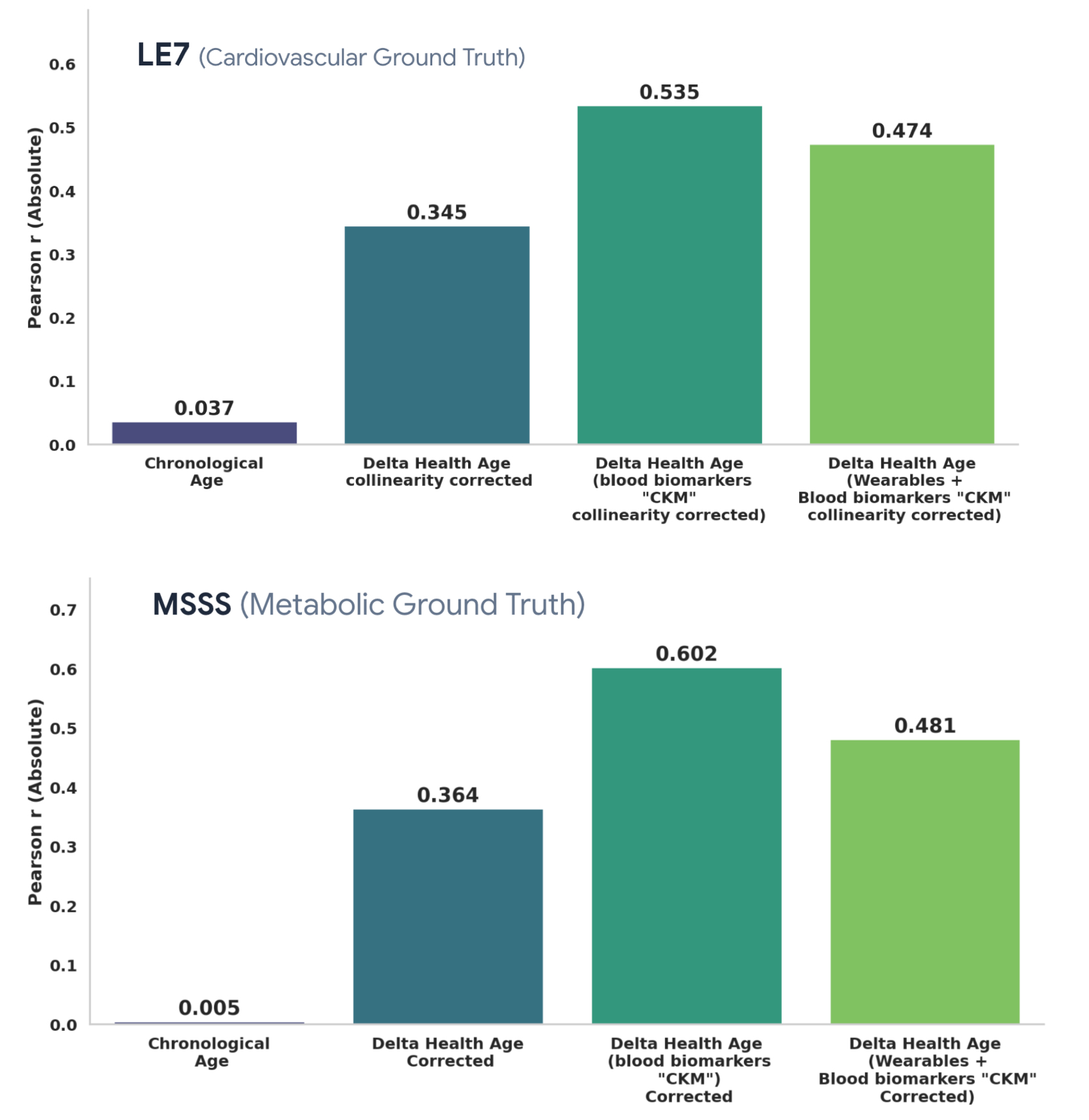}
    \label{fig:supplement_15}
\end{figure*}

\FloatBarrier % Figures cannot go below this, Tables cannot go above this

\section*{Supplementary Tables}
\setcounter{table}{0}
\renewcommand{\thetable}{S\arabic{table}}

\begin{table}[ht]
\centering
\caption{Features Pearson correlation with cardiovascular health status (measured by Chol/HDL ratio)}
\label{tab:s1_pearson_correlation}
\begin{tabularx}{\textwidth}{@{}Xccc@{}}
\toprule
\textbf{Feature} & \textbf{Correlation} & \textbf{p\_value\_corrected} & \textbf{Significance} \\ \midrule
gender number (1 male, 2 female) & -0.19 & <.0001 & TRUE \\ \addlinespace

\multirow{5}{*}{\textbf{mcv}} & -0.17 & <.0001 & TRUE \\
 & -0.15 & <.0001 & TRUE \\
 & -0.14 & <.0001 & TRUE \\
 & -0.12 & <.0001 & TRUE \\
 & -0.11 & <.0001 & TRUE \\ \addlinespace

\multirow{8}{*}{\textbf{mch}} & -0.1 & <.0001 & TRUE \\
 & -0.08 & <.0001 & TRUE \\
 & -0.08 & <.0001 & TRUE \\
 & -0.05 & <.0001 & TRUE \\
 & -0.05 & <.0001 & TRUE \\
 & -0.05 & <.0001 & TRUE \\
 & -0.05 & <.0001 & TRUE \\
 & -0.05 & <.0001 & TRUE \\ \addlinespace

\multirow{6}{*}{\textbf{sodium}} & 0.06 & <.0001 & TRUE \\
 & 0.06 & <.0001 & TRUE \\
 & 0.06 & <.0001 & TRUE \\
 & 0.07 & <.0001 & TRUE \\
 & 0.07 & <.0001 & TRUE \\
 & 0.08 & <.0001 & TRUE \\ \addlinespace

\multirow{6}{*}{\textbf{daily\_deep\_sleep\_std}} & 0.08 & <.0001 & TRUE \\
 & 0.09 & <.0001 & TRUE \\
 & 0.09 & <.0001 & TRUE \\
 & 0.09 & <.0001 & TRUE \\
 & 0.09 & <.0001 & TRUE \\
 & 0.09 & <.0001 & TRUE \\ \addlinespace

\multirow{7}{*}{\textbf{rdw}} & 0.1 & <.0001 & TRUE \\
 & 0.1 & <.0001 & TRUE \\
 & 0.11 & <.0001 & TRUE \\
 & 0.11 & <.0001 & TRUE \\
 & 0.12 & <.0001 & TRUE \\
 & 0.13 & <.0001 & TRUE \\
 & 0.13 & <.0001 & TRUE \\ \addlinespace

\multirow{6}{*}{\textbf{absolute\_lymphocytes}} & 0.14 & <.0001 & TRUE \\
 & 0.14 & <.0001 & TRUE \\
 & 0.14 & <.0001 & TRUE \\
 & 0.15 & <.0001 & TRUE \\
 & 0.19 & <.0001 & TRUE \\
 & 0.2 & <.0001 & TRUE \\ \midrule

crp & 0.2 & <.0001 & TRUE \\
alt & 0.2 & <.0001 & TRUE \\
ggt & 0.25 & <.0001 & TRUE \\
bmi & 0.27 & <.0001 & TRUE \\
hb & 0.27 & <.0001 & TRUE \\
hematocrit & 0.28 & <.0001 & TRUE \\
insulin & 0.32 & <.0001 & TRUE \\
red\_blood\_cell & 0.35 & <.0001 & TRUE \\ \bottomrule
\end{tabularx}
\end{table}

\begin{table}[ht]
\centering
\small % Keeps font readable but compact
\renewcommand{\arraystretch}{0.85} % Tightens row height to ensure single-page fit
\caption{Features Pearson correlation with kidney health status (measured by eGFR)}
\label{tab:S2_kidney_correlation}

\begin{tabularx}{\textwidth}{@{}Xccc@{}}
\toprule
\textbf{Feature} & \textbf{Correlation} & \textbf{p\_value\_corrected} & \textbf{Significance} \\ \midrule

age & -0.16 & <.0001 & TRUE \\ \addlinespace

\multirow{5}{*}{\textbf{calcium}} 
 & -0.16 & <.0001 & TRUE \\
 & -0.15 & <.0001 & TRUE \\
 & -0.14 & <.0001 & TRUE \\
 & -0.14 & <.0001 & TRUE \\
 & -0.13 & <.0001 & TRUE \\ \addlinespace

\multirow{8}{*}{\textbf{mcv}} 
 & -0.12 & <.0001 & TRUE \\
 & -0.11 & <.0001 & TRUE \\
 & -0.1  & <.0001 & TRUE \\
 & -0.09 & <.0001 & TRUE \\
 & -0.09 & <.0001 & TRUE \\
 & -0.09 & <.0001 & TRUE \\
 & -0.08 & <.0001 & TRUE \\
 & -0.07 & 0.0817 & FALSE \\ \addlinespace

\multirow{6}{*}{\textbf{daily\_resp\_rate\_std}} 
 & -0.07 & 0.0930 & FALSE \\
 & -0.07 & 0.0930 & FALSE \\
 & -0.06 & 0.1530 & FALSE \\
 & 0.06  & 0.1570 & FALSE \\
 & 0.07  & 0.0690 & FALSE \\
 & 0.08  & 0.0591 & FALSE \\ \addlinespace

\multirow{6}{*}{\textbf{white\_blood\_cell}} 
 & 0.08 & 0.0580 & FALSE \\
 & 0.08 & 0.0560 & FALSE \\
 & 0.08 & 0.0542 & FALSE \\
 & 0.09 & <.0001 & TRUE  \\
 & 0.1  & <.0001 & TRUE  \\
 & 0.1  & <.0001 & TRUE  \\ \addlinespace

\multirow{7}{*}{\textbf{daily\_rhr\_std}} 
 & 0.1  & <.0001 & TRUE \\
 & 0.11 & <.0001 & TRUE \\
 & 0.12 & <.0001 & TRUE \\
 & 0.14 & <.0001 & TRUE \\
 & 0.14 & <.0001 & TRUE \\
 & 0.15 & <.0001 & TRUE \\
 & 0.16 & <.0001 & TRUE \\ \addlinespace

\multirow{5}{*}{\textbf{daily\_resp\_rate\_mean}} 
 & 0.16 & <.0001 & TRUE \\
 & 0.2  & <.0001 & TRUE \\
 & 0.22 & <.0001 & TRUE \\\bottomrule

\end{tabularx}
\end{table}

% Required in preamble:
% \usepackage{booktabs}
% \usepackage{tabularx}

\begin{table}[ht]
\centering
\small % Keeps the text compact to ensure it fits on one page
\renewcommand{\arraystretch}{0.9} % Slightly tightens the rows
\caption{Features Pearson correlation with metabolic health status (measured by HOMA-IR)}
\label{tab:S3_metabolic_correlation}

\begin{tabularx}{\textwidth}{@{}Xccc@{}}
\toprule
\textbf{Feature} & \textbf{Pearson correlation} & \textbf{p\_value\_corrected} & \textbf{Significance} \\ \midrule
hdl                           & -0.31 & <.0001 & TRUE \\
daily\_steps\_mean            & -0.24 & <.0001 & TRUE \\
albumin/globulin              & -0.22 & <.0001 & TRUE \\
daily\_sleep\_duration\_mean  & -0.18 & <.0001 & TRUE \\
total bilirubin               & -0.14 & <.0001 & TRUE \\
basophils                     & -0.14 & <.0001 & TRUE \\
daily\_entropy\_mean          & -0.14 & <.0001 & TRUE \\
mcv                           & -0.14 & <.0001 & TRUE \\
mch                           & -0.13 & <.0001 & TRUE \\
daily\_deep\_sleep\_mean      & -0.11 & <.0001 & TRUE \\
daily\_rem\_sleep\_std        & 0.1   & <.0001 & TRUE \\
daily\_hrv\_std               & 0.1   & <.0001 & TRUE \\
absolute\_eosinophils         & 0.11  & <.0001 & TRUE \\
total protein                 & 0.12  & <.0001 & TRUE \\
platelet                      & 0.12  & <.0001 & TRUE \\
daily\_deep\_sleep\_std       & 0.12  & <.0001 & TRUE \\
red\_blood\_cell              & 0.13  & <.0001 & TRUE \\
rdw                           & 0.14  & <.0001 & TRUE \\
ggt                           & 0.15  & <.0001 & TRUE \\
daily\_entropy\_std           & 0.15  & <.0001 & TRUE \\
daily\_resp\_rate\_mean       & 0.16  & <.0001 & TRUE \\
alt                           & 0.16  & <.0001 & TRUE \\
daily\_sleep\_duration\_std   & 0.17  & <.0001 & TRUE \\
alp                           & 0.18  & <.0001 & TRUE \\
daily\_light\_sleep\_std      & 0.18  & <.0001 & TRUE \\
daily\_resp\_rate\_std        & 0.2   & <.0001 & TRUE \\
absolute\_monocytes           & 0.2   & <.0001 & TRUE \\
globulin                      & 0.21  & <.0001 & TRUE \\
absolute\_neutrophils         & 0.22  & <.0001 & TRUE \\
absolute\_lymphocytes         & 0.23  & <.0001 & TRUE \\
daily\_rhr\_mean              & 0.25  & <.0001 & TRUE \\
chol/hdl                      & 0.26  & <.0001 & TRUE \\
crp                           & 0.27  & <.0001 & TRUE \\
white\_blood\_cell            & 0.28  & <.0001 & TRUE \\
triglycerides                 & 0.4   & <.0001 & TRUE \\
bmi                           & 0.44  & <.0001 & TRUE \\
glucose                       & 0.53  & <.0001 & TRUE \\ \bottomrule
\end{tabularx}
\end{table}

% Ensure these are in your preamble:
% \usepackage{booktabs}
% \usepackage{tabularx}

\begin{table}[ht]
\centering
\small % Keeps the table compact
\renewcommand{\arraystretch}{0.85} % Tightens row spacing to ensure single-page fit
\caption{Features Pearson correlation with cardiovascular health status (measured by Chol/HDL ratio) adjusted for Age and gender}
\label{tab:S4_adjusted_cardio}

\begin{tabularx}{\textwidth}{@{}Xccc@{}}
\toprule
\textbf{Feature} & \textbf{Pearson correlation} & \textbf{p\_value\_corrected} & \textbf{Significance} \\ \midrule
daily\_steps\_mean            & -0.22 & <.0001   & TRUE  \\
daily\_AZM\_mean              & -0.16 & <.0001   & TRUE  \\
mcv                           & -0.15 & <.0001   & TRUE  \\
albumin/globulin              & -0.12 & <.0001   & TRUE  \\
total bilirubin               & -0.12 & <.0001   & TRUE  \\
daily\_steps\_std             & -0.11 & <.0001   & TRUE  \\
mch                           & -0.1  & <.0001   & TRUE  \\
co2                           & -0.11 & 1.05E-02 & FALSE \\
daily\_light\_sleep\_mean      & -0.09 & 2.30E-02 & FALSE \\
daily\_AZM\_std               & -0.08 & 3.53E-02 & FALSE \\
daily\_hrv\_mean              & -0.08 & 3.95E-02 & FALSE \\
daily\_rmssd\_mean            & -0.08 & 5.00E-02 & FALSE \\
daily\_sleep\_duration\_mean  & -0.07 & 6.76E-02 & FALSE \\
daily\_entropy\_mean          & -0.06 & 1.48E-01 & FALSE \\
sodium                        & -0.06 & 1.53E-01 & FALSE \\
chloride                      & -0.05 & 2.22E-01 & FALSE \\
daily\_light\_sleep\_std      & 0.05  & 2.14E-01 & FALSE \\
daily\_deep\_sleep\_std       & 0.07  & 7.30E-02 & FALSE \\
daily\_sleep\_duration\_std   & 0.09  & 2.04E-02 & FALSE \\
potassium                     & 0.09  & 2.04E-02 & FALSE \\
absolute\_basophils           & 0.09  & 1.64E-02 & FALSE \\
hba1c                         & 0.1   & <.0001   & TRUE  \\
rdw                           & 0.1   & <.0001   & TRUE  \\
daily\_resp\_rate\_mean       & 0.11  & <.0001   & TRUE  \\
alp                           & 0.11  & <.0001   & TRUE  \\
glucose                       & 0.13  & <.0001   & TRUE  \\
alt                           & 0.13  & <.0001   & TRUE  \\
absolute\_monocytes           & 0.13  & <.0001   & TRUE  \\
absolute\_neutrophils         & 0.13  & <.0001   & TRUE  \\
platelet                      & 0.13  & <.0001   & TRUE  \\
absolute\_lymphocytes         & 0.13  & <.0001   & TRUE  \\
total protein                 & 0.13  & <.0001   & TRUE  \\
globulin                      & 0.15  & <.0001   & TRUE  \\
daily\_rhr\_mean              & 0.17  & <.0001   & TRUE  \\
ggt                           & 0.17  & <.0001   & TRUE  \\
white\_blood\_cell            & 0.17  & <.0001   & TRUE  \\
hb                            & 0.18  & <.0001   & TRUE  \\
hematocrit                    & 0.18  & <.0001   & TRUE  \\
red\_blood\_cell              & 0.23  & <.0001   & TRUE  \\
crp                           & 0.25  & <.0001   & TRUE  \\
insulin                       & 0.28  & <.0001   & TRUE  \\
bmi                           & 0.29  & <.0001   & TRUE  \\ \bottomrule
\end{tabularx}
\end{table}

\begin{table}[ht]
\centering
\small % Keeps the text consistent with your other tables
\caption{Features Pearson correlation with kidney health status (measured by eGFR) adjusted for Age and gender}
\label{tab:S5_adjusted_kidney}

\begin{tabularx}{\textwidth}{@{}Xccc@{}}
\toprule
\textbf{Feature} & \textbf{Pearson correlation} & \textbf{p\_value\_corrected} & \textbf{Significance} \\ \midrule
calcium                       & -0.16 & 0.0001 & TRUE  \\
hematocrit                    & -0.14 & 0.0019 & TRUE  \\
hb                            & -0.12 & 0.0188 & FALSE \\
red\_blood\_cell              & -0.1  & 0.0719 & FALSE \\
ast                           & -0.09 & 0.0857 & FALSE \\
total testosterone            & -0.08 & 0.3411 & FALSE \\
daily\_light\_sleep\_mean      & -0.07 & 0.4327 & FALSE \\
daily\_light\_sleep\_std       & -0.06 & 0.5099 & FALSE \\
daily\_wake\_sleep\_std        & -0.06 & 0.5099 & FALSE \\
daily\_sleep\_duration\_std    & -0.06 & 0.5383 & FALSE \\
absolute\_basophils           & -0.05 & 0.5634 & FALSE \\
absolute\_lymphocytes         & 0.05  & 0.5634 & FALSE \\
glucose                       & 0.06  & 0.5383 & FALSE \\
daily\_resp\_rate\_mean        & 0.06  & 0.5383 & FALSE \\ \bottomrule
\end{tabularx}
\end{table}

% Ensure these are in your preamble:
% \usepackage{booktabs}
% \usepackage{tabularx}

\begin{table}[ht]
\centering
\footnotesize % Slightly smaller than \small to ensure all 54 rows fit one page
\renewcommand{\arraystretch}{0.82} % Tighten row spacing
\caption{Features Pearson correlation with metabolic health status (measured by HOMA\_IR) adjusted for Age and gender}
\label{tab:S6_metabolic_adjusted}

\begin{tabularx}{\textwidth}{@{}Xccc@{}}
\toprule
\textbf{Feature} & \textbf{Pearson correlation} & \textbf{p\_value\_corrected} & \textbf{Significance} \\ \midrule
hdl                           & -0.33 & <.0001   & TRUE  \\
daily\_steps\_mean            & -0.25 & <.0001   & TRUE  \\
albumin/globulin              & -0.24 & <.0001   & TRUE  \\
daily\_sleep\_duration\_mean  & -0.18 & <.0001   & TRUE  \\
total bilirubin               & -0.17 & <.0001   & TRUE  \\
basophils                     & -0.15 & <.0001   & TRUE  \\
daily\_entropy\_mean          & -0.14 & <.0001   & TRUE  \\
mcv                           & -0.14 & <.0001   & TRUE  \\
mch                           & -0.14 & 1.00E-04 & TRUE  \\
albumin                       & -0.12 & 1.40E-03 & TRUE  \\
daily\_AZM\_mean              & -0.11 & 2.90E-03 & TRUE  \\
daily\_steps\_std             & -0.11 & 3.74E-03 & TRUE  \\
co2                           & -0.11 & 4.11E-03 & TRUE  \\
daily\_deep\_sleep\_mean      & -0.1  & 5.99E-03 & TRUE  \\
monocytes                     & -0.1  & 8.33E-03 & TRUE  \\
daily\_light\_sleep\_mean      & -0.09 & 2.33E-02 & FALSE \\
daily\_hrv\_mean              & -0.08 & 4.60E-02 & FALSE \\
mchc                          & -0.07 & 6.17E-02 & FALSE \\
daily\_rem\_sleep\_mean       & -0.06 & 9.19E-02 & FALSE \\
chloride                      & -0.06 & 1.31E-01 & FALSE \\
bun                           & -0.06 & 1.35E-01 & FALSE \\
egfr                          & 0.05  & 1.88E-01 & FALSE \\
calcium                       & 0.06  & 1.38E-01 & FALSE \\
hematocrit                    & 0.07  & 8.38E-02 & FALSE \\
non hdl                       & 0.07  & 5.74E-02 & FALSE \\
daily\_wake\_sleep\_std       & 0.08  & 4.39E-02 & FALSE \\
daily\_rmssd\_std             & 0.08  & 2.49E-02 & FALSE \\
daily\_rhr\_std               & 0.1   & 4.63E-03 & TRUE  \\
daily\_hrv\_std               & 0.11  & 3.78E-03 & TRUE  \\
daily\_rem\_sleep\_std        & 0.11  & 3.43E-03 & TRUE  \\
absolute\_eosinophils         & 0.11  & 3.43E-03 & TRUE  \\
total protein                 & 0.12  & 8.83E-04 & TRUE  \\
daily\_deep\_sleep\_std       & 0.12  & 7.58E-04 & TRUE  \\
red\_blood\_cell              & 0.13  & 4.72E-04 & TRUE  \\
rdw                           & 0.14  & 1.61E-04 & TRUE  \\
ggt                           & 0.14  & 1.02E-04 & TRUE  \\
daily\_entropy\_std           & 0.15  & 2.90E-05 & TRUE  \\
alt                           & 0.15  & 2.58E-05 & TRUE  \\
platelet                      & 0.16  & 1.79E-05 & TRUE  \\
daily\_sleep\_duration\_std   & 0.18  & 6.95E-07 & TRUE  \\
daily\_resp\_rate\_mean       & 0.18  & 6.95E-07 & TRUE  \\
alp                           & 0.18  & 6.85E-07 & TRUE  \\
daily\_light\_sleep\_std      & 0.18  & 4.01E-07 & TRUE  \\
daily\_resp\_rate\_std        & 0.19  & 1.29E-07 & TRUE  \\
absolute\_monocytes           & 0.2   & 6.46E-08 & TRUE  \\
globulin                      & 0.23  & 3.01E-10 & TRUE  \\
absolute\_neutrophils         & 0.23  & 5.35E-11 & TRUE  \\
absolute\_lymphocytes         & 0.24  & 7.33E-12 & TRUE  \\
chol/hdl                      & 0.26  & 2.33E-13 & TRUE  \\
daily\_rhr\_mean              & 0.28  & 4.64E-15 & TRUE  \\
crp                           & 0.29  & 3.56E-16 & TRUE  \\
white\_blood\_cell            & 0.3   & 5.21E-17 & TRUE  \\
triglycerides                 & 0.39  & 3.81E-30 & TRUE  \\
bmi                           & 0.46  & 1.77E-42 & TRUE  \\
glucose                       & 0.53  & 4.53E-60 & TRUE  \\ \bottomrule
\end{tabularx}
\end{table}

\begin{table}[ht]
\centering\footnotesize
\renewcommand{\arraystretch}{0.82}
\caption{Description of each of the 26 features used as input for the Wearable Foundation Model.}
\label{tab:S7_wfm_features}
\begin{tabularx}{\textwidth}{@{}llX@{}}
\toprule
\textbf{Feature} & \textbf{Unit} & \textbf{Definition} \\ \midrule
Heart Rate & Beats/Min & Mean of instantaneous heart rate. \\
Heart Rate at Rest & Beats/Min & Mean of heart rate at rest. \\
RR Percent Valid & \% & percentage of 5-minute window with valid RR intervals. \\
RR 80th Percentile & Msec & 80th percentile of 5-minute window of RR ints. \\
RR 20thPercentile & Msec & 20th percentile of RR ints. \\
RR Median & Msec & Median RR interval. \\
RMSSD & Msec & Root mean squared st. dev. of RR ints. \\
SDNN & Msec & Standard deviation of RR intervals. \\
Shannon Ent. RR & Nats & Shannon entropy of the RR intervals. \\
Shannon Ent. RR Diffs & Nats & Shannon entropy of the RR interval differences. \\
Step Count & Steps & Number of steps. \\
Jerk Autocorrelation Ratio & a.u. & Ratio of lag=1 autocorrelation to energy in 1st 3-axis principal component. \\
Log Energy & a.u. & Log of sum of 3-axis root mean squared magnitude. \\
Covariance Condition & a.u. & Estimate of condition number for the 3-axis covariance. \\
Log Energy Ratio & a.u. & Log of ratio of sum of energy in 1st 3-axis principal component over energy of 3-axis root mean squared magnitude. \\
Zero Crossing St.Dev & Seconds & Standard deviation of time between zero crossing of 1st 3-axis principal component. \\
Zero Crossing Avg & Seconds & Mean of time between zero crossing of 1st 3-axis principal component. \\
Axis Mean & a.u. & Mean of 3-axis. \\
Kurtosis & a.u. & Kurtosis of 3-axis root mean squared magnitude. \\
Sleep Coefficient & a.u. & Sum of 3-axis max-min range with 16 log-scaled bins. \\
Skin Conductance Value & $\mu$Siemens & Center of linear tonic SCL value fit. \\
Skin Conductance Slope & $\mu$S/Min & Intraminute slope of SCL values. \\
Lead Contact Counts & Counts & Number of times sensor leads contacted the wrist in a minute. \\
Skin Temperature Value & $^\circ$C & Mean value of skin temperature. \\
Skin Temperature Slope & $^\circ$C/Min & Slope of skin temperature. \\
Altitude St.Dev. Norm & hPa & Standard deviation of altimeter readings. \\
\bottomrule
\end{tabularx}
\end{table}

\end{document}

%% file: math_commands.tex
\usepackage{amsmath,amsfonts,bm}

\def\eqref#1{equation~\ref{#1}}
\def\1{\bm{1}}

\DeclareMathAlphabet{\mathsfit}{\encodingdefault}{\sfdefault}{m}{sl}
\SetMathAlphabet{\mathsfit}{bold}{\encodingdefault}{\sfdefault}{bx}{n}

%% file: sections/0-abstract.tex
\begin{abstract}

Cardiovascular-Kidney-Metabolic (CKM) syndrome represents a growing public health crisis, yet its subclinical heterogeneity remains underexplored. Early detection of physiological deviation is critical for preventing irreversible organ damage. Here, we characterize the prevalence and interplay of CKM impairment in a US cohort (N=819) by integrating continuous wearable data with clinical biomarkers. We assessed cardiovascular, kidney via clinical biomarkers, namely Chol/HDL, eGFR, as well as metabolic health risk  through Homeostatic Model Assessment of Insulin Resistance (HOMA-IR). We show that while metabolic and cardiovascular disruptions are significantly associated (r=0.26, p<0.001), early-stage kidney impairment manifests independently. Utilizing a normalized deviance score, we identified significant health impairments in 29.0\% of the cohort. Cardiovascular deviation was the most prevalent singular phenotype (13.3\%), followed by metabolic (9.1\%) and renal (6.25\%) deviations, with dual metabolic-cardiovascular impairment occurring in only 2.2\% of participants. These findings suggest that high system-specific deviance may serve as an indicator for accelerated physiological aging within the respective organ system. Furthermore, Step count, Active Zone Minutes, and resting heart rate were identified as the key wearable predictors of cardiovascular and metabolic decline. Leveraging these and CKM metrics, we developed a Health Age model that tracks physiological health, aligning with established clinical standards (Metabolic syndrome severity and modified AHA Life’s Essential 8) and user-reported overall health. These findings underscore the necessity of a multi-system subtyping approach, demonstrating that wearable-derived phenotypes can facilitate the early, targeted interventions required to manage the complex landscape of CKM syndrome.

\end{abstract}

%% file: sections/1-introduction.tex
\section{Introduction}

Cardiovascular-Kidney-Metabolic (CKM) syndrome represents a critical, interconnected health challenge. It is a progressive condition where dysfunction in any of the three systems: cardiovascular, kidney, or metabolic rapidly cascades, leading to multi-organ damage, premature morbidity, and mortality\citep{Ndumele2023-ck}. This syndrome has high prevalence in the population with recent estimates of approximately 25\% of individuals having at least one condition\citep{Ndumele2023-pa,Ferdinand2024-ov}. Early identification and intervention, particularly focusing on lifestyle modifications and recognizing early markers, are paramount to disrupting this devastating cycle.

The American Heart Association's staging of CKM syndrome highlights its progression, starting from Stage 0 (no risk factors) to Stage 4 (established cardiovascular disease with or without kidney failure)\citep{Ndumele2023-pa}. Crucially, this progression often begins early in life with biological, social, and lifestyle factors leading to dysfunctional adipose tissue. This early metabolic dysfunction, often underestimated by Body mass index (BMI) alone, is a key instigator, preceding and contributing to both cardiovascular and kidney complications\citep{Longo2019-ad,VanBuren2011}. The metabolic abnormalities often precede and contribute to the development of both cardiovascular and kidney complications, making them important early warning signs in the CKM continuum. While CKM syndrome is classified by its metabolic roots, it's crucial to recognize the inherent interdependence of these dysfunctions. Problems in one system inevitably worsen the others. Kidney dysfunction, in particular, acts as a central link between metabolic risk factors and cardiovascular disease. It can initiate, worsen, or even trigger cardiac dysfunction, and the reverse is also true. Specifically, chronic kidney disease impacts heart function through altered blood flow, and by retaining salt and water, which leads to venous congestion and, consequently, heart failure\citep{Damman2015-ki}. The systemic nature of CKM syndrome necessitates comprehensive screening that addresses all components rather than siloed approaches to individual conditions. By identifying early markers of dysfunction clinicians can intervene at stages when preventive measures may have the greatest impact. Thus, identifying early markers for each component of CKM syndrome is crucial for timely intervention before irreversible damage occurs. Each component of CKM syndrome has specific indicators that can signal early dysfunction and increased risk.

The evolution of consumer wearable technology presents a transformative opportunity to realize this goal of proactive CKM risk management. Wearable devices, such as smartwatches and rings, are now capable of continuously and passively collecting high-fidelity physiological signals, including heart rate variability (HRV), resting heart rate (RHR), sleep metrics, and activity levels that serve as digital biomarkers of systemic health \citep{Piwek2016-ri,Shcherbina2017-ac, Torous2021-gr}. This capacity for long-term, real-world data collection, or “digital phenotyping,” overcomes the limitations of episodic clinical assessments by providing a dense, personalized health timeline \citep{Zhang2025-co}. Several studies have demonstrated the utility of these data streams for disease surveillance and prediction \citep{Metwally2026,Esmaeilpour2024-de,Radin2020-ha,Lubitz2022-de}. For instance, wearable vital signs have been successfully utilized for early prediction of conditions like respiratory illness and changes in inflammation\citep{Esmaeilpour2024-de}. In a highly relevant exploration of this utility, Dunn et al. showed that vital sign data derived from smartwatches could predict components of routine clinical laboratory panels, specifically finding utility in estimating hematocrit (HCT) and hemoglobin (HGB)\citep{Dunn2021-we}. Given the inherent heterogeneity of CKM pathology, which can manifest in diverse organ-specific patterns, the critical next step is to leverage these high-resolution wearable signals to precisely characterize distinct subtypes of early CKM disruption. This allows for the generation of valuable, actionable lifestyle signals. 

The present work addresses this gap by undertaking an early-stage analysis of CKM pathology using a novel approach to quantify physiological deviance. We leveraged comprehensive, concurrent data from wearables and blood biomarkers in a large US cohort (N=819) to assess early markers of cardiovascular health (Chol/HDL), metabolic health (HOMA-IR), and renal function (eGFR). Our methodology introduced a deviance score to objectively establish an impairment threshold, enabling a detailed subtyping of CKM presentation in a population not yet defined by advanced clinical disease. This builds on the precedent of using deviance-based phenotyping to identify early shifts in physiological resilience before they manifest as chronic disease\cite{vandenBroek2017}. We present a method for predicting markers of cardiovascular health (Chol/HDL), metabolic health (HOMA-IR), and renal function (eGFR) using (a) signals derived from a consumer smartwatch (e.g., activity level, sleep duration and sleep stages, resting heart rate and heart rate variability), (b) demographics (i.e., Age, Gender and BMI), and (c) commonly measured blood biomarkers (e.g, lipid panel, liver panel, metabolic panel). To quantify the predictive contribution of these diverse factors, a rigorous regression and classification analysis was executed. This modeling approach provides actionable insights by identifying and ranking the most important demographic and lifestyle features influencing these health markers. Building upon these predictive insights, we further developed a composite Health Age metric that translates these high-dimensional wearable and CKM features into an intuitive proxy for CKM physiological health status. We validate this Health Age against robust clinical benchmarks, specifically the Metabolic Syndrome Severity (MSS) score and a modified American Heart Association Life’s Essential 8 (AHA LE8) framework as well as subjective self-reported health outcomes. Ultimately, this framework provides a scalable, clinically aligned paradigm for continuous health status tracking directly from using consumer wearables and or blood biomarkers.

%% file: sections/2-results.tex
\section{Results}
\label{sec:results}

\subsection{Study Design and cohort characteristics}
\label{sec:Study Design and cohort characteristics}
We implemented a prospective observational investigation, recruiting adult participants from the United States to take part in a consented research study \citep{Metwally2026}. This study received approval from Advarra, serving as the institutional review board (IRB) of record (protocol number Pro00074093), and was executed remotely utilizing the Google Health Studies (GHS) application. GHS functions as a consumer-centric, secure platform for conducting digital studies, facilitating participant enrollment, eligibility verification, and the provision of informed consent (Methods). In this particular instance, the platform was configured to enable the acquisition of wearable data from Fitbit and Google Pixel Watch devices (referred to collectively as “wearables” herein), the completion of questionnaires, and the arrangement of blood tests with Quest Diagnostics. Participants were requested to: (1) establish a link between their Fitbit/Google account and the GHS application to facilitate data sharing, (2) complete questionnaires pertaining to demographic details, health history, and general health information,  (3) schedule and undergo a blood draw at a Quest Patient Service Center within 65 days of their enrollment.

We consented 4,416 participants. Depending on the wearable data representation used, the analysis cohorts differed slightly due to data completeness requirements: 819 participants (18\% completion rate) met the criteria for standard daily-level wearable statistics, and 923 participants (21\% completion rate) met the criteria for the high-resolution WFM embeddings\hyperref[sec:Methods]{(Methods)}. The primary clinical and demographic characterization was conducted on the standard wearable cohort (N = 819)\hyperref[sec:Methods]{(Methods)}. Inclusion criteria included U.S. residents between the ages of 21 and 80 who wear a Fitbit wearable device or a Pixel Watch with heart rate sensing capabilities and are willing to go to a Quest Diagnostics location for blood draws. Participants were asked to do blood draws early in the morning after fasting for at least 8 hours to minimize the effect of the solar diurnal cycle. In this data collection, wearable data serves as a retrospective analysis of blood lab results. Specifically, we collect three months of data from the wearable devices leading up to the date of the participant's lab visit. \hyperref[tab:cohort_characterization]{Table 1} summarizes our cohort demographic and socioeconomic factors. All collected data are listed in the \hyperref[sec:Methods]{methods section}. 

% \usepackage{booktabs}
% \usepackage{multirow}
\begin{table}[htbp]
\centering
\footnotesize
\renewcommand{\arraystretch}{0.80}
\setlength{\tabcolsep}{4pt}
\caption{\textbf{Cohort Characteristics.} Detailed characterization of the total population, categorized by key demographic, socioeconomic, and clinical factors such as gender, BMI, race/ethnicity, age, employment status, highest education status, household income, and comorbidities. Each category includes the number and corresponding percentage of individuals within the cohort.}
\label{tab:cohort_characterization}
\begin{tabular}{@{} l l r @{}}
%\toprule
\label{tab:cohort_characterization} \\ \midrule
\multicolumn{2}{@{}l}{\textbf{Total Population}} & 819 \\ \midrule
 & Male & 408 (49.80) \\
 & Female & 393 (48.00) \\
 & Genderqueer/Gender Nonconforming & 11 (1.31) \\
 & Trans Female or male/Trans Woman, men & 3 (0.36) \\
\multirow{-5}{*}{\textbf{Gender}} & Different identity or choose not to answer & 4 (0.48) \\ \midrule
 & Underweight (BMI < 18.5) & 6 (0.73) \\
 & Healthy Weight (18.5 $\le$ BMI < 25) & 237 (28.94) \\
 & Overweight (25.0 $\le$ BMI < 30) & 278 (33.94) \\
\multirow{-4}{*}{\textbf{BMI}} & Obesity (BMI $\ge$ 30) & 298 (36.39) \\ \midrule
 & African-American & 30 (3.66) \\
 & Asian-Eastern & 25 (3.05) \\
 & Asian-Indian & 37 (4.52) \\
 & Hispanic & 47 (5.74) \\
 & Mixed Race & 22 (2.68) \\
 & Native American & 2 (0.24) \\
 & White/Caucasian & 642 (78.39) \\
\multirow{-8}{*}{\textbf{Race / Ethnicity}} & Other & 14 (1.70) \\ \midrule
 & [20 - 29] & 41 (5.01) \\
 & [30 - 39] & 226 (27.59) \\
 & [40 - 49] & 238 (29.06) \\
 & [50 - 59] & 167 (20.39) \\
 & [60 - 69] & 107 (13.06) \\
\multirow{-6}{*}{\textbf{Age}} & Above 70 & 40 (4.88) \\ \midrule
 & Full-time & 560 (68.38) \\
 & Part-time & 63 (7.69) \\
 & Unable to work & 22 (2.69) \\
 & Unemployed & 120 (14.65) \\
 & Contract / temporary & 18 (2.20) \\
\multirow{-6}{*}{\textbf{Employment status}} & Other / choose not to answer & 36 (4.40) \\ \midrule
 & Associate degree & 71 (8.67) \\
 & Bachelor's degree & 317 (38.71) \\
 & Graduate degree & 269 (32.84) \\
 & High school or equivalent & 28 (3.42) \\
 & Less than high school & 2 (0.24) \\
 & Some college, no degree & 116 (14.16) \\
\multirow{-7}{*}{\textbf{Highest education status}} & Other & 16 (1.95) \\ \midrule
 & Less than \$25,000 & 18 (2.20) \\
 & \$25,000--\$50,000 & 93 (11.36) \\
 & \$50,000--\$100,000 & 202 (24.66) \\
 & \$100,000--\$200,000 & 284 (34.68) \\
 & More than \$200,000 & 161 (19.66) \\
\multirow{-6}{*}{\textbf{Household income}} & Other / choose not to answer & 61 (7.45) \\ \midrule
 & CVD & 26 (3.17) \\
 & Hyperlipidemia & 169 (20.63) \\
 & Diabetes & 46 (5.62) \\
 & Respiratory & 100 (12.21) \\
 & Hypertension & 167 (20.39) \\
 & Kidney Disease & 15 (1.83) \\
 & Aspirin & 46 (5.64) \\
 & Statin & 149 (18.19) \\
 & Blood Pressure Medications & 163 (19.90) \\
 & Heart Medications & 12 (1.47) \\
 & Insulin & 10 (1.22) \\
 & Antidepressant & 175 (21.37) \\
 & Beta Blockers & 34 (4.15) \\
\multirow{-14}{*}{\textbf{Comorbidities}} & Anxiety Medication & 175 (21.37) \\ \bottomrule
\end{tabular}
\end{table}

For evaluating the health status of each CKM subsystem, we chose a system-specific biomarker based on literature and our cohort data distribution, rather than a general marker. Broad systemic indices like the Neutrophil-to-Lymphocyte Ratio (NLR) were deliberately omitted to ensure the specificity of our subsystem-level assessments\citep{Carollo2025}.For cardiovascular health, the ratio of total cholesterol to HDL cholesterol (Chol/HDL) was chosen. Although therapeutic guidelines focus on absolute non-HDL cholesterol for lipid-lowering targets\citep{Jacobson2015}, we modeled the Chol/HDL ratio (algebraically equivalent to non-HDL/HDL+1) to capture the balance between atherogenic and anti-atherogenic particles, a key driver of CKM-associated cardiometabolic risk. Predicting this ratio directly aligns with validated clinical risk calculators, including the QRISK3 and 2024 AHA PREVENT equations, which continue to utilize total cholesterol and HDL as primary parameters to estimate cardiovascular events\citep{HippisleyCox2017,Khan2024}.Large-scale epidemiological studies continue to validate the predictive power of the ratio over individual cholesterol markers like LDL-C alone\citep{Kinosian1994}. Additionally, the selection of Chol/HDL as preferred biomarker was further supported by its statistical distribution within the study cohort \hyperref[fig:supplement_1]{Supplementary Figure S1}.  In our study cohort, the median Chol/HDL ratio was 3.3. A clinical normal threshold of 3.5 was adopted to delineate elevated risk\citep{calling2019ratio},with lower values indicating a more favorable cardiovascular profile. The distribution of the Chol/HDL ratio within the dataset is presented in \hyperref[fig:fig_2]{Figure 2} and more closely in \hyperref[fig:supplement_2]{Supplementary Figure S2, A}.

\begin{figure}[htbp]
    \centering
    % Force the image to fit within the text width AND 80% of the page height
    \includegraphics[angle=-90, width=\textwidth, totalheight=0.7\textheight, keepaspectratio]{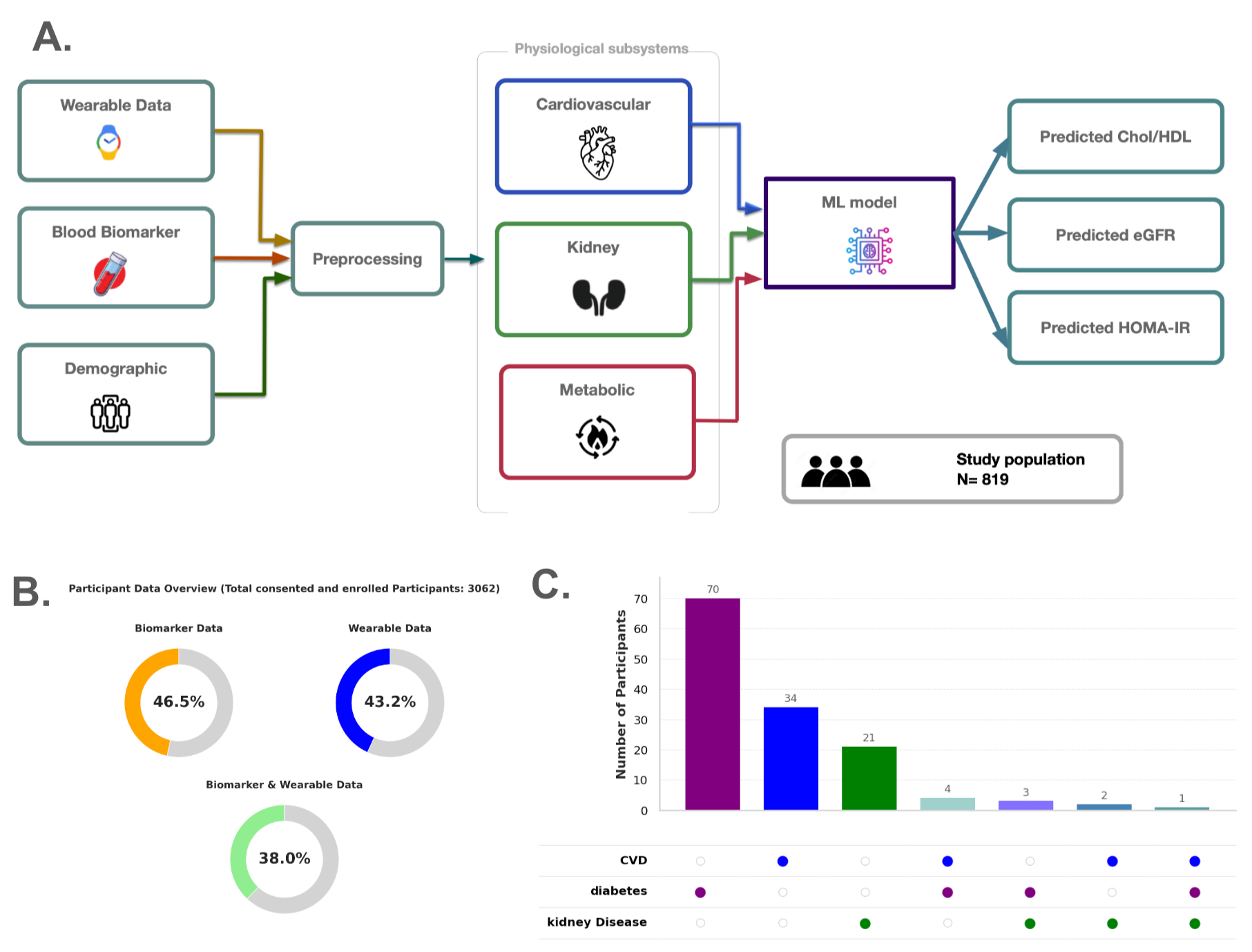}
    \label{fig:1}

    \caption{\textbf{Study design and summary of Dataset.} (A) The study combines wearables data, demographics information and blood lab results to understand how wearables can be used to assess Cardiovascular-kidney-Metabolic Health. (B) Distribution of Data Modalities Among Study Participants. The three donut plots display the percentage of individuals contributing wearable data, those providing only lab biomarker data, and percentage of participants with both wearable and lab biomarker data. (C) Number of study participants with Cardiovascular disease (CVD), Diabetes and Kidney disease per self reporting of participants and all comorbid combinations of these conditions, as recorded in each survey.}
    \label{fig:fig_1}
\end{figure}

(\hyperref[fig:fig_1]{Figure 1}).
% If you are using sidewaysfigure, change it back to just figure:

Estimated Glomerular Filtration Rate (eGFR) is a cornerstone in the assessment of kidney health, widely used by clinicians to detect kidney disease, track its progression, and guide treatment decisions. This calculated measure provides an estimate of how well the kidneys are filtering waste products from the blood\citep{odler2024egfr}. In this study we used eGFR as a metric to evaluate kidney health status. In the study population, the median eGFR was 96.5 mL/min/1.73m2. A clinical threshold of 90 mL/min/1.73 m2 was adopted to delineate elevated risk, with higher values indicating a more favorable kidney health profile\citep{kdigo2024clinical,johnson2012chronic}. The distribution of the eGFR within the dataset is presented in\hyperref[fig:fig_2]{Figure 2} and more closely in, \hyperref[fig:supplement_2]{Supplementary Figure S2, B}.

The Homeostatic Model Assessment of Insulin Resistance (HOMA-IR) is commonly used as an indirect measure for quantifying insulin resistance and $\beta$-cell function, which are key components of metabolic health. It is calculated using fasting plasma glucose and insulin levels\citep{GayosoDiz2013-ir,Matthews1985-ho}.Here, we used HOMA-IR as a metric representing metabolic health. In the study population, the median HOMA-IR was 1.81. A clinical threshold of 2 was adopted to delineate elevated risk, with lower values indicating a more favorable metabolic health\citep{Metwally2026}. The distribution of the HOMA-IR within the dataset is presented in \hyperref[fig:fig_2]{Figure 2} and more closely in \hyperref[fig:supplement_2]{Supplementary Figure S2, C}.
\begin{figure*}[!ht]
    \centering
    \includegraphics[angle=-90, width=\textwidth, totalheight=0.8\textheight, keepaspectratio]{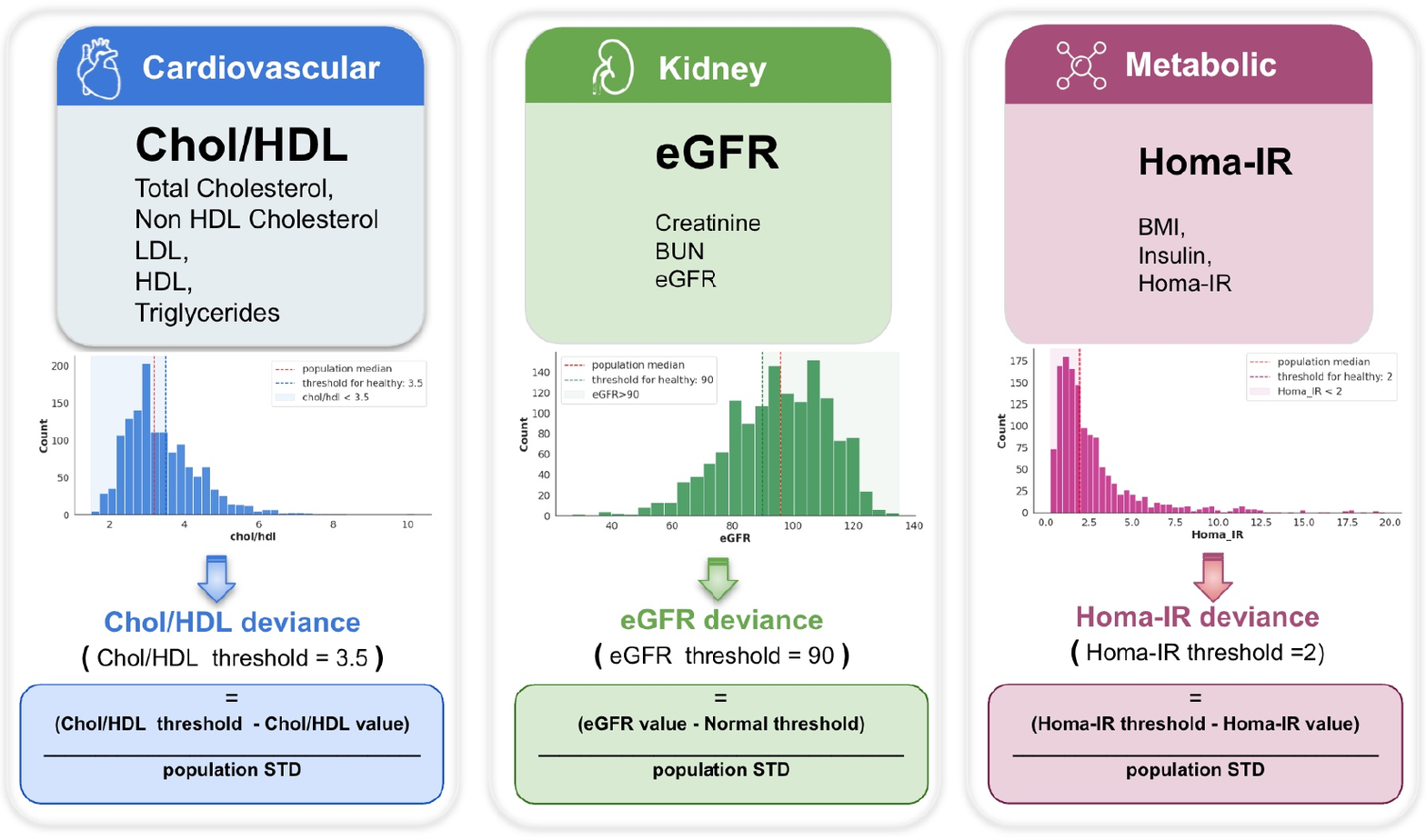}
    \caption{\textbf{Subsystem-Specific Biomarkers and a Standardized Deviance Metric for Integrated Cardio-Kidney-Metabolic Health Monitoring.} Identification of a key biomarker for CKM subsystems: the Chol/HDL ratio for cardiovascular health, eGFR for kidney health, and Homa-IR for metabolic health. Introduction of health status deviance determined by subtracting the normal threshold from the observed value and normalizing the difference by the population standard deviation (STD). A positive deviance value corresponds to a healthy condition in the respective subsystem, whereas a negative value indicates a decline in health.}
    \label{fig:fig_2}
\end{figure*}
Collectively, these three biomarkers Chol/HDL ratio for cardiovascular health, eGFR for kidney health, and HOMA-IR for metabolic health were selected to provide a comprehensive and clinically relevant assessment of the CKM axis within the study cohort. These metrics, along with their established clinical thresholds, formed the basis for evaluating individual subsystem health status in subsequent analyses 
\subsection{Individual Heterogeneity in CKM Health as a Reflectance of System-Specific Accelerated Aging}
\label{sec:Individual Heterogeneity in CKM Health as a Reflectance of System-Specific Accelerated Aging}
To quantify the departure from a healthy state, we utilized a direction-adjusted deviance score. For biomarkers where higher values indicate better health (e.g., eGFR), deviance was calculated by subtracting the clinical reference value from the observed value. For biomarkers where lower values indicate better health (e.g., Chol/HDL and HOMA-IR), the observed value was subtracted from the clinical reference. The resulting difference was then normalized by the population standard deviation (STD). In this framework, a positive deviance value corresponds to a healthy condition in the respective subsystem, whereas a negative value indicates a decline in health status\hyperref[fig:fig_2]{Figure 2}. The median values for each CKM biomarker (i.e., Chol/HDL, eGFR, and HOMA-IR) in our dataset closely align with clinically established reference thresholds, indicating that our data accurately represents the general population\hyperref[fig:fig_2]{Figure 2} and \hyperref[fig:supplement_2]{Supplementary Figure S2}. Specifically, the cohort exhibited a median eGFR of 96.5 mL/min/m2 (clinical threshold: 90 mL/min/m2), a median HOMA-IR of 1.81 (clinical threshold for insulin resistance:  2), and a median Chol/HDL ratio of 3.3 (clinical threshold : 3.5).

\begin{figure*}[!ht]
    \centering
    \includegraphics[width=1.0\textwidth]{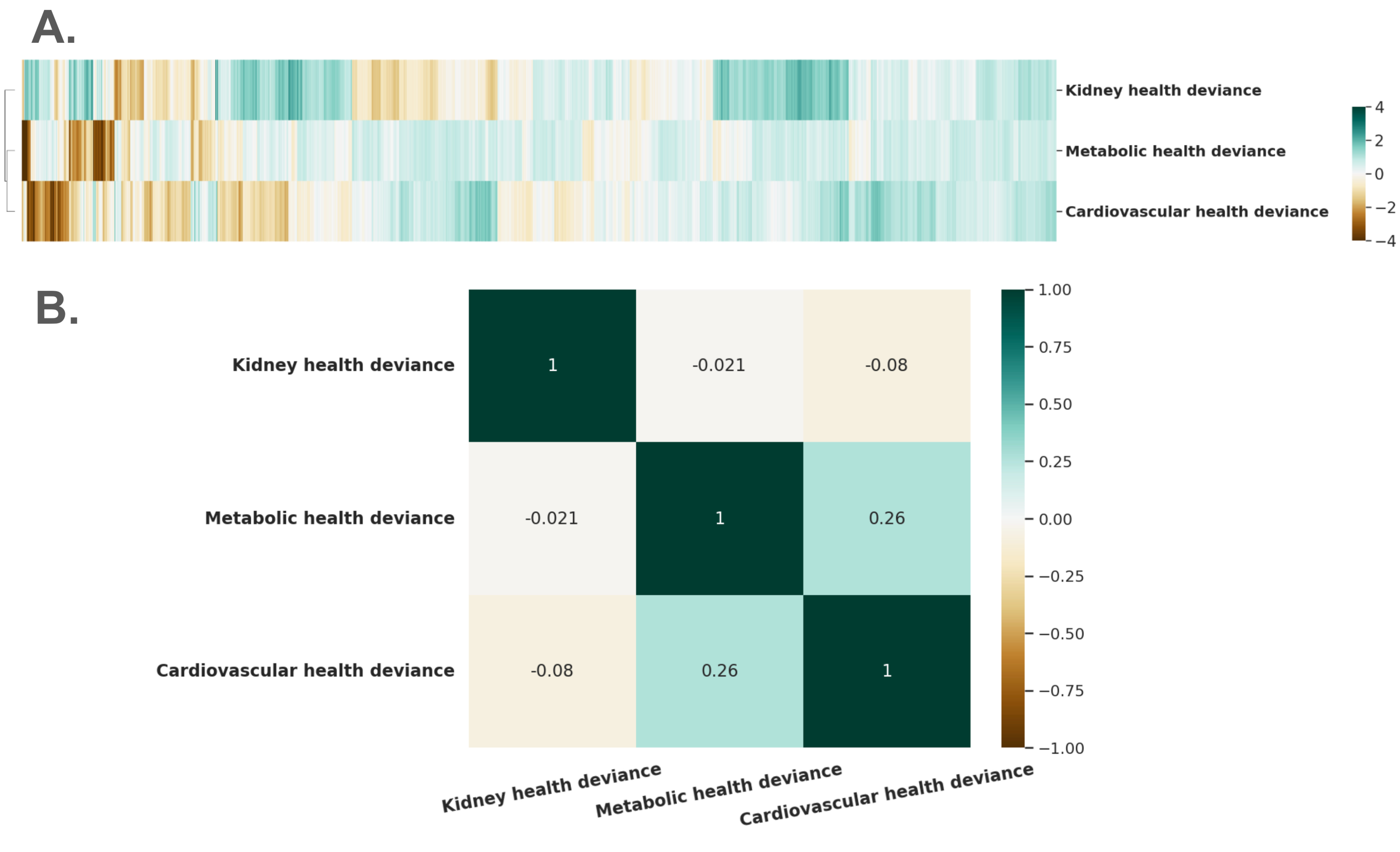}
    \caption{\textbf{Cardiovascular-Kidney-Metabolic health heterogeneity among individuals.} Health statuses across cardiovascular, kidney, and metabolic subsystems within the study population. (A) Heatmap of the calculated health deviance for all participants across the CKM subsystems, allowing for the identification of individual health profiles and patterns of health across CKM subsystems. (B) Correlation matrix heatmap illustrating the inter-correlations among the health statuses of the three subsystems within the study population.}
    \label{fig:fig_3}
\end{figure*}

\hyperref[fig:fig_3]{Figure 3A}  visually represents the calculated deviance from normal for each CKM subsystem across participants. Each cell in the heatmap is color-coded, with the intensity and hue directly corresponding to the deviance value; where positive deviance values indicate healthier status and negative deviance values indicate less healthy status. This visualization allows for a rapid assessment of an individual's overall health profile across the targeted CKM subsystems and facilitates the identification of patterns, such as individuals exhibiting consistently healthy or less healthy states across multiple subsystems, or those with specific subsystem impairments.

A deviance value below $-1$ was subsequently used as the threshold for identifying departure from the healthy status within a specific subsystem (i.e., cardiovascular, kidney or metabolic). This threshold is statistically robust because it utilizes standardized units (Z-scores) to ensure comparability across different biological subsystems, while the $-1$ cut-off provides the necessary sensitivity to detect subclinical shifts before they reach extreme outlier status. This methodology of using a $1-SD$ threshold on standardized continuous risk scales is widely validated in the cardiometabolic literature for identifying departure for healthy status in populations \citep{ruiz2007high,haggstrom2013metabolic}. Translating these standardized $-1$ cut-off  to absolute clinical values, a $-1$ deviance threshold corresponds to a HOMA-IR of 4.68, an eGFR of 73.05 mL/min/1.73m², and a Total Cholesterol/HDL ratio of 4.47. In clinical stratification, the eGFR and lipid ratio thresholds align precisely with early-stage, subclinical pathology, capturing individuals transitioning into Stage G2 renal decline and moderate cardiovascular risk categories, respectively. Conversely, due to the right-skewed distribution of insulin resistance in the study population, the HOMA-IR threshold of 4.68 bypasses mild metabolic drift to individuals with severe, manifest insulin resistance. Consequently, this $-1$ deviance framework successfully balances clinical sensitivity by capturing early kidney and cardiovascular shifts while simultaneously preserving high specificity for advanced metabolic risk. Applying this threshold to the cohort allowed for the identification of the prevalence of singular and concurrent subsystem issues. The highest singular prevalence was observed in the cardiovascular subsystem, with 13.35\% of the population exhibiting a Chol/HDL deviance below $-1$. This high prevalence aligns with global epidemiological data underscoring the ubiquity of dyslipidemia as a leading cardiovascular risk factor, often preceding overt disease for many years 32. Metabolic health impairment (HOMA-IR deviance below $-1$) was observed in 9.1\% of the population, followed by kidney health impairment (eGFR deviance below $-1$) at 6.25\%. The combined assessment revealed that 29\% of the participants exhibited at least one of the CKM conditions, emphasizing the significant burden of interconnected subclinical risk within the population.

Dual-system issues were most common between the metabolic and cardiovascular subsystems, affecting 2.18\% of the population. This link is biologically plausible given that insulin resistance and dyslipidemia often coexist and drive the atherosclerotic process\citep{Visseren2021-es,Chan2002-ir}. The prevalence of dual issues between the cardiovascular and kidney subsystems (cardiorenal axis) was 0.9\%, while the metabolic and kidney pairing (reno-metabolic axis) was the least frequent, affecting 0.6\% of the population. Although less frequent, the existence of these cardio-kidney and kidney-metabolic axes is consistent with the established physiological cross-talk where chronic dysfunction in one system imposes a mechanical or humoral burden on the others, often leading to a vicious cycle of progressive organ damage.

Importantly, no individuals were identified with significant deviance (below $-1$) across all three health conditions simultaneously. Early identification of a single-system deviance thus presents a critical window for intervention to prevent the cascading failure into multi-system disease.

To quantify the inter-relationship in health status among the three health subsystems, calculate the correlation coefficient between the deviance of each subsystem (\hyperref[fig:fig_3]{Figure 3, B}), which revealed the strength and direction of linear relationships between cardiovascular-kidney-metabolic health status within our study population. Specifically, we observed a correlation coefficient of 0.26 (p-value=<0.001) between the metabolic health subsystem (represented by HOMA-IR) and cardiovascular health subsystem. Conversely, no significant correlation was identified between the kidney and cardiovascular subsystems (r=0.07, p value=0.5) and between metabolic and kidney health (r=0.016, p value=0.8). This highlights the heterogeneity in CKM development.

\subsection{Association between CKM health metrics, wearables and blood biomarkers features}
\label{sec:Association between CKM health metrics, wearables and blood biomarkers features}
\begin{figure*}[!ht]
    \centering
    \includegraphics[width=1.0\textwidth]{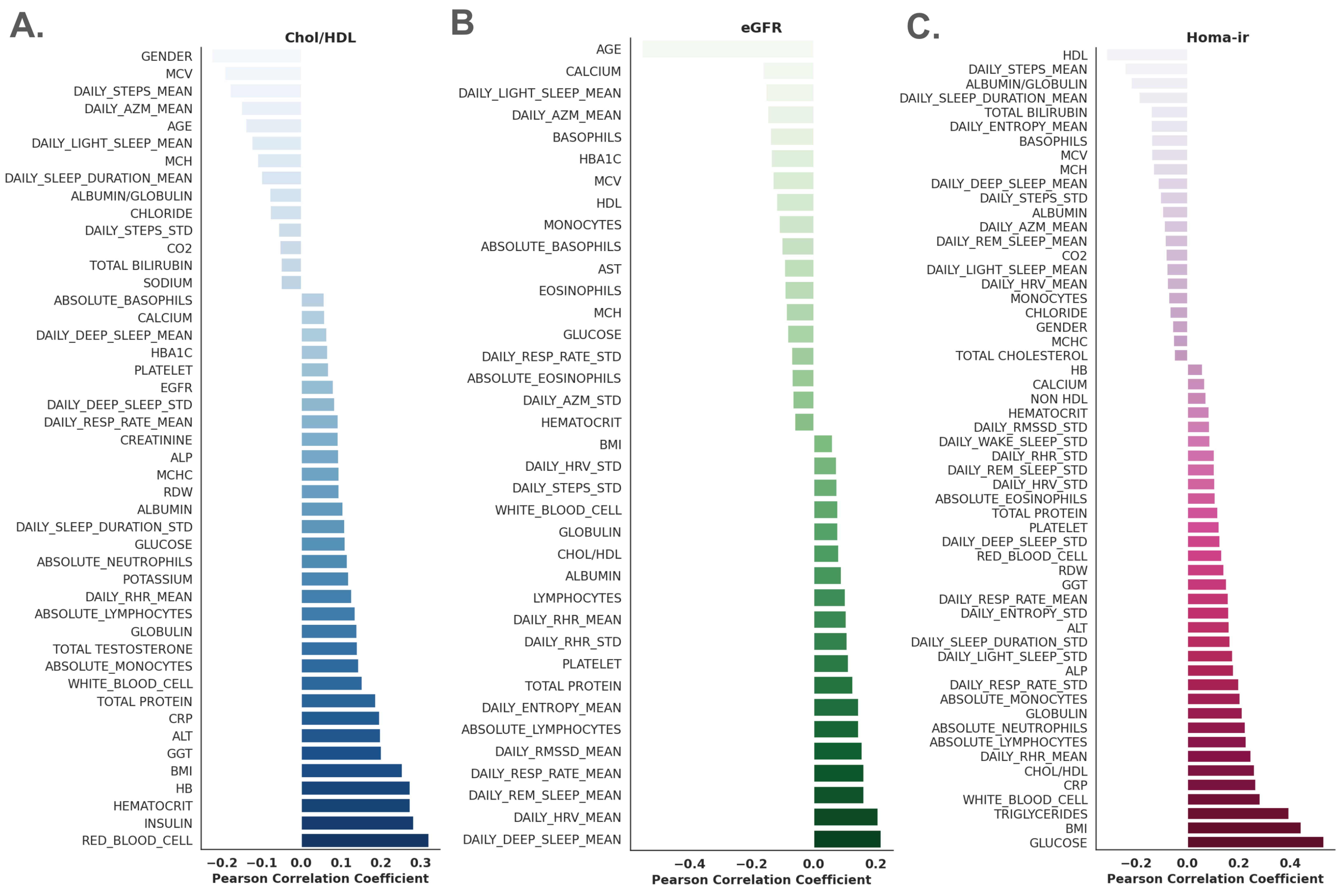}
    \caption{\textbf{Correlated features in Cardiovascular, Kidney and Metabolic subsystems.} The Pearson correlation (correlation coefficient > 0.05) between lifestyle, demographic and blood biomarkers and the health status of Cardiovascular, Kidney and Metabolic subsystems. (A) Illustrates the critical correlates for the cardiovascular health subsystem. (B) Presents the significant relationships for the kidney health subsystem. (C) Outlines the primary correlated factors for the metabolic health subsystem.}
    \label{fig:fig_4}
\end{figure*}

To identify features from the wearable, demographic, and blood biomarker that are most strongly associated with the health status of the CKM subsystems, we calculated the Pearson correlation coefficient for all available features against the three primary health indicators: Cardiovascular Health (Chol/HDL ratio), Kidney Health (eGFR), and Metabolic Health (HOMA-IR). This analysis was conducted to establish foundational univariate relationships and to prioritize potential predictors for subsequent modeling (\hyperref[fig:fig_4]{Figure 4A-C}, \hyperref[fig:supplement_1]{Supplementary Table 1},  \hyperref[fig:supplement_2]{Supplementary Table 2}, \hyperref[fig:supplement_3]{Supplementary Table 3}).

\textbf{Cardiovascular health subsystem (Chol/HDL Ratio)}

Our investigation into cardiovascular health, as quantified by the Chol/HDL ratio, revealed several significant associations across feature categories (\hyperref[fig:fig_4]{Figure 4A}). The demographic analysis pointed to BMI as a positive correlate ($r=0.25$, $p_{\text{corrected}} < 0.001$), suggesting that higher BMI is associated with a less favorable Chol/HDL ratio. Gender (encoded as Male = 1, Female = 2) showed a moderate negative link ($r = -0.22$, $p_{\text{corrected}} < 0.001$), and age had a weaker negative link ($r = -0.14$, $p_{\text{corrected}} < 0.001$).

Focusing on wearable features, metrics reflecting physical activity and sleep quality dominated the top correlates. Average daily step count ($r = -0.17$, $p_{\text{corrected}} < 0.001$) and average active zone minutes ($r = -0.15$, $p_{\text{corrected}} < 0.001$) both showed negative correlations, reinforcing the established benefits of activity for cardiovascular markers. Conversely, average daily resting heart rate showed a positive correlation ($r = 0.125$, $p_{\text{corrected}} < 0.001$). Sleep metrics also played a role, with average daily light sleep ($r = -0.123$, $p_{\text{corrected}} < 0.001$) and average daily sleep duration std ($r = 0.1$, $p_{\text{corrected}} < 0.001$) rounding out the top five.

Among blood lab results, the strongest associations were found with markers related to liver function and blood cell components. The top five correlates included insulin ($r=0.28$, $p_{\text{corrected}}=<0.001$), followed closely by hematocrit ($r = 0.27$, $p_{\text{corrected}}=<0.001$) and red blood cells ($r=0.23$, $p_{\text{corrected}}=<0.001$). Liver enzymes GGT ($r=0.20$, $p_{\text{corrected}}=<0.001$) and ALT ($r=0.19$, $p_{\text{corrected}}=<0.001$) also exhibited significant positive correlations. A full list of coefficients for cardiovascular health is detailed in\hyperref[tab:s1_pearson_correlation]{Supplementary Table S1}.

\textbf{Kidney health subsystem (eGFR)}
The analysis of the kidney subsystem, represented by eGFR, revealed a distinct pattern of key associations (\hyperref[fig:fig_4]{Figure 4B}). Notably, age emerged as the single most influential demographic factor, exhibiting a strong negative correlation ($r = -0.55$, $p_{\text{corrected}} < 0.001$) with eGFR. Unlike the cardiovascular subsystem, gender and BMI did not show significant correlations with eGFR in this cohort.

For wearable features, the strongest links were primarily tied to sleep structure and autonomic function. Average daily deep sleep duration ($r=0.21$, $p_{\text{corrected}}=<0.001$) and average daily heart rate variability ($r=0.20$, $p_{\text{corrected}}=<0.001$) were the top positive correlates, suggesting that robust sleep quality and healthier autonomic balance are associated with better kidney function. Other significant predictors included average daily REM sleep minutes ($r = 0.16$, $p_{\text{corrected}}=<0.001$), average daily respiration rate ($r = 0.16$, $p_{\text{corrected}}=<0.001$), and average daily light sleep mean ($r = 0.15$, $p_{\text{corrected}}=<0.001$).

The highest correlated blood lab results included an inverse relationship with calcium ($r = -0.16$, $p_{\text{corrected}}=<0.001$) and hba1c ($r = -0.136$, $p_{\text{corrected}}=<0.001$). Positively correlated immune markers were absolute lymphocyte count ($r = 0.14$, $p_{\text{corrected}}=<0.001$), while basophils ($r = -0.138$, $p_{\text{corrected}}=<0.001$) and MCV ($r = -0.130$, $p_{\text{corrected}}=<0.001$) showed negative associations. A full list of coefficients is provided in \hyperref[tab:S2_kidney_correlation]{Supplementary Table S2}.

\textbf{Metabolic health subsystem (HOMA-IR)}

In the metabolic health subsystem, assessed using HOMA-IR (\hyperref[fig:fig_4]{Figure 4C}), BMI stood out as the only significantly correlated demographic feature ($r = 0.44$, $p_{\text{corrected}}=<0.001$), indicating its strong relationship with insulin resistance. Age and gender were not identified as important demographic predictors in this specific analysis.
The blood lab results showed the most pronounced correlations, led by Glucose ($r = 0.53$, $p_{\text{corrected}}=<0.001$). Other top positive correlates were triglycerides ($r = 0.39$ , $p_{\text{corrected}}=<0.001$), white blood cell count ($r = 0.28$, $p_{\text{corrected}}=<0.001$), and CRP ($r = 0.26$, $p_{\text{corrected}}=<0.001$), highlighting the expected centrality of glucose and inflammatory markers in metabolic dysfunction. As anticipated, HDL showed a negative correlation ($r = -0.31$ , $p_{\text{corrected}}=<0.001$).
Finally, wearable features associated with metabolic health showed that average daily resting heart rate ($r = 0.25$, $p_{\text{corrected}}=<0.001$) and average daily respiration rate ($r = 0.19$ , $p_{\text{corrected}}=<0.001$) had positive associations with HOMA-IR, consistent with poorer metabolic state. Conversely, metrics of physical activity and quantity of sleep, such as average daily step count ($r = -0.24$, $p_{\text{corrected}}=<0.001$) and average daily sleep duration ($r = -0.18$ , $p_{\text{corrected}}=<0.001$), were negatively correlated. Average daily light sleep std also showed a positive correlation ($r = 0.18$, $p_{\text{corrected}}=<0.001$). A full list of coefficients is provided in
 \hyperref[tab:S3_metabolic_correlation]{Supplementary Table S3}.

The univariate correlation analysis successfully identified numerous wearable, demographic, and blood biomarker features significantly associated with the CKM health indicators (Chol/HDL, eGFR, and HOMA-IR). Notably, demographic variables such as age and gender exhibited strong correlations with several indicators (e.g., age with eGFR, gender with Chol/HDL). Because these factors are known confounders in cardiovascular and metabolic epidemiology, and their influence could mask or distort the true associations of the other features, it is necessary to account for them. Therefore, the subsequent analysis focuses on adjusting the observed correlations for the effects of age and gender to better isolate the independent contributions of the wearable and blood biomarker features.

\subsection{Characterization of Modifiable CKM Health Determinants Independent of Age and Gender}
To mitigate the strong confounding effects of non modifiable demographic attributes, we used partial Pearson correlation analysis to adjust for confounding effects. Specifically, the correlation analysis described in this section systematically controlled age and gender. The correlation with age and gender are presented in \hyperref[fig:supplement_4]{supplementary figure S4}. This methodology was crucial for stripping away the variance explained by demographics, thereby revealing the underlying and modifiable influences of CKM metrics, wearables data, and blood biomarkers, which represent critical targets for future health interventions. Following the partial correlation calculations, we applied a multiple comparisons correction to the resulting p-values to mitigate the risk of Type I errors (false positives). The Benjamini-Hochberg (FDR) procedure was used for this correction, with a false discovery rate (FDR) controlled at a significance level of $\alpha=0.05$. Only features with a partial correlation coefficient with a corrected p-value below the significance level were considered statistically significant and included in the final analysis. 

In the assessment of factors related to cardiovascular health (Chol/HDL ratio), the strongest association within the demographic and anthropometric features was observed with BMI ($r =  0.29, p_{\text{corrected}}=<0.001$).
Analysis of the wearable-derived features revealed a spectrum of significant correlations, primarily reflecting measures of physical activity and physiological status. The most strongly associated feature was the average resting heart rate ($r =  0.16, p_{\text{corrected}}=<0.001$), followed closely by inverse correlations with markers of physical activity: average daily active zone minutes ($r = -0.15, p_{\text{corrected}}=<0.001$) and average daily step counts ($r = -0.11, p_{\text{corrected}}=<0.001$). Notably, average daily respiration rate also demonstrated a small but significant positive correlation ($r = 0.10, p_{\text{corrected}}=<0.001$).
Among the blood-based laboratory features, measures related to insulin and inflammation showed the highest magnitudes of correlation. The top three significant associations included insulin ($r = 0.29, p_{\text{corrected}}=<0.001$), CRP ($r =  0.28, p_{\text{corrected}}=<0.001$), red blood cell count ($r = 0.23, p_{\text{corrected}}=<0.001$).
A comprehensive list of all correlation coefficients is provided in \hyperref[tab:S4_adjusted_cardio]{Supplementary Table S4} and \hyperref[fig:supplement_3]{Supplementary Figure S3}.

\textbf{Correlates of Kidney Health (eGFR)}:The investigation of factors associated with kidney health status (eGFR) employed a covariate-adjusted analysis. After accounting for age and gender across the three feature categories (demographics, wearables, and blood biomarkers), a significant attenuation of many initial correlations was observed. Only calcium ($r = -0.16, p_{\text{corrected}}=<0.001$) and hematocrit ($r = -0.14, p_{\text{corrected}}=<0.001$) retained significant independent associations with eGFR. A comprehensive list of all correlation coefficients is provided in \hyperref[tab:S5_adjusted_kidney]{Supplementary Table S5} and \hyperref[fig:supplement_3]{Supplementary Figure S3}.

A critical finding was the complete attenuation of significance for all wearable-derived sleep features following adjustment for age and gender. This effect is likely attributable to the strong, well-documented physiological dependency between age and sleep architecture (\hyperref[fig:supplement_5]{Supplementary Figure S5},\hyperref[fig:supplement_6]{Supplementary Figure S6},\hyperref[fig:supplement_7]{Supplementary Figure S7})\citep{Visseren2021-es,Chan2002-ir,Ohayon2004-me,Li2018-sl,Duffy2015-ag}. Since age itself is a major confounder that directly influences the sleep variables captured by wearable devices, the significance of sleep features as independent predictors of health outcomes (such as eGFR) is often absorbed by the age variable during multivariate regression or correlation analysis.

\textbf{Correlates of Metabolic Health (HOMA-IR)}: For metabolic health status, quantified by the Homeostasis Model Assessment of Insulin Resistance (HOMA-IR), numerous features maintained robust, significant correlations even after adjusting for all covariates. Within the demographic and anthropometric category, BMI emerged as a significant correlate ($r = 0.29, p_{\text{corrected}}=<0.001$), underscoring its central role in metabolic dysfunction.

The strongest associations were observed among the blood-based biomarkers, particularly those directly related to glucose and lipid metabolism: glucose ($r =  0.53, p_{\text{corrected}}=<0.001$), triglycerides ($r = 0.39, p_{\text{corrected}}=<0.001$), HDL ($r = -0.33,  p_{\text{corrected}}=<0.001$), and the Chol/HDL ratio ($r = 0.26, p_{\text{corrected}}=<0.001$).

Crucially, several wearable-derived features maintained significant, independent correlations with HOMA-IR, suggesting their utility as non-invasive indicators of metabolic stress. These included an inverse association with average resting heart rate ($r = -0.28, p_{\text{corrected}}=<0.001$) and average daily steps count ($r = -0.25,  p_{\text{corrected}}=<0.001$). Furthermore, markers of physiological variability standard deviation of respiration rate and standard deviation of light sleep as well as average daily respiration rate ($r = 0.19,  p_{\text{corrected}}=<0.001$) also retained significant associations. A comprehensive list of all correlation coefficients is provided in \hyperref[fig:supplement_3]{Supplementary Figure S3},\hyperref[tab:S6_metabolic_adjusted]{Supplementary Table S6}.  

\subsection{Prediction of Cardiovascular-Kidney-Metabolic Health markers Using Wearable Statistics, WFM embeddings and Routine Blood Biomarkers}

To identify the impact of various features on predicting the key biomarkers within the CKM subsystems, XGBoost models were employed using a 5-fold cross-validation approach (\hyperref[tab2:target_performance]{Table 2}). For each CKM subsystem, continuous values for Chol/HDL, eGFR, and HOMA-IR were predicted. Additionally, classification thresholds were applied to each subsystem's predictions to identify healthy individuals within the dataset (\hyperref[fig:supplement_8]{Supplementary figure S8}, \hyperref[fig:supplement_9]{Supplementary figure S9}, \hyperref[fig:supplement_10]{Supplementary figure S10}). Due to varying data completeness requirements between the high-resolution data needed for WFM embeddings and the daily summaries used for standard wearable statistics, the analysis cohorts differed slightly (N=923 for WFM models; N=819 for wearable statistics models; see Methods,\hyperref[fig:supplement_13]{Supplementary Figure S13} and \hyperref[fig:supplement_14]{Supplementary Figure 14})

\begin{sidewaystable}[p]
\centering
\scriptsize
\renewcommand{\arraystretch}{0.72}
\setlength{\tabcolsep}{6pt}
\caption{\textbf{Model Performance Across Biomarker Targets.} Evaluation metrics ($R^2$, RMSE, and MAE) across unimodal and multimodal feature sets for predicting Chol/HDL ratio, eGFR, and HOMA-IR.}
\label{tab2:target_performance}
\vspace{4pt}
\begin{tabular}{@{} l l r r r @{}}
\toprule
\textbf{Target} & \textbf{Feature Set} & \boldmath$R^2$ \textbf{Score} & \textbf{RMSE} & \textbf{MAE} \\ \midrule

% ----------------- Chol/HDL -----------------
\multirow{20}{*}{\textbf{Chol/HDL}} 
 & demo & $0.08 \pm 0.04$ & $0.95 \pm 0.06$ & $0.73 \pm 0.03$ \\
 & wearable stat & $0.02 \pm 0.02$ & $0.91 \pm 0.06$ & $0.71 \pm 0.05$ \\
 & WFM embeddings & $0.08 \pm 0.03$ & $0.94 \pm 0.04$ & $0.73 \pm 0.02$ \\
 & renal\_function\_panel & $0.02 \pm 0.02$ & $0.98 \pm 0.06$ & $0.78 \pm 0.04$ \\
 & CBC & $0.06 \pm 0.08$ & $0.96 \pm 0.06$ & $0.75 \pm 0.05$ \\
 & basic\_metabolic\_panel & $-0.06 \pm 0.03$ & $1.02 \pm 0.05$ & $0.79 \pm 0.03$ \\
 & liver\_panel & $0.13 \pm 0.02$ & $0.95 \pm 0.06$ & $0.74 \pm 0.05$ \\
 & endocrine\_panel & $0.07 \pm 0.05$ & $0.99 \pm 0.07$ & $0.75 \pm 0.06$ \\
 & demo + wearable stat & $0.10 \pm 0.03$ & $0.88 \pm 0.07$ & $0.68 \pm 0.04$ \\
 & demo + WFM embeddings & $0.11 \pm 0.04$ & $0.93 \pm 0.04$ & $0.71 \pm 0.01$ \\
 & demo + wearable + CBC & $0.14 \pm 0.05$ & $0.94 \pm 0.07$ & $0.73 \pm 0.06$ \\
 & demo + WFM embeddings + CBC & $0.14 \pm 0.04$ & $0.90 \pm 0.06$ & $0.69 \pm 0.04$ \\
 & demo + wearable stat + CBC + liver\_panel & $0.18 \pm 0.02$ & $0.89 \pm 0.04$ & $0.68 \pm 0.04$ \\
 & demo + WFM embeddings + CBC + liver\_panel & $0.12 \pm 0.05$ & $0.92 \pm 0.07$ & $0.69 \pm 0.03$ \\
 & demo + wearable + CBC + liver\_panel + endocrine\_panel & $0.18 \pm 0.08$ & $0.88 \pm 0.05$ & $0.68 \pm 0.05$ \\
 & demo + WFM embeddings + CBC + liver\_panel + endocrine\_panel & $0.16 \pm 0.05$ & $0.92 \pm 0.05$ & $0.70 \pm 0.03$ \\
 & demo + wearable stat + CBC + liver\_panel + endocrine\_panel + basic\_metabolic\_panel & $0.18 \pm 0.06$ & $0.88 \pm 0.05$ & $0.68 \pm 0.04$ \\
 & demo + WFM embeddings + CBC + liver\_panel + endocrine\_panel + basic\_metabolic\_panel & $0.15 \pm 0.05$ & $0.92 \pm 0.05$ & $0.70 \pm 0.03$ \\
 & demo + wearable + CBC + liver\_panel + endocrine\_panel + basic\_metabolic\_panel + renal\_function\_panel & $0.19 \pm 0.05$ & $0.87 \pm 0.05$ & $0.67 \pm 0.04$ \\
 & demo + WFM embeddings + CBC + liver\_panel + endocrine\_panel + basic\_metabolic\_panel + renal\_function\_panel & $0.17 \pm 0.04$ & $0.91 \pm 0.05$ & $0.69 \pm 0.03$ \\ \midrule

% ----------------- eGFR -----------------
\multirow{20}{*}{\textbf{eGFR}} 
 & demo & $0.26 \pm 0.04$ & $14.67 \pm 0.63$ & $12.33 \pm 0.69$ \\
 & wearable stat & $0.01 \pm 0.06$ & $17.08 \pm 0.78$ & $13.92 \pm 0.71$ \\
 & WFM embeddings & $0.18 \pm 0.06$ & $14.95 \pm 0.57$ & $12.31 \pm 0.47$ \\
 & lipid\_panel & $-0.15 \pm 0.06$ & $18.12 \pm 0.73$ & $14.75 \pm 0.71$ \\
 & CBC & $-0.01 \pm 0.02$ & $17.10 \pm 0.56$ & $13.98 \pm 0.41$ \\
 & basic\_metabolic\_panel & $-0.04 \pm 0.04$ & $17.37 \pm 0.75$ & $14.40 \pm 0.76$ \\
 & liver\_panel & $-0.04 \pm 0.05$ & $17.50 \pm 0.75$ & $14.29 \pm 0.62$ \\
 & endocrine\_panel & $0.04 \pm 0.03$ & $16.26 \pm 0.92$ & $13.16 \pm 0.88$ \\
 & demo + wearable & $0.18 \pm 0.05$ & $15.57 \pm 0.84$ & $12.85 \pm 0.69$ \\
 & demo + WFM embeddings & $0.28 \pm 0.06$ & $14.03 \pm 0.69$ & $11.67 \pm 0.44$ \\
 & demo + wearable stat + CBC & $0.21 \pm 0.03$ & $15.02 \pm 0.69$ & $12.49 \pm 0.50$ \\
 & demo + WFM embeddings + CBC & $0.30 \pm 0.06$ & $14.24 \pm 0.73$ & $11.80 \pm 0.53$ \\
 & demo + wearable + CBC + liver\_panel & $0.22 \pm 0.03$ & $15.01 \pm 0.68$ & $12.49 \pm 0.51$ \\
 & demo + WFM embeddings + CBC + liver\_panel & $0.29 \pm 0.05$ & $14.05 \pm 0.56$ & $11.60 \pm 0.43$ \\
 & demo + wearable stat + CBC + liver\_panel + endocrine\_panel & $0.28 \pm 0.06$ & $14.78 \pm 0.82$ & $12.36 \pm 0.82$ \\
 & demo + WFM embeddings + CBC + liver\_panel + endocrine\_panel & $0.26 \pm 0.09$ & $14.28 \pm 0.48$ & $11.77 \pm 0.12$ \\
 & demo + wearable + CBC + liver\_panel + endocrine\_panel + basic\_metabolic\_panel & $0.29 \pm 0.07$ & $14.63 \pm 0.83$ & $12.24 \pm 0.79$ \\
 & demo + WFM embeddings + CBC + liver\_panel + endocrine\_panel + basic\_metabolic\_panel & $0.24 \pm 0.08$ & $14.47 \pm 0.54$ & $11.93 \pm 0.23$ \\
 & demo + wearable stat + CBC + liver\_panel + endocrine\_panel + basic\_metabolic\_panel + lipid\_panel & $0.30 \pm 0.04$ & $14.07 \pm 0.59$ & $11.63 \pm 0.35$ \\
 & demo + WFM embeddings + CBC + liver\_panel + endocrine\_panel + basic\_metabolic\_panel + lipid\_panel & $0.31 \pm 0.01$ & $14.66 \pm 0.70$ & $12.18 \pm 0.51$ \\ \midrule

% ----------------- HOMA-IR -----------------
\multirow{20}{*}{\textbf{HOMA-IR}} 
 & demo & $0.16 \pm 0.04$ & $2.40 \pm 0.12$ & $1.42 \pm 0.10$ \\
 & wearable stats & $-0.00 \pm 0.19$ & $2.30 \pm 0.48$ & $1.42 \pm 0.21$ \\
 & WFM embeddings & $0.11 \pm 0.03$ & $2.88 \pm 0.27$ & $1.73 \pm 0.18$ \\
 & renal\_function\_panel & $-0.02 \pm 0.06$ & $2.64 \pm 0.16$ & $1.64 \pm 0.13$ \\
 & CBC & $0.05 \pm 0.04$ & $2.24 \pm 0.18$ & $1.41 \pm 0.03$ \\
 & basic\_metabolic\_panel & $0.25 \pm 0.07$ & $2.27 \pm 0.18$ & $1.35 \pm 0.15$ \\
 & liver\_panel & $0.13 \pm 0.03$ & $2.48 \pm 0.15$ & $1.48 \pm 0.11$ \\
 & lipid\_panel & $0.12 \pm 0.04$ & $2.41 \pm 0.18$ & $1.38 \pm 0.06$ \\
 & demo + wearable stats & $0.13 \pm 0.18$ & $2.14 \pm 0.44$ & $1.34 \pm 0.18$ \\
 & demo + WFM embeddings & $0.16 \pm 0.05$ & $2.80 \pm 0.27$ & $1.63 \pm 0.13$ \\
 & demo + wearable stats + CBC & $0.21 \pm 0.08$ & $2.35 \pm 0.26$ & $1.40 \pm 0.12$ \\
 & demo + WFM embeddings + CBC & $0.09 \pm 0.05$ & $2.75 \pm 0.47$ & $1.69 \pm 0.20$ \\
 & demo + wearable stats + CBC + liver\_panel & $0.21 \pm 0.10$ & $2.35 \pm 0.28$ & $1.37 \pm 0.11$ \\
 & demo + WFM embeddings + CBC + liver\_panel & $0.15 \pm 0.04$ & $2.66 \pm 0.42$ & $1.63 \pm 0.22$ \\
 & demo + wearable stats + CBC + liver\_panel + lipid\_panel & $0.25 \pm 0.09$ & $2.11 \pm 0.25$ & $1.22 \pm 0.12$ \\
 & demo + WFM embeddings + CBC + liver\_panel + lipid\_panel & $0.26 \pm 0.07$ & $2.55 \pm 0.26$ & $1.60 \pm 0.10$ \\
 & demo + wearable stats + CBC + liver\_panel + lipid\_panel + basic\_metabolic\_panel & $0.43 \pm 0.09$ & $1.85 \pm 0.24$ & $1.07 \pm 0.11$ \\
 & demo + WFM embeddings + CBC + liver\_panel + lipid\_panel + basic\_metabolic\_panel & $0.47 \pm 0.08$ & $2.14 \pm 0.22$ & $1.27 \pm 0.09$ \\
 & demo + wearable stats + CBC + liver\_panel + lipid\_panel + basic\_metabolic\_panel + renal\_function\_panel & $0.45 \pm 0.09$ & $1.81 \pm 0.26$ & $1.03 \pm 0.11$ \\
 & demo + WFM embeddings + CBC + liver\_panel + lipid\_panel + basic\_metabolic\_panel + renal\_function\_panel & $0.46 \pm 0.09$ & $2.15 \pm 0.24$ & $1.29 \pm 0.10$ \\ \bottomrule
\end{tabular}
\end{sidewaystable}

To systematically dissect the predictive contribution of different physiological domains, features were grouped into distinct clinical panels (Demographic, Lipid, Renal, Basic Metabolic, Liver, Endocrine, and Complete Blood Count) alongside two alternative representations of physical activity, sleep and lifestyle: standard hand-engineered wearable statistics (e.g., mean and standard deviation of wearable metrics) and WFM embeddings. To prevent circularity and multicollinearity, target-associated biomarker panels were excluded from their respective models (e.g., the lipid panel was omitted when predicting Chol/HDL; the renal and endocrine panels were excluded for eGFR and HOMA-IR, respectively). 
For the prediction of the Cholesterol-to-HDL ratio (Chol/HDL), a multi-domain data approach utilizing a sequential addition of feature categories generally resulted in a progressive enhancement of model performance. Demographics alone provided a moderate baseline predictive capacity (regression R2= 0.081 ± 0.040 in the wearable stats cohort; classification ROC AUC = 0.665 ± 0.024, respectively). Integrating wearable features with demographics yielded small changes. In the regression task, adding WFM embeddings to demographics resulted in a minor increase in performance (R2=0.112 ± 0.026), whereas adding standard wearable stats achieved a similar effect (R2=0.096 ± 0.032). In isolation, however, standalone WFM embeddings demonstrated significantly higher predictive capacity than standard wearable statistics (regression R2=0.083 ± 0.027 vs. 0.019 ± 0.016; classification ROC AUC = 0.633 ± 0.027 vs. 0.590 ± 0.041). The addition of clinical panels led to a steady increase in variance explained. For the WFM-based models, the sequential addition of CBC and Endocrine panels yielded a peak regression performance of R2=0.167 ± 0.041 when combined with Liver, BMP, and Renal panels\hyperref[tab2:target_performance]{Table 2}. For the models leveraging standard wearable statistics, the peak regression performance reached R2=0.190 ± 0.053 with the same comprehensive feature set. In the classification task, the peak ROC AUC was achieved using standard wearable statistics combined with demographics, CBC, Liver, and Endocrine panels (ROC AUC = 0.727 ± 0.055;\hyperref[fig:supplement_10]{Supplementary Figure S10A}), compared to a peak of 0.658 ± 0.058 for the corresponding fully integrated WFM model \hyperref[fig:supplement_14]{Supplementary Figure Figure S9A}. 

Predicting estimated Glomerular Filtration Rate (eGFR) exhibited a distinct pattern, where demographic features served as the dominant driver of model performance, reflecting their role in clinical eGFR calculation formulas \hyperref[tab2:target_performance]{Table 2}. Demographics alone achieved a highly robust baseline regression performance (R2= 0.257 ± 0.039 in the wearable stats cohort) and classification performance (ROC AUC = 0.720 ± 0.020, respectively). The choice of wearable representation significantly affected the demographic baseline. Adding standard wearable statistics to demographics degraded regression performance (R2dropped to 0.180 ± 0.052;), and classification ROC AUC decreased to 0.699±0.031 \hyperref[fig:supplement_10]{Supplementary Figure S10B}. Conversely, adding WFM embeddings to demographics improved regression performance (R2=0.283 ± 0.056) and maintained stable classification performance (ROC AUC = 0.733 ± 0.051). In isolation, WFM embeddings were highly superior to standard wearables (regression R2=0.185 ± 0.061 vs. 0.012 ± 0.061; classification ROC AUC = 0.687±0.036 vs. 0.610±0.058; \hyperref[fig:supplement_11]{Supplementary Figure S11}). The integration of additional clinical panels (CBC, Liver, Endocrine, BMP, and Lipid) yielded the peak overall performance. The optimal regression model incorporating WFM embeddings achieved an R2 of 0.310 ± 0.012 (Table 2), outperforming the equivalent model using standard wearable statistics (R2=0.290 ± 0.028, \hyperref[tab2:target_performance]{Table 2}). The peak classification performance was also achieved by the WFM-integrated multimodal model (ROC AUC = 0.751 ± 0.022; Supplementary Figure S9B), surpassing the standard wearable statistics model (ROC AUC = 0.727 ± 0.031;\hyperref[fig:supplement_10]{ Supplementary Figure S10,B}).

Models predicting Homa-IR for Insulin Resistance (HOMA-IR) demonstrated the strongest overall performance, characterized by clear incremental gains as clinical and physiological domains were integrated \hyperref[tab2:target_performance]{Table 2}. Demographics alone provided a baseline regression performance of R2= 0.169 ± 0.025 in the wearable stats cohort, with classification ROC AUCs of 0.776 ± 0.022, respectively. Similar to the eGFR models, adding standard wearable statistics to demographics reduced regression performance (R2=0.133 ± 0.181;\hyperref[tab2:target_performance]{Table 2}), whereas integrating WFM embeddings with demographics increased performance (R2=0.155 ± 0.050; \hyperref[tab2:target_performance]{Table 2}). In isolation, WFM embeddings performed comparably to demographics alone (R2=0.107 ± 0.034) and far exceeded standard wearable statistics, which explained virtually no variance (R2=0.001 ± 0.184; \hyperref[fig:supplement_8]{Supplementary Figure S8}). This pattern was also evident in classification, where WFM embeddings alone achieved a ROC AUC of 0.777 ± 0.015 (compared to 0.662 ± 0.043 for standard wearables; \hyperref[fig:supplement_11]{Supplementary Figure S11}).
The sequential addition of blood panels resulted in substantial performance gains. For WFM, a notable drop in regression performance occurred when adding the CBC panel alone (R2=0.090 ± 0.047), but subsequent integration of Liver, Lipid, and BMP panels yielded the highest overall variance explained in the study (R2=0.473 ± 0.079; \hyperref[tab2:target_performance]{Table 2}). For standard wearable stats, the peak regression performance was R2=0.452 ± 0.092 when adding CBC, Liver, Lipid, BMP, and Renal panels\hyperref[tab2:target_performance]{(Table 2)}. Classification models achieved very high accuracy, with peak performance reaching a ROC AUC of 0.871 ± 0.017 for standard wearable statistics combined with the full clinical panel (\hyperref[fig:supplement_10]{Supplementary Figure S10C}), and a comparable ROC AUC of 0.850 ± 0.009 for the WFM-integrated model (\hyperref[fig:supplement_9]{Supplementary Figure S9,C}).

\subsection{Development and Validation of a Wearable and CKM driven Health Age Models}
\begin{figure*}[!htbp] 
\centering
\includegraphics[width=\textwidth, height=0.8\textheight, keepaspectratio]{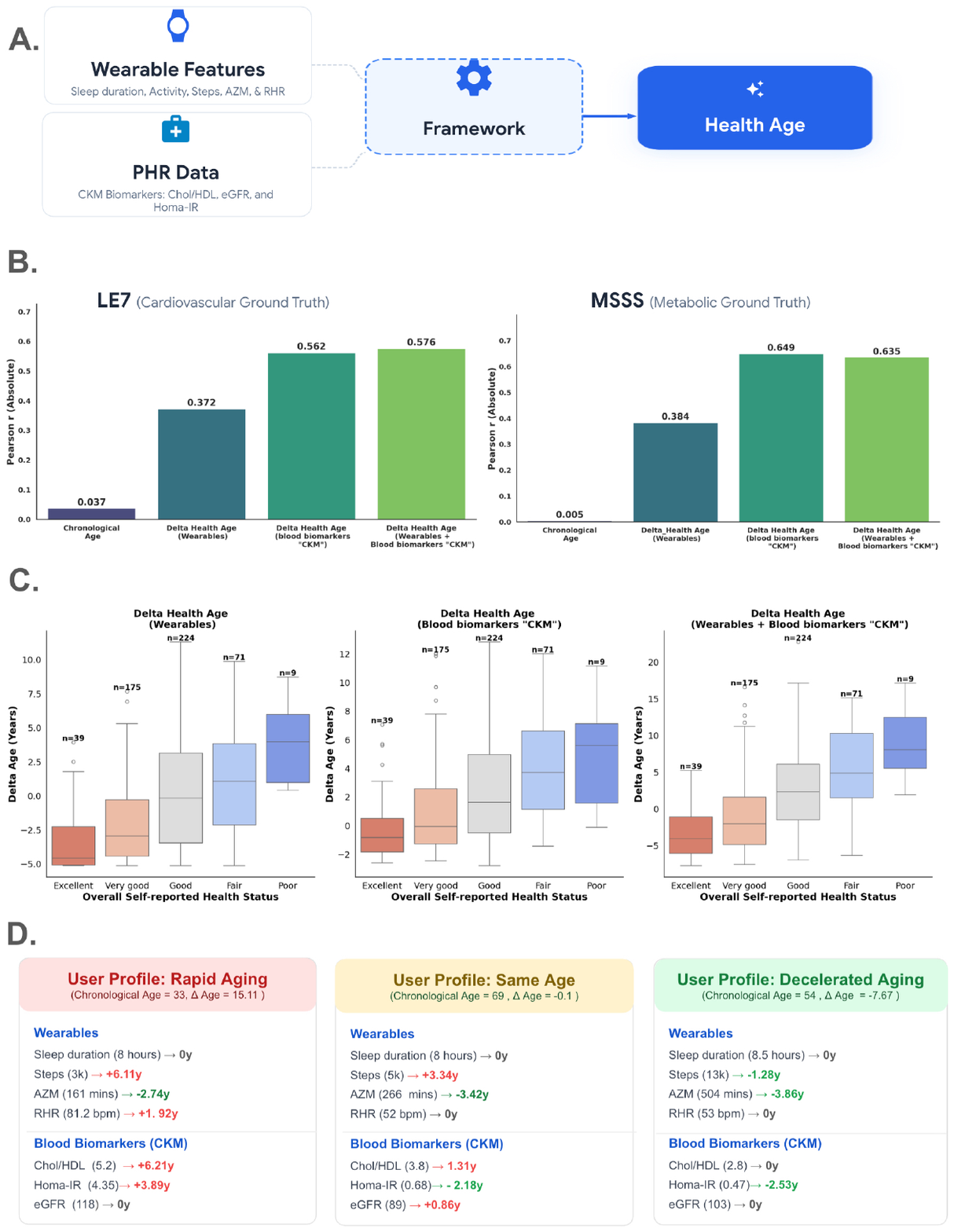} 

    \caption{\textbf{Framework and validation of wearable and biomarker-driven Health Age models.}  (A) Methodological schematic of the Health Age framework. (B) Comparative validation against clinical ground truths. Pearson correlation coefficients (absolute values, |r|) benchmarking baseline Chronological Age against three iterations of the delta Health Age model, Wearables-Based, CKM-Based (blood biomarkers), and a Combined (Wearables + PHR) model. (C) Real-world association of the three Health Age models with self-reported health status.}
    \label{fig:fig_5}
\end{figure*}

To translate multi-system CKM health status and consumer wearable signals into an intuitive, actionable metric, we developed a composite Health Age model \hyperref[fig:fig_5]{Figure 5}. To isolate the predictive utility of digital health metrics versus clinical data, we developed and evaluated three distinct iterations of the Health Age model: (1) a Wearables-Based model leveraging digital features (RHR, step count, and Active Zone Minutes), (2) a CKM-Based model built on clinical biomarkers (Chol/HDL, Homa-IR and eGFR), and (3) a Combined model integrating both wearable features and CKM metrics. Chronological age was utilized as the baseline control, given its established status as a standard indicator of overall risk of morbidity and mortality\citep{beltran2022modeling}.

To establish clinical and real-world validity, we cross-referenced the derived Delta Health Age against recognized clinical gold standards namely Metabolic syndrome severity score (MSSS) and a modified version of AHA life essential 8 (i.e, LE7 excluding nutrition data) as well as user self-reported overall health status\hyperref[fig:fig_5]{(Figure 5B \&C)}. Utilizing the delta Health Age (Health Age - Chronological age) effectively standardizes the metric across the lifespan, allowing us to evaluate whether a participant's accelerated or decelerated aging phenotype is truly associated with their cardiometabolic and behavioral health status independently of their actual birth year.

While baseline chronological age correlated weakly with clinical both gold standards, the Combined mode and CKM-Based demonstrated the strongest overall alignment across standards. Specifically, the Combined model (|r| = 0.635, p <0.001) and CKM-Based (|r| = 0.649, p <0.001) achieved strong Positive correlation with the Metabolic Syndrome Severity (MSS) score, outperforming the Wearables-Based (|r| = 0.384, p <0.001) models. A similar trend was observed when validating against the modified LE 8 score (i.e., LE7), where the Combined model showed the most robust inverse correlation (|r| = 0.575, p <0.001), followed by the CKM-Based (|r|= 0.562, p <0.001) and the Wearables-Based model (|r| = 0.372, p <0.001), highlighting the unique capacity of continuous behavior tracking to capture lifestyle-driven health factors \hyperref[fig:fig_5]{(Figure 5)}.

To enhance model stability and ensure statistical rigor, we instituted an additional step to our Health Age frameworks to algorithmically correct for internal feature multicollinearity\hyperref[fig:supplement_12]{(Supplementary figure S12)}. 

In contrast, standard validation instruments like the MSSS and LE7 do not adjust for internal collinearity, inherently retaining overlapping and mathematically redundant feature variance. When cross-referencing our collinearity-corrected Health Age against these uncorrected clinical benchmarks, we observed a systematic minor attenuation in the raw correlation coefficients\hyperref[fig:supplement_12]{(Supplementary figure S12)}. This expected drop in correlation does not indicate a lack of alignment; rather, it reflects the mathematical divergence between our highly regularized, orthogonalized model architecture and the unadjusted, variance-inflated structures of traditional clinical scores. 

Beyond clinical benchmarks, all three models of Health Age reflected the participants' subjective well-being, demonstrating a significant association with user self-reported overall health status (p <0.001). Taken together, these results indicate that a wearable-derived Health Age serves as a valid, multi-dimensional proxy for systemic physiological health status, offering a passive, continuous and scalable method for tracking health status outside traditional clinical settings.

%% file: sections/3-discussion.tex
\section{Discussion}
\label{sec:discussion}

This study demonstrated that CKM syndrome is a heterogeneous condition characterized by multi-system physiological disruptions. To rigorously quantify this departure from a healthy physiological state across the interacting subsystems, we introduced a novel system-specific deviance score, by subtracting the observed value of each of CKM markers from its established healthy clinical threshold and normalizing the result by the marker's standard deviation (SD). By leveraging a large cohort of 819 participants, this modeling approach successfully mapped how distinct yet overlapping pathways of decline manifest across different organ systems. The resulting subclinical heterogeneity strongly challenges the traditional view of aging as a synchronous, uniform decline \citep{Prattichizzo2024}.  Our findings indicate that individual organ systems undergo independent trajectories of deterioration, with the cardiovascular system displaying the highest prevalence of early subclinical impairment (deviance below $-1$) at 13.35\%, followed by metabolic health at 9.1\% and kidney health at 6.25\%. Crucially, while dual-system issues, most notably the cardiometabolic dyad affecting 2.18\% of the population were observed, no single participant exhibited significant concurrent subclinical deficits across all three systems simultaneously. This ultimately underscores a critical reality: CKM subclinical risk accumulates asynchronously, offering a strategic window for targeted, preventative interventions before multi-system collapse occurs. The recognition of these distinct, yet overlapping, pathways of decline underscores the necessity of a multidisciplinary approach as emphasized by recent frameworks\citep{Carollo2025}.

To effectively target these early, organ-specific subclinical declines, the frameworks should rely on continuous biomarkers that reflect daily physiological stress, a demand increasingly met by consumer-grade wearable devices. In our models, continuous behavioral and autonomic features derived from wearables emerged as powerful independent correlates of cardiovascular and metabolic health\citep{Kaminsky2022-im}. Specifically, elevated resting heart rate (RHR), reduced daily step counts, diminished active zone minutes, and fragmented sleep architecture were correlated to Chol/HDL and HOMA-IR. The integration of these continuous, objective measures aligns with a growing body of literature advocating for the use of digital phenotyping in enhancing traditional cardiovascular risk stratification models\citep{Hughes2023-we}. A high RHR is a recognized, independent risk factor for cardiovascular disease (CVD) and is frequently associated with an adverse metabolic profile, including dyslipidemia\citep{Wu2019-re}. Specifically, studies have reported that elevated RHR is associated with an increased prevalence of metabolic syndrome\citep{Kaminsky2022-im}, and its components, such as high blood glucose and elevated triglycerides (TG)\citep{Wu2019-re}. Accelerometer-based metrics, such as daily step count and measures of moderate-to-vigorous physical activity (i.e., active zone minutes), are strongly and inversely associated with adverse cardiometabolic markers. Higher daily step counts are consistently linked to a more favorable lipid profile, including lower levels of total cholesterol and higher levels of HDL-C\citep{Hughes2023-we,Wu2019-re,Wang2025-as}. The American Heart Association (AHA) and other major health organizations emphasize that increased physical activity significantly improves cardiovascular health outcomes by improving metabolic fitness and reducing insulin resistance, which in turn positively regulates lipid metabolism\citep{Bahls2025-ph}. Wearables capture homeostatic resilience with a temporal granularity impossible during sporadic clinical visits. Mechanistically, the inverse relationship between physical activity and atherogenic lipids is driven by improved insulin sensitivity and optimized lipid metabolism, while the protective effects of sleep regularity are mediated through stable circadian regulation of metabolic pathways\citep{Bian2025-de,Bahls2025-ph,Huang2022-sl}. These digital biomarkers were further complemented by traditional laboratory panels and visceral adiposity proxies; BMI served as a dominant, independent driver of both unfavorable lipid shifts and metabolic stress, while hematological markers (such as hematocrit and hemoglobin) and systemic inflammatory markers (including C-reactive protein and white blood cell counts) highlighted the chronic, subclinical inflammatory milieu that fuels insulin resistance (HOMA-IR) \citep{Oda2010-co,Sebekova2024-al,McLaughlin2002-di}. 

Heterogeneity in these tracking mechanisms reveals that the predictive utility of wearable data is not uniform across all organ systems, presenting a critical nuance in digital nephrology. While autonomic metrics like heart rate variability (HRV) and deep sleep duration initially demonstrated strong univariable correlations with kidney function (eGFR), these associations were entirely extinguished during multivariate adjustment for age and gender. This statistical vanishing act indicates that wearable sleep and autonomic metrics are fundamentally confounded by foundational demographic variables when predicting kidney health. Following rigorous adjustment, only the core physiological biomarkers of calcium and hematocrit remained independent, significant predictors of eGFR.  This robust, inverse relationship confirms Hct as a core physiological marker directly linked to the consequences of diminishing renal function, independent of demographic factors\citep{Gutierrez2016-re,Go2006-he}. Given the asymptomatic and silent nature of CKD in its early stages, a proactive approach to prevention and surveillance is essential. Our findings confirm the established physiological reality that eGFR exhibits an age-dependent decline in the general population, even in the absence of overt kidney disease\citep{Guppy2024-ra}. The typical rate of decline in healthy adults is estimated to be between 0.37 and 1.07 mL/min/1.73 m2 per year. This observation supports the current clinical practice guidelines, such as those advocating to utilize age and gender specific eGFR thresholds to accurately stage and monitor patients, thus avoiding over-diagnosis of CKD in elderly individuals.

The performance of our predictive models highlights the clinical utility of multimodal data integration, demonstrating that the best representation of CKM subsystem health requires combining behavioral, demographic, and clinical data. This is particularly evident in the prediction of insulin resistance (HOMA-IR), where the multimodal model achieved a peak regression $R^2$ of $0.473 \pm 0.079$ and a classification ROC AUC of $0.850 \pm 0.009$ (and $0.871 \pm 0.017$ with standard statistics). A screening accuracy of this magnitude utilizing passive, continuous wearable tracking combined with basic clinical panels represents a highly scalable, non-invasive mechanism for the early detection of metabolic dysfunction, which frequently remains asymptomatic for years \citep{Metwally2026}. However, our results also reveal important nuances in how different wearable representations interact with clinical data. While standalone WFM embeddings consistently outperformed standard statistics across all subsystems, this advantage was not uniform when clinical panels were added. In eGFR and HOMA-IR models, WFM embeddings maintained their superiority, suggesting that the WFM embeddings capture complex, non-linear physiological signatures (such as subtle autonomic fluctuations or sleep architecture changes) that are complementary to standard blood tests. In contrast, for Chol/HDL prediction, standard wearable statistics yielded slightly better performance in the fully integrated model. This suggests that for cardiovascular lipid risk, simple, interpretable behavioral metrics (like step counts and sleep volume) may capture the necessary lifestyle-driven signals as effectively as high-dimensional embeddings once standard blood biomarkers are present, indicating a degree of redundancy between the WFM features and the clinical lipid panels. Finally, the limits of eGFR prediction ($R^2 \approx 0.31$, ROC AUC $\approx 0.75$) reinforce the clinical reality that digital screening is not a replacement for direct biochemical tracking. Because eGFR is clinically calculated based on demographics, a large portion of its variance is mathematically fixed. Nevertheless, the fact that WFM embeddings provided an independent, additive boost to the demographic baseline ($R^2$ increasing from $0.227$ to $0.283$) demonstrates that wearables can detect subclinical kidney stress that demographics alone miss, serving as an effective early warning system to prompt formal clinical evaluation.

To contextualize the practical implications of our findings, it is essential to mention that demographics and established blood biomarkers are undeniably powerful metrics for CKM risk assessments. However, these metrics fundamentally represent static snapshots in time, capturing the individual's physiological state only at the specific moment of clinical encounter. In contrast, wearable sensors offer a paradigm shift through the continuous, longitudinal tracking of real-world physiology and lifestyle behaviors. CKM health is not static; it is dynamically influenced by daily physical activity, sleep architecture, and autonomic stress responses. While a static blood biomarker reflects downstream pathology, wearable data captures the upstream behavioral and lifestyle drivers that actively modulate CKM health status over time. The incremental utility of this approach is clearly demonstrated by our model’s performance. The integration of wearable foundational models yielded superior predictive performance compared to hand-crafted statistical wearable features (i.e., mean and STD of wearable features). This performance boost signifies that foundational models can capture complex, non-linear physiological signatures within dense time-series data that traditional hand-crafted wearable features. Rather than replacing established biomarkers, these advanced digital models unlock the full predictive potential of consumer sensors, translating raw telemetry into highly nuanced risk phenotypes. 

To translate multi-system CKM health status and consumer wearable signals into an intuitive, actionable metric, we integrated these digital and blood biomarker data verticals to develop a novel Health Age metric, extending previous paradigms of biological age estimation\citep{levine2018epigenetic}. Rather than looking at isolated, asynchronous system deviances, this composite score holistically unifies core wearable streams, and CKM blood biomarkers into a single, intuitive phenotypic index. The clinical validity of this Health Age metric is validated by its strong concordance with established, gold-standard clinical ground truths, including the modified version of American Heart Association’s Life’s Essentials 8\citep{lloyd2022lifes}(i.e., LE7) and the Metabolic Syndrome Severity Score (MSSS)\citep{deboer2017clinical} as well as self-reported health status. Beyond its clinical concordance, a primary advantage of this wearable and CKM-infused Health Age is its utility as a passive, continuous monitoring tool that can actively drive preventive behavior change. Traditional risk indices like the MSSS and LE8 are fundamentally episodic and reactive, requiring resource-intensive clinic visits, active laboratory draws, or manual self-reporting that limits their longitudinal scalability. In contrast, our framework leverages a passive data stream that continually updates to reflect a user’s immediate physiological state. This continuous feedback loop provides a highly responsive window for timely, personalized digital coaching interventions\citep{larsen2022effectiveness}. By translating complex, multi-system CKM signals into a singular, intuitive metric, the Health Age can serve as an actionable biofeedback mechanism for the user. Fluctuations in the score can visually demonstrate the immediate physiological consequences of acute lifestyle choices such as a week of poor sleep or sedentary behavior thereby fostering sustained engagement and powering targeted behavior change strategies that are crucial for early cardiometabolic prevention.

We acknowledge, however, an inherent metric overlap between our Health Age framework and these clinical validation benchmarks, which introduces a potential for circular validation that warrants discussion. Specifically, core inputs such as fasting blood glucose, lipids, and physical activity levels or close proxy of them are foundational to both the calculation of MSSS and LE8 scores, as well as our wearable and CKM-centric digital phenotype. Crucially, our Health Age metric does not treat these variables as static, isolated thresholds. While clinical indices like MSSS or LE8 utilize linear, cross-sectional modeling of conventional biomarkers, our framework non-linearly unifies these blood biomarkers with continuous, longitudinal wearable streams in a passive manner.

Finally, we address the architectural boundary conditions embedded within our Health Age framework, which prevent calculated reductions of more than 5.0 years younger than chronological age for wearable streams alone, and 8.4 years younger when combined with CKM biomarkers. Rather than representing an arbitrary truncation, this younger-side cap is fundamentally rooted in the non-linear, asymptotic relationships observed between lifestyle exposures and chronic disease risk. Large-scale epidemiological data from the All of Us Research Program utilizing commercial wrist-worn trackers demonstrate that the protective benefits of behavioral variables do not scale infinitely; for instance, the risk reduction for critical cardiometabolic conditions like type 2 diabetes and hypertension completely plateaus after reaching a threshold of approximately 8,000 to 9,000 steps per day\citep{master2022association}. Consequently, a participant averaging 20,000 daily steps yields negligible additional longevity returns compared to one maintaining a highly consistent 10,000-step routine. By incorporating these biological saturation points into our model features, the Health Age avoids mathematical overfitting that would otherwise generate clinically implausible rejuvenations. This design ensures that the metric accurately reflects true homeostatic limits: while optimal lifestyle habits and CKM blood profiles drastically mitigate risk, they ultimately reach a physiological plateau that cannot bypass the intrinsic, non-modifiable cellular baselines of chronological aging.

While these findings present a promising path forward for digital medicine, certain cohort and hardware limitations must be carefully contextualized. We enrolled a diverse group of participants spanning various ages and genders. We are aware that our final cohort may have overrepresented those with cardiometabolic disease due to the study name (i.e. Wearables for Metabolic Health Study) or proactive in health management due to the requirement for blood lab tests and use of wearable devices. However, many basic wearable fitness trackers with the necessary measurement capabilities are becoming increasingly affordable and widely available, and we expect this trend to continue. In our models, we incorporated features commonly found on lower-cost devices, such as resting heart rate, daily steps, sleep stages, and heart rate variability. Additionally, recognizing the limitations of affordable wearables in providing high-resolution data. Literature suggests that while single-day measurements can vary by 20–30\% due to motion artifacts, transient behavioral shifts, or acute physiological stressors \citep{fennell2024inter,lederer2023importance,germini2022accuracy}, a 14-day window is sufficient to establish a stable baseline. By focusing on the 14-day mean, we minimize the impact of transient measurement errors and outlier days, providing a more robust representation of a participant's true physiological state and ensuring the applicability of our models to a wider range of devices while maintaining their relevance and practical utility. We acknowledge that consumer smartwatch brands, sensors, and firmware may provide different accuracy, precision, and temporal resolution for measuring digital biomarkers (heart rate, steps, sleep) compared to research-grade or clinical-grade devices. In our study, we utilized data from seven different smartwatches and four trackers, all from the same manufacturer, Google/Fitbit, and employing similar algorithms. We implemented extensive quality control measures to ensure the data used for the model were of high fidelity and accuracy, and that the model received sufficient wearable data input before generating a prediction. While we expect the results to generalize to other brands, further investigation and more specific quality control steps to mitigate known accuracy issues are warranted.

%% file: sections/4-methods.tex
\section{Method}
\label{sec:Methods}
\subsection{Participant recruitment via Google Health studies (GHS)}
The study protocol was approved by the institutional review board at Advarra (Columbia, MD, Protocol Pro00074093). All participants provided informed consent. In addition to the consent, participants were also asked to sign the Quest HIPAA authorization as part of the consent flow within the GHS app. Participants were recruited using the “Google Health Studies” (GHS) Application and Fitbit App. GHS is a platform to run digital studies that allows participants to enroll into studies, check eligibility and provide informed consent. GHS enables the collection of wearable features using Fitbit devices or Pixel watches and allows participants to complete questionnaires and order blood tests with Quest Diagnostics. This prospective, observational study resulted in consent of 4418 participants in the United States, where a subset of those (N=841) had complete data and thus were included in our analysis. The study was conducted remotely with one visit to a Quest Patient Service Center for a blood draw. Participants were asked to continuously wear their wrist-worn wearable devices, schedule and complete a blood draw, and answer questionnaires. This study used an electronic informed consent process managed by the GHS App using electronic signatures.

Inclusion criteria included U.S. residents between the ages of 21 - 80, Fitbit users in Android that wear a Fitbit wearable device or a Pixel Watch with heart rate sensing capabilities, users who have at least 3 months of existing data where they have used the device for at least 75\% of the days to track their activity and sleep, participants willing to update their Fitbit Android App, participants willing to install or update their Google Health Studies (GHS) app on their Android phone, participants willing to link (or create) their Quest Diagnostics account to the GHS app, participants able to speak and read English and provide informed consent and HIPAA authorization to participate in study, participants who have access to and be willing to go to a Quest Diagnostics location for blood draws. Wearable devices were limited to any model of a Fitbit device or Pixel watch with heart rate (HR) sensing capabilities (e.g., Pixel Watch, Charge HR, Charge models 2 - 5, Sense, Sense 2, Versa, Versa 2, 3 \& 4, Versa Lite, Inspire HR, Inspire 2, and Luxe). 

Exclusion criteria included participants living in Alaska, Arizona, Hawaii and the US territories since Quest Diagnostics cannot provide blood tests in those states, participants with uncontrolled disease (for example, a recent change in treatment in the past 6 weeks, awaiting review to trigger a change of treatment, a treating physician has indicated the condition is not yet controlled, or where symptoms of the condition are not responding to treatment), or participants with conditions that might make collection of blood samples through venipuncture impractical.

\textbf{Cohort Definition and Data Completeness}: The final study cohort was determined by data availability constraints across the clinical, demographic, and wearable domains. Because standard wearable statistics and the Wearable Foundational Model (WFM) process data at different granularities, they require distinct preprocessing pipelines. For the standard wearable statistics, participants were included if they had daily-level summaries meeting the inclusion criteria, resulting in a cohort of 819 participants\hyperref[fig:supplement_13]{(Supplementary Figure S13)}. For the WFM embeddings, which require continuous minutely resolution data and use median pooling to aggregate representations across all available days per participant, the data completeness criteria yielded a slightly larger cohort of 923 participants \hyperref[fig:supplement_14]{(Supplementary Figure S14)}. Clinical panels and demographic data were subsetted to match the respective wearable cohorts for downstream multimodal modeling.

\subsection{Study Design}
As part of this study, participants were asked to link their Fitbit account with the GHS app. They were also asked to grant GHS permission to collect Fitbit data throughout the study, including data for up to 3 months before study enrollment. Once participants were enrolled in the study they were asked to perform the following: (1) wear their Fitbit device or Pixel watch during the day and while they sleep (at least 3 out of every 4 days) for the duration of the study; (2) complete four questionnaires which will request demographic information, health history and health information such as sleep and exercise habits, participant perception of health, and blood test interpretation (see below for details); (3) schedule an appointment to complete the lab test orders that were placed for them and go to a Quest Patient Service Center for a blood draw within 65 days of enrollment; (4) complete a blood draw at the Quest Patient Service Center; (5) review blood test results in the GHS App when available. 
\subsection{Metadata collection}
Demographics (e.g., age, gender, ethnicity, weight, height), optional measures, such as medical conditions (e.g. diabetes, hyperlipidemia, cardiovascular disease, kidney disease, hypertension, etc.), blood pressure, waist circumference, medications, self-reported health management and habits were collected through a baseline questionnaire that participants completed immediately after enrollment through the GHS app.
\subsection{Blood biomarkers via Quest diagnostics}
Eligible participants were asked to schedule and complete a visit at a Quest Diagnostics Patient Service Center of their choice in their local area. This visit included a standard blood draw to measure the following: Complete Blood Count, Comprehensive Metabolic Panel, insulin, Lipid panel, HbA1c, hs-CRP, GGT and Total Testosterone. Participants were asked to have their blood drawn while fasting for at least 8 hours, early morning, 7am-10am local time, to minimize the effect of the solar diurnal cycle. This study also had clinical oversight by a physician network partner. Lab results from the blood draw were returned to participants and made available for participant review in the GHS app for the duration of the study; however, these were pulled directly from Quest Diagnostics each time it is requested by the participant and not stored in the GHS app. Participants provided consent and HIPAA authorization to grant GHS permission to collect the corresponding results from Quest Diagnostics. Data transferred from Quest Diagnostics were retrieved securely using encrypted protocols. Once the study was completed, lab results remained in the participant’s Quest Account in accordance with Quest's standard practices. 

\subsection{Wearables data via Fitbit devices and Pixel Watches}
Participants were asked to wear their own Fitbit device or Pixel Watch that may include sensors such as PPG optical heart rate tracker, multipurpose electrical sensors, gyroscope, altimeter, accelerometer, on-wrist skin temperature, blood oxygen saturation (SpO2), and ambient light sensor. Participants consented to share the following data from these devices: (1) heart rate metrics: heart rate (HR), resting heart rate captured daily, interbeat interval (IBI, also known as RR interval) calculated from the PPG sensor, and HRV metrics (such as RMSSD, SDNN,Entropy, SDRR, pNN50, etc.) (2) physical activity metrics: Measures of physical activity, including steps, floors, and active zone minutes (AZMs), (3) sleep metrics: bedtime, wake time, sleep duration, sleep stages, sleep quality, sleep coefficients, and sleep logs, (4) respiration and skin temperature: Respiration rate values during the day/night and skin temperature (if the sensor is available on the wearable), (5) SpO2: Blood oxygen saturation values during the day/night (if available on the wearable), (6) weight: Measure of weight that may be logged in the Fitbit account, (7) exercise and activity data: Daily total exercise sessions completed and logs of activities that have been classified. Fitbit daily RHR is calculated from periods of stillness throughout the day, as determined by the on-device accelerometer. If a person wears their device while sleeping, their sleeping heart rate is also included in the calculation. Additional details are described in\citep{Russell2019-in,Speed2023-me}. Daily RMSSD HRV is calculated from pulse intervals measured during sleep periods greater than 3 hours 58.Daily entropy of heart rate variability is a nonlinear time domain measurement (Shannon entropy) computed using the histogram of RR intervals over the entire night. AZMs is a feature that tracks the time a person spends in different heart rate zones during physical activity. A person receives 1 AZM for every minute spent in the “moderate” zone, and 2 AZMs for every minute spent in the “vigorous” or “peak” zones\citep{Fitbit2024-az}. The heart rate zones are based on percentage of heart rate reserve achieved, where the heart rate reserve is the difference between maximum heart rate and resting heart rate Moderate zone is defined from 40-59\% of heart rate reserve, vigorous is from 60-84\%, and peak is 85-100\% \citep{Fitbit2024-az,Fitbit2024-zp}.

\subsection{Wearable Foundational Model (WFM)}
We took a holistic approach to modeling the wearable device data by leveraging a foundation model. These models have become essential in scientific analysis, offering a powerful way to extract robust, high-dimensional feature representations from complex, unlabeled datasets via self-supervised pretraining\citep{liang2024foundation}.To evaluate the utility of such large-scale pretraining on the task of CKM prediction, we used the Large Sensor Model (LSM)\citep{xu2025lsm2}, a foundation whose embeddings have shown utility for a number of downstream health tasks, including activity recognition, mood classification, hypertension detection, anxiety detection, and insulin resistance\citep{Metwally2026,xu2025lsm2}. For CKM prediction, we trained an LSM-V2-style Wearable Foundation Model (WFM), following the exact architecture, objective functions, and pretraining protocol described previously by\citep{Metwally2026,xu2025lsm2}. Briefly, the model was pretrained on 40 million person-hours sampled at minute data resolution from Fitbit users. Each data sample contains 26 minutely aggregated features from a set of 5 sensors (photoplethysmography, accelerometer, skin conductance, altimeter, and temperature) for a time span of 1,440 min (one day). The 26 features used as input for the model embedding function are described in\hyperref[tab:S7_wfm_features]{ Supplementary Table S7}. 

Our pretrained wearable foundational model (WFM) extracted representations from wearables input data at one-minute resolution. The input window size was one day (1,440 min) by 26 signals. Given that each participant has many days of data from wearables, we used median pooling to generate a single embedding per participant. These embeddings have a dimensionality of 384. To reduce dimensionality, the pooled embeddings were first standardized to zero mean and unit variance. Principal Component Analysis (PCA) was then fitted on the training set, and the learned mapping was applied to the test set. To retain the majority of the information while minimizing data redundancy, we retained the minimum number of principal components required to explain 95\% of the total variance.

\subsection{Selection of thresholds for Chol/HDL, eGFR and HOMA-IR}
Based on medical guidelines and information from reputable health organizations, a Chol/HDL ratio of $3.5:1$ or lower is generally considered optimal or very good, indicating a lower risk of heart disease\citep{Bailey2022-bi,Kinosian1994-ch}. Consequently, for the deviance analysis and classification tasks, a threshold of 3.5 was utilized to define a healthy cardiovascular subsystem. According to the National Kidney Foundation, a normal eGFR is generally above $90 \ mL/min/1.73m^2$, but values as low as $60 \ mL/min/1.73m^2$ are considered normal if there is no other evidence of kidney disease. A threshold of $\ge 90 mL/min/1.73m^2$ for eGFR was utilized to define a healthy kidney subsystem in the deviance analysis and classification tasks, consistent with established clinical guidelines for normal kidney function\citep{Stevens2013-ev}. For the HOMA-IR threshold, we used 2.9 which is used in earlier studies and a midpoint between the NHANES-derived threshold for the U.S. population (2.77) and the maximum value in the paper by Geyoso-Diaz et al\cite{Metwally2026,GayosoDiz2013-ir}.

\subsection{Categorization of Clinical and Physiological Features}
To thoroughly evaluate the impact of various clinical and physiological factors on the prediction of CKM subsystem biomarkers, we grouped the comprehensive set of features into distinct categories. These categories and their respective constituents are detailed as follows: Demographic Data: including age, gender (encodes as Male =1, Female =2 for analysis), and BMI; Wearable data: mean and standard deviation of the follow wearable metrics: daily steps count, daily sleep duration, daily heart rate variability (HRV), daily entropy of heart rate variability, daily resting heart rate, daily RMSSD, daily sleep duration, daily deep sleep duration, daily light sleep duration, daily respiration rate, daily rem sleep. Lipid Panel: Features within this panel comprised total cholesterol, high-density lipoprotein (HDL), triglycerides, low-density lipoprotein (LDL), Chol/HDL ratio, and non-HDL cholesterol; Renal Function Panel: including blood urea nitrogen (BUN), creatinine, estimated glomerular filtration rate (eGFR), sodium, potassium, chloride, carbon dioxide (CO2), and the BUN/creatinine ratio; Basic Metabolic Panel: Key markers in this panel were glucose, calcium, total protein, albumin, globulin, and the albumin/globulin ratio; Liver Panel: This category included albumin, globulin, total bilirubin, alkaline phosphatase (ALP), aspartate aminotransferase (AST), alanine aminotransferase (ALT), and gamma-glutamyl transferase (GGT); Endocrine Panel: This panel consisted of HbA1c, insulin, and total testosterone; Complete Blood Count (CBC): The CBC provided a detailed breakdown including white blood cell count, red blood cell count, hemoglobin (Hb), hematocrit, mean corpuscular volume (MCV), mean corpuscular hemoglobin (MCH), mean corpuscular hemoglobin concentration (MCHC), red cell distribution width (RDW), platelet count, mean platelet volume (MPV), absolute neutrophils, absolute lymphocytes, absolute monocytes, absolute eosinophils, absolute basophils, neutrophils (percentage), lymphocytes (percentage), monocytes (percentage), eosinophils (percentage), and basophils (percentage).

\subsection{Partial correlation analysis and confounder adjustment}
Both wearable metrics, such as resting heart rate, heart rate variability, and sleep, as well as lab metrics, like eGFR, naturally change with age and vary between genders\citep{Natarajan2020-hr,Astley2025-as}. To assess the relationship between various wearable and demographic features and the outcome variable for each CKM subsystem while controlling for confounding effects of age and gender, we utilized partial correlation analysis. This method determines the correlation between two variables after the influence of age and gender variables has been removed. The analysis was performed using Python, leveraging the statsmodels and scipy libraries.
The procedure for calculating the partial correlation coefficient between a feature and the outcome variable, adjusting for age and gender, was as follows:
\begin{itemize}
    \item A separate linear regression model was constructed for both the target outcome variable and each individual feature. In each model, the outcome variable was regressed on the two controlling variables, age and gender.
    
    \item The residuals were extracted from each of these regression models. The residuals represent the part of each variable (the target and the feature) that cannot be explained by age and gender.
    
    \item A standard Pearson correlation coefficient was then computed between the residuals of the target outcome and the residuals of each feature. This correlation between the residuals is the partial correlation coefficient, representing the unique association between the feature and the target after removing the shared linear effect of age and gender.

\end{itemize}

Following the partial correlation calculations, we applied a multiple comparisons correction to the resulting p-values to mitigate the risk of Type I errors (false positives). The Benjamini-Hochberg (FDR) procedure was used for this correction, with a false discovery rate (FDR) controlled at a significance level of $\alpha=0.05$. Only features with a partial correlation coefficient above a predefined threshold and a corrected p-value below the significance level were considered statistically significant and included in the final analysis.

\subsection{Modeling and Computational Pipeline}
As shown in Figure 1A, our method consists of three stages: (i) Data preprocessing, (ii) modeling and training, (iii) Prediction and classification. We describe each of these components below.

\subsubsection{Data preprocessing}

\textbf{Age and BMI (Demographics)}: From the user self-reported data on recruiting surveys, we extract a user’s age and compute BMI from height and weight. As a quality control process, Users were excluded if their BMI was greater than 65 or less than 12, resulting in the exclusion of 326 users from the analysis.

\textbf{Digital Markers Derived from Wearables (Wearables Features)}: Using the estimated digital markers from Fitbit algorithms, we aggregated users’ digital markers using mean, standard deviation for the past 14 days prior to blood test collection. 

\textbf{Biomarkers from Blood Biochemistry (Blood Tests)}: As a first filter, we removed any participant who was not fasting as reported in a survey at the time of blood test collection. Additionally, for each experiment, we exclude participants with missing values from the input feature set. Lastly, to remove outliers, we use the true HOMA-IR value ($HOMA-IR = [Fasting Insulin (µU/ml) × Fasting Glucose (mg/dL) ] / 405$) and exclude any participant whose HOMA-IR value is greater than or equal to 20. This exclusion criteria removed 1343 users from the study.   

\textbf{Data Normalization}: The data used for modeling is a concatenation of demographics, aggregated digital markers and blood biomarkers. In order to create a consistent modeling data that is agnostic to the learning model, we standardized input features to have zero mean and unit variance. For each training fold, our “normalizer” object was fit to the data in the training subset (not including samples in the testing subset). The fitted object was then used to transform the samples in both the training and testing subsets. The standardized data was used for all modeling tasks and evaluations.

\subsubsection{Predictive Regression Modeling Framework}
\textbf{Study Targets and Cohort Preparation}: To evaluate the capability of passive digital tracking data (wearable sensors) and clinical laboratory panels in predicting CKM as continuous values, we developed a series of machine learning regression models. We targeted three continuous clinical markers: eGFR, Chol/HDL and Homa-IR. For each target task, we constructed models across a sequence of feature subsets ranging from basic demographics to full clinical panels, combined with either daily aggregated wearable statistics or Wearable Foundational Model (WFM) embeddings.
\textbf{Feature Preprocessing and Dimensionality Reduction}
\begin{itemize}
    \item \textbf{Demographics and Clinical Panels:}Categorical variables were preprocessed using one-hot encoding. Continuous variables were standardized using a z-score scaler ($mu=0,sigma=1$) fit exclusively on the training partition of each split to prevent data leakage.
    \item \textbf{WFM Embeddings}:Daily sensor embeddings were aggregated per participant using median pooling over the tracking window. To prevent overfitting and manage the high-dimensionality of the latent space, Principal Component Analysis (PCA) was applied to the aggregated embeddings, retaining the minimum number of principal components required to explain 95\% of the total variance.

\end{itemize}
\textbf{Model Architecture:} Gradient Boosting Machines via the XGBoost framework were used \cite{Chen2016-xg} since they are highly suited for tabular clinical datasets due to their native ability to handle non-linear decision boundaries, collinear features, and interaction effects without assuming normal distributions.

\textbf{Robust Validation Framework (Repeated Holdout):} To obtain unbiased estimates of model performance and assess stability against split variance, we utilized a Repeated Holdout Validation strategy over five independent runs (using random seeds 42, 100, 200, 300, and 400). For each feature combination and each seed:

\begin{enumerate}
    \item \textbf{Partitioning:} The cohort was randomly split into a Training Set (80\%) and an independent Holdout Test Set (20\%). The test partition was completely set aside and kept blind to all downstream preprocessing, scaling, and tuning steps.
    
    \item \textbf{Inner Hyperparameter Optimization:} To optimize model parameters without leaking test data, a 5-fold Cross-Validation (CV) Grid-like Random Search (\texttt{RandomizedSearchCV}, 20 iterations) was conducted strictly within the 80\% training set. The search space covered:
    \begin{itemize}
        \item \textbf{Number of estimators (trees):} $\{50, 100, 200, 300\}$
        \item \textbf{Learning rate (shrinkage):} $\{0.01, 0.05, 0.1, 0.2\}$
        \item \textbf{Maximum tree depth:} $\{3, 4, 5, 6\}$
        \item \textbf{Subsample and colsample ratios:} $\{0.7, 0.8, 0.9, 1.0\}$
        \item \textbf{L1 and L2 regularization coefficients:} $(\alpha, \lambda)$
    \end{itemize}
    
    \item \textbf{Refitting:} The parameter combination yielding the lowest mean squared error across the 5 validation folds was selected. The final model for that seed was then trained on the entire 80\% training set using these optimized parameters.
    
    \item \textbf{Holdout Evaluation:} The finalized model was evaluated on the unseen 20\% holdout test set.
\end{enumerate}

This process was repeated independently for all five seeds. We report the final performance as the Mean $\pm$ Standard Deviation (SD) across the five test splits.

This process was repeated independently for all five seeds. We report the final performance as the Mean ± Standard Deviation (SD) across the five test splits.

\subsubsection{Predictive Classification Modeling Framework}
\noindent \textbf{Binary Target Definitions and Clinical Justification:} To evaluate the clinical utility of our models in identifying individuals with sub-clinical or clinical metabolic and renal dysfunction, continuous markers were dichotomized into binary targets based on established clinical guidelines: 

\begin{enumerate}
    \item \textbf{Kidney function ($\text{eGFR} < 90\text{ mL/min/1.73m}^2$):} An eGFR threshold of $90\text{ mL/min/1.73m}^2$ was selected to isolate individuals with mild-to-moderate impairment from those with normal kidney function (Stage 1).
    \begin{equation}
    \text{Target} = 
    \begin{cases} 
    1 & \text{if eGFR} < 90 \\ 
    0 & \text{if eGFR} \ge 90 
    \end{cases}
    \end{equation}
    
    \item \textbf{Elevated Cardiovascular Risk ($\text{chol/HDL} \ge 3.5$):} A total cholesterol to HDL ratio threshold of $3.5$ was selected based on clinical dyslipidemia guidelines. Ratios $\ge 3.5$ represent an unfavorable lipid profile associated with moderate-to-high cardiovascular risk.
    \begin{equation}
    \text{Target} = 
    \begin{cases} 
    1 & \text{if chol/HDL} \ge 3.5 \\ 
    0 & \text{if chol/HDL} < 3.5 
    \end{cases}
    \end{equation}
    
    \item \textbf{Insulin Resistance ($\text{HOMA-IR} \ge 2.0$):} A HOMA-IR threshold of $2.0$ was established to identify early metabolic dysfunction and insulin resistance. A HOMA-IR score $\ge 2.0$ indicates significant peripheral insulin insensitivity, serving as a precursor to type 2 diabetes.
    \begin{equation}
    \text{Target} = 
    \begin{cases} 
    1 & \text{if HOMA-IR} \ge 2.0 \\ 
    0 & \text{if HOMA-IR} < 2.0 
    \end{cases}
    \end{equation}
\end{enumerate}

\noindent \textbf{Model Architecture and Handling Class Imbalance:} We trained Gradient Boosted Decision Tree classifiers using the XGBoost framework (\texttt{XGBClassifier} with a binary logistic loss objective function). Because clinical populations often exhibit skewed distributions (e.g., a higher proportion of healthy individuals than those with severe dysfunction), we addressed potential class imbalance dynamically during model training. Specifically, we incorporated the \texttt{scale\_pos\_weight} parameter in the hyperparameter search grid. This parameter adjusts the gradient updates to scale the loss of the minority (positive) class, defined as:
\begin{equation}
\texttt{scale\_pos\_weight} = \frac{\text{Number of negative class samples}}{\text{Number of positive class samples}}
\end{equation}
The search grid sampled \texttt{scale\_pos\_weight} values of $[1, 2, 3, 5]$ to optimize minority class sensitivity without excessively penalizing specificity.

\noindent \textbf{Stratified Validation Framework:} To preserve the prevalence of positive clinical cases across the training and test splits, we utilized a Repeated Stratified Holdout Validation framework. As in the regression pipeline, this was repeated across 5 independent seeds ($42, 100, 200, 300, 400$). For each subset and seed:
\begin{enumerate}
    \item \textbf{Stratified Split:} The dataset was partitioned into an 80\% Training Set and a 20\% Holdout Test Set using stratified sampling (\texttt{train\_test\_split} with \texttt{stratify=y}). This guarantees that the positive-to-negative class ratio remains identical in both the training and test sets.
    \item \textbf{Stratified Inner Tuning:} Hyperparameters were tuned strictly within the training set using 5-fold Stratified Cross-Validation (\texttt{StratifiedKFold}). The parameter search space covered max depth ($[3, 4, 5, 6]$), learning rate ($[0.01, 0.05, 0.1, 0.2]$), estimators ($[50, 100, 200, 300]$), subsampling, regularization, and \texttt{scale\_pos\_weight} options.
    \item \textbf{Optimization Metric:} Hyperparameters were optimized by maximizing the Area Under the Receiver Operating Characteristic curve (ROC AUC) on the inner validation folds.
    \item \textbf{Holdout Evaluation:} The optimized model was refitted on the full 80\% training partition and evaluated once on the unseen 20\% holdout test partition.
\end{enumerate}
We report performance as the $\text{Mean} \pm \text{Standard Deviation (SD)}$ across the five test splits.

\noindent \textbf{Evaluation Metrics for Classification:} To provide a comprehensive assessment of the models' diagnostic utility and performance under varying class balances, we evaluated the models on the holdout test set using six distinct metrics: 
\begin{itemize}
    \item \textbf{Area Under the Receiver Operating Characteristic curve (ROC AUC):} Measures the model's ability to discriminate between healthy and unhealthy states across all decision thresholds. It is independent of class prevalence.
    \item \textbf{Area Under the Precision-Recall curve (PR AUC / Average Precision):} Evaluates classification performance on the minority class. Unlike ROC AUC, PR AUC is highly sensitive to false positives and class imbalance, making it a critical metric for clinical screening utility in low-prevalence cohorts.
\end{itemize}

\subsubsection{Health Age Estimation Utilizing Wearable and Clinical CKM Biomarkers}

\textbf{Model Framework and Theoretical Foundation}: We implemented a parametric physiological aging model rooted in Gompertz-Makeham mortality dynamics\citep{spiegelhalter2016how}. The fundamental assumption of this framework is that an individual's log-transformed risk of mortality or adverse health events increases linearly with chronological age. By leveraging hazard ratios for incidents of cardio-metabolic disease derived from epidemiological literature, an individual's accumulated physiological deviance can be mapped to a corresponding shift in chronological age, yielding a metric termed Health Age.

The biological age delta ($\Delta \text{Age}$), representing the years of accelerated or decelerated aging relative to chronological age, is calculated as follows:
\begin{equation}
\Delta \text{Age} = \frac{\sum_{i} w_i \ln(\text{HR}_i)}{G}
\end{equation}
Where $\text{HR}_i$ represents the hazard ratio (or relative risk) contributed by a specific biomarker $i$, $w_i$ represents an empirical scaling weight applied to the log-hazard space, and $G$ is the standard Gompertz aging rate constant, set at $0.1$ (approximating a 10\% exponential increase in mortality risk per year of life).

The final Biological Health Age ($A_h$) is computed as:
\begin{equation}
A_h = \max(18, A_c + \Delta \text{Age})
\end{equation}
Where $A_c$ represents the chronological age of the individual. A lower bound of $18$ years is structurally enforced to prevent biologically implausible outputs in highly optimized phenotypes.

\textbf{Hazard Ratio Parameterization and Functional Forms}
We developed three different health ages. The first one is based on wearable data. The model integrates four continuous digital biomarkers derived from wearable sensors such as Resting heart rate, Steps counts, Active Zone minutes (AZM) and sleep duration. To capture the epidemiological relationships documented in literature, features are modeled via non-linear, or piecewise linear relationships. The second one is based on CKM metrics. The model integrates the CKM biomarkers, namely Cho/HDL, eGFR and Homa-IR. And the third mode is based on both wearables and CKM metrics. The following is how each metric was modeled in the Health Age framework.

\textbf{Wearable Biomarkers}

\begin{itemize}
    \item \textbf{Resting Heart Rate (RHR):} Modeled using a piecewise linear log-hazard relationship targeting incident heart failure risk\citep{zhang2016resting}. Risk was assumed to be flat at the baseline reference level for highly optimized rates ($< 60$~bpm). For the intermediate range ($60$--$80$~bpm), the log-hazard scales continuously, yielding a hazard ratio (HR) of $1.08$ at $70$~bpm (equivalent to an HR of approximately $1.17$ at $80$~bpm). Above $80$~bpm, the risk accelerates sharply at a rate corresponding to an HR of $1.33$ per $10$~bpm increase.
    
    \item \textbf{Daily Step Counts:} Daily physical activity was modeled using a log-linear relationship between daily step count and the log-hazard of the clinical outcome, incorporating a saturation threshold to account for diminishing returns at higher activity volumes\citep{master2022association}. An optimized baseline of $8{,}000$~steps/day was designated as the reference standard ($\text{HR} = 1.0$). Below the saturation point, each additional $1{,}000$ daily steps was associated with a $12\%$ reduction in risk ($\text{HR} = 0.88$). To prevent overestimating the benefits of extreme activity volumes, a step plateau was enforced at $9{,}000$~steps/day. Consequently, the maximum risk reduction achieved via step volume was capped at a hazard ratio of $0.88$ for all activity levels at or exceeding $9{,}000$~steps/day.
    
    \item \textbf{Sleep Duration:} Sleep duration was modeled using a piecewise linear log-hazard relationship to capture the well-established U-shaped association with cardiovascular disease (CVD) risk\citep{daghlas2019sleep}. A normal sleep window of $6.0$ to $9.0$ hours per night was designated as the reference category ($\text{HR} = 1.0$, $\ln(\text{HR}) = 0$). Outside of this optimal range, risk penalties were applied symmetrically as linear departures from the threshold boundaries. For short sleep duration ($< 6.0$ hours), each hourly deficit below the $6$-hour threshold log-linearly increased risk, corresponding to a $20\%$ increase in CVD risk per hour of sleep lost ($\text{HR} = 1.20$). Conversely, for long sleep duration ($> 9.0$ hours), each excess hour above the $9$-hour threshold incurred a log-linear penalty, translating to a $34\%$ increase in risk per additional hour ($\text{HR} = 1.34$).
    
    \item \textbf{Active Zone Minutes (AZM):} Weekly Active Zone Minutes (AZM) were modeled using a continuous piecewise linear log-hazard spline to evaluate the dose-response relationship with incident cardiometabolic disease \citep{kany2024associations}. The lowest activity bracket ($< 115$ minutes/week) served as the reference category ($\text{HR} = 1.0$). For activity levels between $115$ and $500$ minutes/week, risk reductions scaled log-linearly across successive threshold segments, dropping to an HR of $0.76$ at $245$ minutes/week, $0.71$ at $418$ minutes/week, and reaching a maximum benefit of $0.68$ at $500$ minutes/week. Beyond $500$ minutes/week, the risk profile plateaued, with no additional clinical benefit assumed ($\text{HR} = 0.68$).
\end{itemize}

\subsection{Clinical CKM Biomarkers}

\begin{itemize}
    \item \textbf{Total Cholesterol / HDL-C Ratio (TC/HDL):} Modeled as a piecewise constant step function associated with incident cardiometabolic disease\citep{emerging2009major,gu2025total,tohidi2010lipid}. A clinically optimized ratio of $< 3.5$ was designated as the reference category ($\text{HR} = 1.0$). For ratios exceeding this baseline, risk increased in discrete, step-wise tiers (see Table~\ref{tab:tc_hdl_risk}).
    
    \item \textbf{Insulin Resistance (HOMA-IR):} HOMA-IR was modeled using a piecewise linear log-hazard relationship to evaluate its dose-response association with incident cardiometabolic disease\citep{gast2012insulin}. A HOMA-IR value of $2.0$ was designated as the clinical reference standard ($\text{HR} = 1.0$). For values below the saturation threshold ($< 5.0$), risk scales continuously, corresponding to an $18\%$ increase in cardiometabolic risk per unit increase in HOMA-IR ($\text{HR} = 1.18$ per unit). At a HOMA-IR value of $5.0$, the risk trajectory transitions continuously into a right-censored plateau, capping the maximum relative risk penalty at an HR of $1.64$ for all values exceeding this threshold.
    
    \item \textbf{Kidney Function (eGFR):} eGFR was parameterized as a piecewise constant step function associated with incident cardiometabolic disease\citep{wang2014kidney,lees2019glomerular}. Normal to high kidney function ($> 90\text{ mL/min/1.73m}^2$) was designated as the reference standard ($\text{HR} = 1.0$). For lower eGFR values, cardiometabolic risk stepped up progressively within discrete clinical strata (see Table~\ref{tab:egfr_risk}).
\end{itemize}

\begin{table}[htbp]
\centering
\caption{Total Cholesterol / HDL-C Ratio (TC/HDL) Risk Stratification}
\label{tab:tc_hdl_risk}
\begin{tabular}{llc}
\toprule
\textbf{Clinical Risk Tier} & \textbf{TC/HDL Range} & \textbf{Hazard Ratio (HR)} \\
\midrule
Reference Standard & $< 3.5$ & $1.00$ \\
Borderline High & $3.5$ -- $4.0$ & $1.14$ \\
High Risk & $4.0$ -- $5.0$ & $1.46$ \\
Severely Elevated & $> 5.0$ & $1.86$ \\
\bottomrule
\end{tabular}
\end{table}

\begin{table}[htbp]
\centering
\caption{Estimated Glomerular Filtration Rate (eGFR) Risk Stratification}
\label{tab:egfr_risk}
\begin{tabular}{llc}
\toprule
\textbf{Clinical Stratum} & \textbf{eGFR Range ($\text{mL/min/1.73m}^2$)} & \textbf{Hazard Ratio (HR)} \\
\midrule
Normal to High & $> 90$ & $1.00$ \\
Mild Reduction & $75$ -- $89$ & $1.09$ \\
Mild-to-Moderate Insufficiency & $60$ -- $74$ & $1.10$ \\
Moderate-to-Severe Impairment & $30$ -- $59$ & $1.31$ \\
Severe Impairment / ESRD & $< 30$ & $2.20$ \\
\bottomrule
\end{tabular}
\end{table}

\subsubsection{Compounding Mathematical Saturation and Architectural Health Age Plateaus}
Because each individual biomarker within our frameworks incorporates discrete plateaus, right-censored saturation thresholds, or bounded reference ranges, the resulting aggregate Health Age metrics naturally exhibit structural ceilings on maximum age reduction. These ceilings differ across our three proposed models based on the compounding effects of their underlying parameterizations:

\begin{itemize}
    \item \textbf{The Wearable-Based Framework:} In this model, an individual can only achieve the maximum possible age reduction (capped at 5.13 years younger than chronological age) if they simultaneously optimize all four digital inputs. This requires maintaining an RHR under 60 bpm (reference risk level), exceeding 9,000 daily steps (capping the step HR at 0.88), tracking between 6.0 and 9.0 hours of sleep per night (avoiding U-shaped risk penalties), and logging 500 or more weekly Active Zone Minutes (capping the AZM HR at 0.68). Because further behavioral extensions (e.g., 20,000 steps/day or 1,000 AZM/week) hit explicit log-hazard plateaus rooted in epidemiological saturation points, the wearable-only model mathematically prevents an individual from bypassing this lifestyle-driven rejuvenation floor\citep{master2022association}.
    
    \item \textbf{The Clinical CKM-Based Framework:} This framework dictates a separate biological ceiling by evaluating metabolic and renal stability. Maximum age reduction is achieved when an individual maintains an optimized cholesterol profile ($\text{Chol/HDL} < 3.5$), optimal kidney function ($\text{eGFR} > 90 \text{ mL/min/1.73m}^2$), and baseline insulin sensitivity ($\text{HOMA-IR} < 2.0$). Because these clinical markers are modeled using piecewise constant step functions or right-censored log-hazard relationships, achieving sub-threshold values yields no additional mathematical weight. If assuming the minimum HOMA-IR is 0.35 then the minimum CKM Health Age is 2.73 years younger.
    
    \item \textbf{The Combined Wearable-CKM Framework:} When these two distinct data verticals are integrated, the mathematical saturation points compound non-linearly. An individual who exhibits completely optimized behavioral markers (maximum wearable risk reduction) alongside pristine clinical biomarkers (maximum CKM risk reduction) hits the definitive algorithmic floor of 8.4 years younger (hypothetically assuming $\text{HOMA-IR} = 0$) than chronological age. This combined cap represents the absolute biological saturation point of our system architecture. It reflects the clinical reality that while the synergistic alignment of an optimal lifestyle and ideal blood chemistry maximizes homeostatic resilience and minimizes CKM risk, it ultimately plateaus against the intrinsic, unmodifiable baseline of chronological cellular aging.
\end{itemize}

\subsubsection{Multivariate regression for collinearity in Health Age}
To account for potential confounding and multicollinearity among the continuous digital health metrics—Active Zone Minutes (AZM), daily step count, resting heart rate (RHR), and sleep duration—we adapted a generalized least squares (GLS) summary-statistic framework to approximate mutually adjusted hazard ratios from univariable effects. First, the baseline continuous features were transformed into clinical effect scores based on literature-derived non-linear hazard functional forms. Univariable log-hazard ratios ($\beta_{\text{uni}} = \ln(\text{HR})$) and their empirical sample standard deviations ($\sigma_i$) were established for each feature. Because evaluating highly correlated digital biomarkers simultaneously in a standard regression model can inflate variance and bias estimates, we reconstructed the joint variance-covariance structure ($\mathbf{\Sigma}$) via matrix multiplication of the diagonal standard deviation matrix ($\mathbf{D}$) and the empirical sample correlation matrix ($\mathbf{R}$):

\begin{equation}
\mathbf{\Sigma} = \mathbf{D} \mathbf{R} \mathbf{D}^T
\end{equation}

Where $\mathbf{D} = \text{diag}(\sigma_1, \sigma_2, \dots, \sigma_n)$. Mutual adjustment was then achieved by solving the system of linear equations mapping the crude variance-weighted coefficients to the joint covariance structure:

\begin{equation}
\beta_{\text{adjusted}} = \mathbf{\Sigma}^{-1} \mathbf{D}^2 \beta_{\text{uni}}
\end{equation}

The resulting adjusted beta coefficients ($\beta_{\text{adjusted}}$) were exponentiated ($\exp(\beta_{\text{adjusted}})$) to yield multivariable-corrected hazard ratios, effectively isolating the independent contribution of each digital biomarker while eliminating collinearity-induced bias.

\subsection{Metabolic Syndrome Severity Score (MSSS) Calculation}
To quantify continuous metabolic risk and establish a metabolic ground truth baseline, we calculated a continuous Metabolic Syndrome Severity Score (MSSS) for each participant. Traditional clinical definitions of metabolic syndrome rely on a binary threshold (e.g., meeting at least three out of five discrete risk criteria), which lacks the sensitivity to track incremental physiological changes or subclinical risk gradients. Following the established, validated methodology outlined by Gurka et al.\citep{gurka2014examination,deboer2015severity},we utilized an equation that treats metabolic syndrome severity as a continuous latent variable derived via confirmatory factor analysis (CFA) from five primary clinical components: Body Mass Index (kg/m2), systolic blood pressure (mmHg), fasting plasma glucose (mg/dL), triglycerides (mg/dL), and high-density lipoprotein (HDL) cholesterol (mg/dL).

Because the relative contribution and manifestation of these risk factors vary significantly across demographic groups, the MSSS operates mathematically as a standard Z-score (mean = 0, standard deviation = 1) that is strictly stratified by biological sex and racial/ethnic sub-populations. For each participant, regression coefficients were assigned according to their self-reported sex (male or female) and ethnicity (Non-Hispanic Black, Hispanic, or Non-Hispanic White/General baseline) to calculate the continuous score. Higher Z-scores correspond to higher percentiles of metabolic severity relative to the U.S. population baseline, with a score above 0 signifying greater risk than the average national adult. Participants with missing values in any of the five requisite component metrics were excluded from the analysis to ensure complete-case data integrity.

\subsubsection{Modified American Heart Association Life’s Essential 8 (AHA LE8) Score}
To evaluate overall cardiovascular health and provide a clinical baseline validation for our Health Age models, we adapted the American Heart Association’s (AHA) Life’s Essential 8 framework\citep{lloyd2022lifes}.  The standardized LE8 construct is a comprehensive metric scoring individual cardiovascular health from 0 to 100 based on four health behaviors (diet, physical activity, nicotine exposure, and sleep) and four health factors (BMI, blood lipids, blood glucose, and blood pressure).
Because comprehensive, standardized nutritional survey data was unavailable for this cohort, we implemented a modified 7-metric version of the LE8 framework (effectively reflecting the core clinical updates of LE8 while computing the composite score across the remaining seven components). Missing components are handled by taking the unweighted mean of all available, valid individual category scores to produce a final normalized composite score ranging from 0 to 100.
\begin{itemize}
    \item \textbf{Physical Activity:} Scored using continuous wearable data. Weekly weighted moderate-to-vigorous physical activity was approximated by the average of the participant's daily Active Zone Minutes (AZM) for a week. Maximum points ($100$) were awarded for $\ge 150\text{ minutes/week}$, scaling down linearly to $0\text{ points}$ for $0\text{ active minutes}$.
    
    \item \textbf{Sleep Health:} Derived from continuous wearable sleep metrics. The daily sleep duration mean was converted to hours, with optimal sleep ($7 \le \text{hours} \le 9$) scoring $100\text{ points}$, and extreme short sleep ($< 4\text{ hours}$) or long sleep ($> 10\text{ hours}$) scoring $0\text{ points}$.
    
    \item \textbf{Nicotine Exposure:} Derived via self-reported surveys. Active smokers or those with nicotine exposure were scored at $0\text{ points}$, while non-smokers received $100\text{ points}$.
    
    \item \textbf{Body Mass Index (BMI):} Calculated from clinical measurements ($\text{kg/m}^2$). Ideal BMI ($< 25\text{ kg/m}^2$) was awarded $100\text{ points}$, scaling downward to $0\text{ points}$ for severe obesity ($> 40\text{ kg/m}^2$).
    
    \item \textbf{Blood Pressure:} Scored using a combination of clinical systolic/diastolic blood pressure readings and self-reported hypertension treatment status. Optimal untreated blood pressure ($< 120 / < 80\text{ mmHg}$) received $100\text{ points}$, while a penalty was applied to controlled blood pressure under active antihypertensive treatment (maximum of $80\text{ points}$).
    
    \item \textbf{Blood Lipids:} Evaluated using clinical Non-HDL cholesterol ($\text{mg/dL}$) alongside self-reported hyperlipidemia treatment status. Untreated optimal levels ($< 130\text{ mg/dL}$) scored $100\text{ points}$, with a treatment penalty ceiling of $80\text{ points}$ applied to controlled lipid profiles under medication.
    
    \item \textbf{Blood Glucose:} Scored using Glycated Hemoglobin ($\text{HbA1c}$) thresholds and self-reported clinical diabetes status. For individuals without diabetes, an $\text{HbA1c} < 5.7\%$ scored $100\text{ points}$. For individuals with diagnosed diabetes, scoring was adjusted dynamically based on glycemic control, with a maximum score of $40\text{ points}$ ($\text{HbA1c} < 7.0\%$) down to $10\text{ points}$ for poorly controlled states ($\text{HbA1c} > 9.0\%$).
\end{itemize}